\RequirePackage{fix-cm}

\documentclass[preprint]{ptephy_v1} 

\usepackage{graphicx}
\usepackage{hyperref}
\usepackage[english]{babel} 
\usepackage{amssymb}
\usepackage{siunit}

\usepackage{txfonts}
\usepackage{mathdots}
\begin{document}

\title{Theory of $d$-wave high temperature superconductivity in the cuprates involving non-linear lattice modes}





\author{B. S. Lee}
\affil{School of Physics, Universiti Sains Malaysia, 11800 USM, Penang, Malaysia \email{leebsim@gmail.com}}

\author{T. L. Yoon}
\affil{School of Physics, Universiti Sains Malaysia, 11800 USM, Penang, Malaysia \email{tlyoon@usm.my}}

\author{R. Abd-Shukor}
\affil{School of Applied Physics, Universiti Kebangsaan Malaysia (National University of Malaysia), 43600 Bangi, Selangor, Malaysia
\email{ras@ukm.edu.my}}



\date{Received: date / Accepted: date}

\begin{abstract}
The transition mechanism in high temperature cuprate superconductors is an outstanding puzzle. A previous suggestion on the role of non-linear local lattice instability modes on the microscopic pairing mechanism in high temperature cuprate superconductors \cite{Lee:JSNM09} is re-examined to 
provide a viable mechanism for superconductivity in these cuprates via an unusual lattice vibration in which an electron is predominantly interacting with a nonlinear $Q_2$ mode of the oxygen clusters in the CuO$_2$ planes. It is shown that the interaction has explicit $d$-wave symmetry
and leads to an indirect coupling of $d$-wave symmetry between electrons. As a follow-up of \cite{Lee:JSNM09}, in this paper, we report detailed derivation of the superconducting gap equation and numerical solutions for the transition temperature as inherently integrated into the so-called Extended Hubbard Model (EHM). A unique feature in the EHM is that the transition temperature has an inherent k-dependence. In addition, superconducting gap solutions are restrained to specific regions in the first Brillouin zone (1BZ). It is very feasible to expect that the EHM naturally inherits a huge parameter space in which experimentally measured results, such as the well-known superconducting dome and the phase diagram from electronic Raman scattering \cite{Sacuto:RPP13} can be accommodated. The EHM model hence offers a viable venue to search for or confirm any signature in k-point-sensitive experimental measurements.

\keywords{Superconducting cuprates  \and Jahn--Teller polaron  \and Non-linear localized modes  \and Transition temperature \and  Superconducting gap function}
\end{abstract}

\maketitle

\section{Introduction}\label{intro}
 Since the discovery of high temperature superconductivity in the cuprates in 1986 \cite{Bednorz:ZP86}, there is still no consensus on the microscopic mechanism despite the numerous theories and ideas put forward  \cite{Norman:RPP03}, \cite{Lee:RPP08} (and references therein). The purpose of this paper is to report numerical computations of a theory of $d$-wave superconductivity involving non-linear lattice modes. The non-linear modes are 2D vibrations of the oxygens in the ${\mathrm{CuO}}_2$ plane known as $Q_2$ modes. In these modes, the vibrations of a pair of oxygens along the $x$-axis of a ${\mathrm{CuO}}_2$ cluster are anti-phase to the vibrations of a pair of oxygens along the $y$-axis. Detailed descriptions have been discussed in the earlier paper \cite{Lee:JSNM09}.

We have also shown \cite{Lee_JSNM11} that such planar non-linear vibrations can produce anomalous small magnetic moments \cite{Lee:JSNM14}. The broken rotational symmetry in each ${\mathrm{CuO}}_2$ unit cell due to the generation of small magnetic fields makes our model a candidate to explain intra-unit-cell electronic nematicity measured recently \cite{Lawler:Nature10}. In summary, in addition to an explanation of the pseudogap and Fermi Arcs \cite{Hufner_RPP08}, our electron lattice model is able to explain, compared to the RVB \cite{ANDERSON:Science87} and t-J model theories \cite{Ogata:RPP08}, quantum critical point fluctuation theories \cite{Varma:RPP16}, the following experimental observations: 

\begin{enumerate}
\item  The electron-lattice model involving a predominant 2-D lattice mode is consistent with the unusual or surprising observation of a ubiquitous single phonon energy in the photo-emission measurements of a large number of samples \cite{Lanzara:Nature01}. We note that the interpretation of that paper tends to favour electron-lattice interaction.

\item  It has been shown \cite{Lee_JSNM11} that such planar non-linear vibrations can produce anomalous small magnetic moments \cite{Lee:JSNM14}.

\item  The broken $p/2$ rotational symmetry in each ${\mathrm{CuO}}_2$ unit cell due to the generation of small magnetic fields makes our model a candidate to explain intra-unit-cell electronic nematicity in the pseudogap phase as has been measured recently \cite{Lawler:Nature10}. A model with fully symmetric electron-lattice interaction will not be able to have this feature. 
\end{enumerate}
Another puzzling feature of high-$T_c$ cuprates comes from electronic Raman scattering measurement as reported in \cite{Sacuto:RPP13}, which suggests that
the density of Cooper pairs is strongly dependent on doping, and for low
doping levels it is confined in k-space where they form ``islands'' around the nodes. We shall show that the model \cite{Lee:JSNM09} by construct provides a 
natural setting in which k-space electronic distribution of the Cooper pairs, such as that observed in \cite{Sacuto:RPP13}, can be understood.

The main purpose of this paper is to report on the superconducting behavior as inherently embedded in the effective model Hamiltonian of \cite{Lee:JSNM09}. We focus attention on this effective Hamiltonian and show that it leads to a $d$-wave gap equation. Numerical solutions of the $d$-wave gap equation have been studied and we report some of the results here.

\section{Model Hamiltonian}
 The basis of the model has been discussed in detail in an earlier paper \cite{Lee:JSNM09}. We shall refer the model as the ``Extended Hubbard Model'' (EHM). We suggested an operating mechanism which we termed as ``hole induced suppression of the Jahn-Teller effect'' for a CuO cluster centered around a Cu ion. This means that when the cuprate is doped to produce holes located in the planar Cu-O bond, the presence of the hole breaks the symmetry of the oxygen coordination which leads to the well-known Jahn-Teller (JT) instability for Cu${}^{2+}$. This causes a hole induced suppression of the JT distortion. When an electron moves to the hole, the conditions for the JT instability are restored and the oxygen clusters will distort to a more stable configuration with lower potential energy. The relaxation energy is quite large and we suggested that this large energy can excite a non-linear mode involving the vibrations of the JT active oxygen modes. It should be noted that our suggestion was based on a qualitative argument involving broken symmetry. It has been shown by other workers studying the problem computationally  that hole induced suppression of JT effect does indeed occur. They termed this suppression as ``anti-Jahn-Teller effect'' \cite{Kamimura_JSNM12}. 

The total model Hamiltonian of the EHM is
\begin{eqnarray}\label{eq1}
H&=&H_{\mathrm{harmonic}}+H_4+H_{\mathrm{ep}} \nonumber \\
&&+\sum_{i,j,\sigma }{t_{ij}c^{\dagger }_{i,\sigma }c_{j,\sigma }}+U\sum_{i,\sigma }{n_{i,\sigma }n_{i,-\sigma }-\mu \sum_{i,\sigma }{n_{i,\sigma }}}.\end{eqnarray}
$n_{i,\sigma }=c^{\dagger }_{i,\sigma }c_{i,\sigma }$ is the electron number operator. The total Hamiltonian is made up of a Hubbard Hamiltonian \cite{Lee:RPP08} with an anharmonic lattice background $H_4$ and the electron-lattice interaction $H_{\mathrm{ep}}$. The last three terms in 
Eq.~\eqref{eq1} represent the standard Hubbard Hamiltonian. The electron-lattice interaction term is written as 
\begin{eqnarray}
\label{eq2}
H_{\mathrm{ep}}=-\sum_{i,\sigma }{gQ_i}c^{\dagger }_{i,\sigma }c_{i,\sigma }.\end{eqnarray}
The coupling parameter is $g$. It is convenient to express $Q_i$ as defined in 
Eq.~\eqref{eq2} in terms of lattice phonon creation (destruction) operators $a^{\dagger }_{\boldsymbol{\mathrm{k}}}\ (a_{\boldsymbol{\mathrm{k}}})$, 
\begin{eqnarray}
\label{eq3}
\nonumber
H_{\mathrm{ep}}=-\frac{1}{\sqrt{N}\ }\sum_{\boldsymbol{\mathrm{k}}}{\sum_{i,\sigma }{g\left(\boldsymbol{\mathrm{k}}\right)}c^{\dagger }_{i,\sigma }c_{i,\sigma }}\left(a_{\boldsymbol{\mathrm{k}},\sigma }+a^{\dagger }_{\boldsymbol{\mathrm{-}}\boldsymbol{\mathrm{k}}\boldsymbol{\mathrm{,-}}\sigma }\right)e^{i\boldsymbol{\mathrm{k}}\cdot {\boldsymbol{\mathrm{R}}}_{{i}}},\end{eqnarray}
where 
\begin{eqnarray}g\left(\boldsymbol{\mathrm{k}}\right)=2g{\left(\frac{\hslash }{8m{\omega }_{\boldsymbol{\mathrm{k}}}}\right)}^{\frac{1}{2}}e_{\alpha }\left(\boldsymbol{\mathrm{k}}\right)\left[{\mathrm{cos} \left(ak_x\right)-\ }{\mathrm{cos} \left(ak_y\right)\ }\right].
\end{eqnarray}
${\omega }_{\boldsymbol{\mathrm{k}}}$ is the phonon frequency, $e_{\alpha }\left(\boldsymbol{\mathrm{k}}\right)$ is the phonon polarization vector along the $\alpha $-axis. ${\bf {\mathrm R}}_i$ is the vector position at site $i$. $g(\boldsymbol{\mathrm{k}})$ has the classic $d$-wave symmetry property, 
\begin{eqnarray}\nonumber
\label{eq4}
g\left(\boldsymbol{\mathrm{k}}\right)&=&0\ \ \mathrm{for}\ k_x=\pm k_y; \\
g\left(0\right)&=&0.
\end{eqnarray}
$t_{ij}$ is the hopping or tunneling parameter for an electron from site $i$ to site $j$, $U$ is the Hubbard on-site repulsion energy and $\mu $ is the chemical potential. The harmonic lattice background is given by $H_{\mathrm{harmonic}}$ and $H_4$ is the fourth order anharmonic term. We choose for simplicity 
\begin{eqnarray}
\label{eq5}
H_4=\frac{1}{4}A\sum_i{Q^{\dagger }_i},
\end{eqnarray}
where $A$ is the fourth order anharmonicity parameter. Detailed arguments for an effective Hamiltonian have been discussed in our earlier paper \cite{Lee:JSNM09}. The following effective extended Hubbard Hamiltonian $H\ensuremath{'}$ is obtained from 
Eq.~\eqref{eq1} via canonical transformation
\begin{eqnarray}
H\ensuremath{'}&=&e^SHe^{-S}, \nonumber \\
S&=&\frac{1}{\sqrt{N}\ }\sum_{\boldsymbol{\mathrm{k}}}{\sum_{i,\sigma }{\frac{g\left(\boldsymbol{\mathrm{k}}\right)}{\hslash {\omega }_{\boldsymbol{\mathrm{k}}}}}c^{\dagger }_{i,\sigma }c_{i,\sigma }}\left(a_{\boldsymbol{\mathrm{k}},\sigma }-a^{\dagger }_{\boldsymbol{\mathrm{-}}\boldsymbol{\mathrm{k}}\boldsymbol{\mathrm{,-}}\sigma }\right)e^{i\boldsymbol{\mathrm{k}}\cdot {\boldsymbol{\mathrm{R}}}_i},\nonumber
\end{eqnarray}
\begin{eqnarray}
\label{eq6}
H'&=&
\underbrace{\sum_{i,j,\sigma }{t_{ij}c^{\dagger }_{i,\sigma }c_{j,\sigma }}}_{H\ensuremath{'}_1}
+\underbrace{U\sum_{i,\sigma }{n_{i,\sigma }n_{i,-\sigma }}}_{H\ensuremath{'}_2}
\underbrace{-\frac{1}{2}\sum_{\left\langle ij\right\rangle ,\sigma {,\sigma }\ensuremath{'}}{V_{ij}n_{i,\sigma }n_{j,{\sigma }\ensuremath{'}}}}_{H\ensuremath{'}_3} \nonumber \\
&&
\underbrace{-\sum_{i,\sigma }{G_in_{i,\sigma }}}_{H\ensuremath{'}_4}
\underbrace{-\mu \sum_{i,\sigma }{n_{i,\sigma }}}_{H\ensuremath{'}_5}.
\end{eqnarray}
Note that we have dropped the background harmonic term $H_{\mathrm{harmonic}}$ from $H'$ as it is not directly involved in the dynamics of superconductivity in the EHM. We treat the strong electron correlation problem as in the Hubbard Model. In the Hubbard model, all repulsive interactions between electrons are projected out except the on-site repulsion for double occupation. The third term represents effective electron-electron interaction. We restrict the interaction $V_{ij}$ to nearest neighbours for simplicity. Explicit expression of $V_{ij}$ is referred to Eq.~\ref{A2} in Section 
\ref{asecA0} of the Appendix. The fourth term, 
\begin{eqnarray}
\label{eq7}
H\ensuremath{'}_4\backsimeq -\sum_i{G_i{c\ }^{\dagger }_{i,\sigma }c_{i,\sigma }},
\end{eqnarray} 
represents phenomenologically the energy of polaron formation caused by the interaction of an electron with the non-linear mode. It is assumed that $G_i$ in $H\ensuremath{'}_4$
and its Fourier transform take the following form 

\begin{eqnarray}
\label{eq8}
G_i&=&\sum_{\boldsymbol{\mathrm{k}}}{G_{\boldsymbol{\mathrm{k}}}e^{i\boldsymbol{\mathrm{k}}\cdot {\boldsymbol{\mathrm{R}}}_{{{i}}}}},\nonumber \\ 
G_{\boldsymbol{\mathrm{k}}}&=&\frac{2A\left\langle Q_{\boldsymbol{\mathrm{k}}}\right\rangle g^2(\boldsymbol{\mathrm{k}})}{{\sqrt{N}g\hslash \omega }_{\boldsymbol{\mathrm{k}}}}.
\end{eqnarray}
Here, $\left\langle Q_{\boldsymbol{\mathrm{k}}}\right\rangle \mathrm{\ }$is a thermal average and defined as 
\begin{eqnarray}\nonumber
\left\langle Q_{\boldsymbol{\mathrm{k}}}\right\rangle =\frac{1}{\sqrt{N}}\sum_i{\left\langle Q^3_i\right\rangle e^{i\boldsymbol{\mathrm{k}}\cdot {\boldsymbol{\mathrm{R}}}_{{{i}}}}}.\end{eqnarray} 
We treat this thermal average as a parameter at this stage of approximation. We have approximated the contributions from anharmonicity by retaining only terms linear in $c^{\dagger }c$. Higher order terms in $c^{\dagger }c$ will be neglected in our first attempt to evaluate the anharmonic effects. Hence it must be noted that the parameter for effective electron-electron interaction includes both harmonic and anharmonic interactions. In the fifth term in 
Eq.~\eqref{eq6}, the chemical potential $\mu $ is determined by the doping level. 

If we examine the terms in the canonically transformed Hamiltonian, 
Eq.~\eqref{eq6}, and compare them with the original terms added to a standard Hubbard Hamiltonian, 
Eq.~\eqref{eq1}, we can justify writing an extended Hubbard Hamiltonian to investigate the thermodynamic properties of the system. We note that there has been a considerable amount of effort \cite{Micnas:RMP90} in the use of extended Hubbard Hamiltonians to study superconductivity in narrow band systems with local attractive interactions. What is new in our study is the role of non-linear lattice modes in the polaron formation. We have shown that this enables the model to produce anomalous magnetic moments and show nematicity in the electron properties in the pseudogap phase \cite{Lee_JSNM11}.

$G_{\boldsymbol{\mathrm{k}}}$ in 
Eq.~\eqref{eq8} possesses $d$-symmetry properties. It could be the JT stabilization or relaxation energy as described above which is quite large. Physically, we can expect the JT polaron to be trapped if thermal excitation $(\sim kT)$ is not sufficient to excite the non-linear mode to free the electron. The electron can move to another site by tunneling. Alternatively, the JT relaxation energy released by a nearby site could possibly excite a non-linear vibration to release the electron. This suggested physical scenario of charge carriers interacting with non-linear modes to form a polaron is different from the well-developed theory of polarons and bipolarons in a harmonic lattice \cite{Mott:WS96}. 

The EHM by construct accommodates a mechanism (the JT $Q_2$ mode) that explains pseudogap phenomena in a natural setting without invoking exotic physics. Our first paper on EHM \cite{Lee:JSNM09} places emphasis on elucidating the pseudogap in the light of the EHM. The 
model can also generate superconducting gaps, as in the case of generic Hubbard models and variants, due to the presence of on-site hoping and repulsive mechanism intrinsic to this class of models. In addition, as will be explained in the  following sections, transition temperature solutions are integrally tied to the dopant concentration (which creates holes in the cuprates) in the EHM. The solutions for transition temperature and supercondicting gap embedded in the model have non-trivial and rich structures. In the following we shall explore the characteristics of high-$T_c$ superconductivity phenomena embedded in the EHM.

\section{Derivation of superconducting gap equation }
We note that the structure of EHM is quite similar (but not entirely so) to the work by \cite{Micnas:PRB88}. To begin with, 
Eq.~\eqref{eq6} is cast into the following form,
\begin{eqnarray}
\label{eq9}
H&=&\sum_{\boldsymbol{\mathrm{k}},\sigma }{\left({\epsilon}_{\mathrm{\bf k}}-\mu \right)c^{\dagger }_{\boldsymbol{\mathrm{k}},\sigma }c_{\boldsymbol{\mathrm{k}},\sigma }} \nonumber \\ 
&& -\frac{1}{2N}\sum_{\sigma ,{\sigma }\ensuremath{'}}{\sum_{\boldsymbol{\mathrm{k}},{\boldsymbol{\mathrm{k}}}\ensuremath{'},\boldsymbol{\mathrm{q}}}{\left(V_{\boldsymbol{\mathrm{q}}}-{2U\delta }_{\sigma ,{-\sigma }\ensuremath{'}}\right)c^{\dagger }_{\boldsymbol{\mathrm{k}} 
+\boldsymbol{\mathrm{q}},\sigma }c_{\boldsymbol{\mathrm{k}},\sigma }}c^{\dagger }_{{\boldsymbol{\mathrm{k}}}\ensuremath{'}-\boldsymbol{\mathrm{q}}{,\sigma }\ensuremath{'}}c_{{\boldsymbol{\mathrm{k}}}\ensuremath{'},{\sigma }\ensuremath{'}}} \nonumber \\
& &- 
\frac{1}{\sqrt{N}}\sum_{\sigma }{\sum_{\boldsymbol{\mathrm{k}},\boldsymbol{\mathrm{q}}}{G_{\boldsymbol{\mathrm{q}}}c^{\dagger }_{\boldsymbol{\mathrm{k}}-\boldsymbol{\mathrm{q}},\sigma }c_{\boldsymbol{\mathrm{k}},\sigma }}}.
\end{eqnarray}
We have dropped the prime superscript on $H'$. Details of casting 
Eq.~\eqref{eq6}
into 
Eq.~\eqref{eq9} are referred to subsection~\ref{asecA1} 
in the Appendix. $\epsilon_{\mathrm{\bf k}}$ is the Fourier transform of $t_{ij}$. To facilitate the derivation of superconducting gap equation from the model Hamiltonian, we begin with a trial Hamiltonian $H_t$,
\begin{eqnarray}
\label{eq10}
H_t&=&\sum_{\boldsymbol{\mathrm{k}},\sigma }{\left({\epsilon }_{\boldsymbol{\mathrm{k}}}-\mu -A_{\boldsymbol{\mathrm{k}}}\right)c^{\dagger }_{\boldsymbol{\mathrm{k}},\sigma }c_{\boldsymbol{\mathrm{k}},\sigma }} \nonumber \\
&& + \frac{1}{2}\sum_{\boldsymbol{\mathrm{k}},\sigma }{\left(B_{\boldsymbol{\mathrm{k}}}c^{\dagger }_{\boldsymbol{\mathrm{k}},\sigma }c^{\dagger }_{-\boldsymbol{\mathrm{k}},-\sigma }+B^*_{\boldsymbol{\mathrm{k}}}c_{-\boldsymbol{\mathrm{k}},-\sigma }c_{\boldsymbol{\mathrm{k}},\sigma }\right),}
\end{eqnarray} 
where two variational parameters are introduced, $A_{\boldsymbol{\mathrm{k}}}$ for pseudogap and $B_{\boldsymbol{\mathrm{k}}}$ for singlet superconductivity. To derive superconducting gap equation, we first diagonalize the trial Hamiltonian, then carry out variational calculation on the trial free energy with respect to the superconductivity variational parameter $B_{\boldsymbol{\mathrm{k}}}$ by neglecting $A_{\boldsymbol{\mathrm{k}}}$ (which is not related to superconductivity). The resultant diagonalized trial Hamiltonian is written in the form 
\begin{eqnarray}\label{eq11}
H_t&=&\sum_{\boldsymbol{\mathrm{k}}}{{\lambda }_{\boldsymbol{\mathrm{k}}}{\alpha }^{\dagger }_{\boldsymbol{\mathrm{k}},\sigma }{\alpha }_{\boldsymbol{\mathrm{k}},\sigma }}+C, \nonumber \\
C&=&\sum_{\boldsymbol{\mathrm{k}}}{E_{\boldsymbol{\mathrm{k}}}}-\sum_{\boldsymbol{\mathrm{k}}}{{\lambda }_{\boldsymbol{\mathrm{k}}}.}
\end{eqnarray}
${\alpha }_{\boldsymbol{\mathrm{k}},\sigma },{\alpha }^{\dagger }_{\boldsymbol{\mathrm{k}},\sigma }$ are Bogolyubov-Valatin transformation operators of $c_{\boldsymbol{\mathrm{k}},\sigma },c^{\dagger }_{\boldsymbol{\mathrm{k}},\sigma }$, as defined in Eq.~\eqref{A19}
in the Appendix. The eigenvalue ${\lambda }_{\boldsymbol{\mathrm{k}}}$ of $H_t$ is given by
\begin{eqnarray}\label{eq12}
{\lambda }_{\boldsymbol{\mathrm{k}}}=\sqrt{E^2_{\boldsymbol{\mathrm{k}}}+B^2_{\boldsymbol{\mathrm{k}}}},
\end{eqnarray} 
\begin{eqnarray}\label{eq13}
E_{\boldsymbol{\mathrm{k}}}={\epsilon }_{\boldsymbol{\mathrm{k}}}-\mu -A_{\boldsymbol{\mathrm{k}}}.
\end{eqnarray} 
Diagonalization of the trial Hamiltonian into the form 
Eq.~\eqref{eq11}, and derivation of the corresponding eigenvalue [Eq.~\eqref{eq12}], are detailed in subsection~\ref{asecA2}
in the Appendix. ${\epsilon }_{\boldsymbol{\mathrm{k}}}$ in 
Eq.~\eqref{eq13}, which originates from the kinetic energy (hopping term) $t_{ij}$ in the Hubbard Hamiltonian, can be approximated by tight-binding band energy,
\begin{eqnarray}
\label{eq14}
{\epsilon }_{\boldsymbol{\mathrm{k}}}=-2t\left[{\mathrm{cos} \left(ak_x\right)+\ }{\mathrm{cos} \left(ak_y\right)\ }\right],\end{eqnarray} 
where $t$ is the nearest neighbor hopping parameter, while $a$ is the lattice parameter of the CuO lattice. This type of notation is used for Hubbard model, t-J model, etc., and reflects the physics of electron hopping from site to site. 

To obtain the free energy $F$, we apply variational principle and write 
\begin{eqnarray}
\label{eq15}
F\le F_v&=&F_t+{\left\langle H-H_t\right\rangle }_t+\mu N_e \nonumber \\
&=&F_t+{\left\langle H\right\rangle }_t-{\left\langle H_t\right\rangle }_t+\mu N_e, \end{eqnarray} 
where $N_e=nN$, $n$ charge carrier density, and $N$ is the total number of lattice sites. The trial free energy $F_t$ is defined in terms of the partition function $Z_t$ for the trial Hamiltonian $H_t$,   
\begin{eqnarray}
\label{eq16}
Z_t&=&\mathrm{Tr}\left(e^{-\beta H_t}\right), \nonumber \\
F_t&=&-k_BT{\mathrm{ln} Z_t},\ \end{eqnarray} 
where $\beta $ is the inverse temperature $\beta =\frac{1}{k_BT}$, and $k_B$ Boltzmann constant. ${\left\langle H\right\rangle }_t,{\left\langle H_t\right\rangle }_t$ which are defined as expectation values in the trial basis set, can be derived using the expectation value of an arbitrary operator in the same basis, as prescribed in statistical mechanics, 
\begin{eqnarray}
\label{eq17}
\mathrm{\ }{\left\langle O\right\rangle }_t=\frac{\mathrm{Tr}\left(Oe^{-\beta H_t}\right)}{\mathrm{Tr}\left(e^{-\beta H_t}\right)}.\end{eqnarray} 
The full expressions of $Z_t,F_t,\ {\left\langle H\right\rangle }_t,{\left\langle H_t\right\rangle }_t$ and $F_v$ are derived in Appendix~\ref{asecA3}.
In particular, we have [from Eq.~\eqref{A58}]
\begin{eqnarray}
\label{eq18}
F_v&=&-\frac{1}{\beta }\sum_{\boldsymbol{\mathrm{k}},\sigma }{{\mathrm{ln\ } \left[\mathrm{2cosh}\left(\frac{\beta {\lambda }_{\boldsymbol{\mathrm{k}}}}{2}\right)\right]\ }}
\nonumber \\&&-\frac{1}{8N}\sum_{\sigma ,\boldsymbol{\mathrm{k}}}
\sum_{\boldsymbol{\mathrm{q}}}
\left(V_{\boldsymbol{\mathrm{q}}\boldsymbol{-}\mathrm{k}}-\mathrm{2}U\right)
\frac{B_{\boldsymbol{\mathrm{q}}}B_{\boldsymbol{\mathrm{k}}}}{{\lambda }_{\boldsymbol{\mathrm{q}}}{\lambda }_{\boldsymbol{\mathrm{k}}}} \cdot
\nonumber \\ &&
\,\,\,\,\,\, {\mathrm{tanh} \left(\frac{\beta {\lambda }_{\boldsymbol{\mathrm{q}}}}{2}\right)\ }
{\mathrm{tanh} \left(\frac{\beta {\lambda }_{\boldsymbol{\mathrm{k}}}}{2}\right)\ }
\nonumber \\ && +\frac{1}{2}\sum_{\boldsymbol{\mathrm{k}},\sigma }{\frac{B^2_{\boldsymbol{\mathrm{k}}}}{{\lambda }_{\boldsymbol{\mathrm{k}}}}{\mathrm{tanh} \left(\frac{\beta {\lambda }_{\boldsymbol{\mathrm{k}}}}{2}\right)\ }}+ \mathrm{ terms\ independent\ of\ }B^2_{\boldsymbol{\mathrm{k}}}. \nonumber \\
\end{eqnarray} 
Minimizing $F_v$ with respect to $B_{\boldsymbol{\mathrm{k}}}$, 
\begin{eqnarray}
\label{eq19}
\nonumber
\frac{\partial F_v}{\partial B_{\boldsymbol{\mathrm{k}}}\mathrm{\ }}=0,\end{eqnarray} 
results in the following equality for $B_{\boldsymbol{\mathrm{k}}}$, 
\begin{eqnarray}B_{\boldsymbol{\mathrm{k}}}=\frac{1}{4N}\sum_{\boldsymbol{\mathrm{q}}}{\left(V_{\boldsymbol{\mathrm{k}}-\boldsymbol{\mathrm{q}}}-2U\right)B_{\boldsymbol{\mathrm{q}}}\frac{{\mathrm{tanh} \left(\frac{\beta {\lambda }_{\boldsymbol{\mathrm{q}}}}{2}\right)\ }}{{\lambda }_{\boldsymbol{\mathrm{q}}}}}.\end{eqnarray} 
Details of the minimization procedure to arrive at 
Eq.~\eqref{eq19} 
is reported in Appendix~\ref{asecA4}.
Note that in arriving at 
Eq.~\eqref{eq19}, the variable $A_{\boldsymbol{\mathrm{k}}}$ is ignored because we wish to concentrate only on the superconductivity sector of the current model. Replacing the symbol $B_{\boldsymbol{\mathrm{k}}}\to {\mathrm{\Delta }}_{\boldsymbol{\mathrm{k}}}$, where ${\mathrm{\Delta }}_{\boldsymbol{\mathrm{k}}}$ denotes the superconducting energy gap, we arrive at the gap equation for $d$-wave superconductivity, 
\begin{eqnarray}
\label{eq20}
{\mathrm{\Delta }}_{\boldsymbol{\mathrm{k}}}=\sum_{\boldsymbol{\mathrm{q}}}{\left(V_{\boldsymbol{\mathrm{k}}-\boldsymbol{\mathrm{q}}}-2U\right){\mathrm{\Delta }}_{\boldsymbol{\mathrm{q}}}F_{\boldsymbol{\mathrm{q}}}\left(\beta ,\mu \right)},\end{eqnarray} 
\begin{eqnarray}
\label{eq21}
F_{\boldsymbol{\mathrm{q}}}\left(\beta ,\mu \right)=\frac{{\mathrm{tanh} \left(\frac{\beta {\lambda }_{\boldsymbol{\mathrm{q}}}\left(\mu \right)}{2}\right)\ }}{4N{\lambda }_{\boldsymbol{\mathrm{q}}}},\end{eqnarray} 
\begin{eqnarray}
\label{eq22}
{\lambda }_{\boldsymbol{\mathrm{q}}}\left(\mu \right)=\sqrt{E^2_{\boldsymbol{\mathrm{q}}}\left(\mu \right)+{\mathrm{\Delta }}^2_{\boldsymbol{\mathrm{q}}}},\end{eqnarray} 
\begin{eqnarray}
\label{eq23}
E_{\boldsymbol{\mathrm{q}}}\left(\mu \right)={\epsilon }_{\boldsymbol{\mathrm{q}}}-\mu .\end{eqnarray} 
The $V_{\boldsymbol{\mathrm{k}}}$ term in 
Eq.~\eqref{eq20} is given as (see the derivation of Eq.~\eqref{A4}
in Appendix 
\ref{asecA0}),
\begin{eqnarray}
\label{eq24}
V_{\boldsymbol{\mathrm{k}}}=\frac{{g\left(\boldsymbol{\mathrm{k}}\right)}^2}{\hslash {\omega }_{\boldsymbol{\mathrm{k}}}}.\end{eqnarray} 
It is the Fourier conjugate for $V_{ij}$ that appears in 
Eq.~\eqref{eq6}. In the EHM, $V_{\boldsymbol{\mathrm{k}}}$ carries a $d$-wave symmetry which is traced back to the $Q_2$-mode as embedded in $G_{\boldsymbol{\mathrm{k}}}$ term in 
Eq.~\eqref{eq8}. Explicitly, $V_{\boldsymbol{\mathrm{k}}-\boldsymbol{\mathrm{q}}}$, as appeared in 
Eq.~\eqref{eq20}, reads
\begin{eqnarray}
\label{eq25}
V_{\boldsymbol{\mathrm{k}}-\boldsymbol{\mathrm{q}}}&=&\frac{g^2}{2m{\omega }_{\boldsymbol{\mathrm{k}}-\boldsymbol{\mathrm{q}}}}{\left[{\mathrm{cos} \left[a\left(k_x-q_x\right)\right]-\ }\left[a\left(k_y-q_y\right)\right]\right]}^2 \nonumber \\ & \equiv& V{\eta }^2_{\boldsymbol{\mathrm{k}}-\boldsymbol{\mathrm{q}}},\end{eqnarray} 
where we have defined
\begin{eqnarray}
\label{eq26}
{\eta }_{\boldsymbol{\mathrm{k}}}={\mathrm{cos} \left(ak_x\right)-\ }{\mathrm{cos} \left(ak_y\right)\ }.\end{eqnarray} 
In 
Eq.~\eqref{eq25}, we have assumed that $V$ is frequency-independent, i.e.,  
\begin{eqnarray}
\label{eq27}
\ V\equiv \frac{g^2}{2m{\omega }_{\boldsymbol{\mathrm{k}}-\boldsymbol{\mathrm{q}}}}\approx \frac{g^2}{2m{\omega }_0}.\end{eqnarray} 
$V$, along with $U$, are treated as free parameters of the model.  

The superconducting gap equation, Eq.~\eqref{eq20},
essentially agrees with Eq. (2.8) in \cite{Micnas:PRB88} and Eqs. (2), (3) in \cite{Micnas:JPCM02}, who employed Hartree-Fock factoring method to derive a mean field Hamiltonian. With nearest-neighbor interaction and no specific symmetry, \cite{Micnas:PRB88} used an ansatz for the gap function which has on-site extended $s$-wave and $d$-wave symmetries. The same results are arrived at in present work using variational method. However, in our case, we start with an interaction $V_{\boldsymbol{\mathrm{k}}}$ [Eq.~\eqref{eq25}] which has $d$-wave symmetry. 

The number of electrons (charge carriers) per ${\mathrm{Cu}}^{2+}$ lattice site, $n$, by definition, is given by 
\begin{eqnarray}
\label{eq28}
Nn=\sum_{\boldsymbol{\mathrm{k}},\sigma }{{\left\langle c^{\dagger }_{\boldsymbol{\mathrm{k}},\sigma }c_{\boldsymbol{\mathrm{k}},\sigma }\right\rangle }_t.}\end{eqnarray} 
Evaluation of the RHS of 
Eq.~\eqref{eq28}
leads to (for derivation, see Eq.~\eqref{A61} of subsection~\ref{asecA3}
in the Appendix)
\begin{eqnarray}
\label{eq29}
\mathrm{\ }\sum_{\boldsymbol{\mathrm{k}},\sigma }{{\left\langle c^{\dagger }_{\boldsymbol{\mathrm{k}},\sigma }c_{\boldsymbol{\mathrm{k}},\sigma }\right\rangle }_t}=N-\sum_{\boldsymbol{\mathrm{k}}}{\frac{E_{\boldsymbol{\mathrm{k}}}}{{\lambda }_{\boldsymbol{\mathrm{k}}}}}{\mathrm{tanh} \left(\frac{\beta {\lambda }_{\boldsymbol{\mathrm{k}}}}{2}\right)\ }.\end{eqnarray} 
Putting 
Eq.~\eqref{eq28}
and 
Eq.~\eqref{eq29}
together results in a constraint condition imposed by the charge carried concentration $n$, 
\begin{eqnarray}
\label{eq30}
n=1-\frac{1}{N}\sum_{\boldsymbol{\mathrm{k}}}{\frac{E_{\boldsymbol{\mathrm{k}}}}{{\lambda }_{\boldsymbol{\mathrm{k}}}}}{\mathrm{tanh} \left(\frac{\beta {\lambda }_{\boldsymbol{\mathrm{k}}}}{2}\right).\ }\end{eqnarray} 
The concentration of charge carriers $n$ can be experimentally correlated to the hole doping concentration in the cuprate, $x$, so that 
\begin{eqnarray}
\label{eq31}
x=n.\end{eqnarray} 
Eq.~\eqref{eq20}, subjected to the constraint from 
Eq.~\eqref{eq30}, determines the superconductivity transition temperature in the present model. The (inverse) transition temperature, denoted ${\beta }_c(=\frac{1}{k_BT_c})$, at which superconductivity is switched on when temperature is reduced from above, can be predicted by the model Hamiltonian by solving Eq.~\eqref{eq20}. 

\begin{figure}
\includegraphics*[width=2.02in, height=1.98in, keepaspectratio=false]{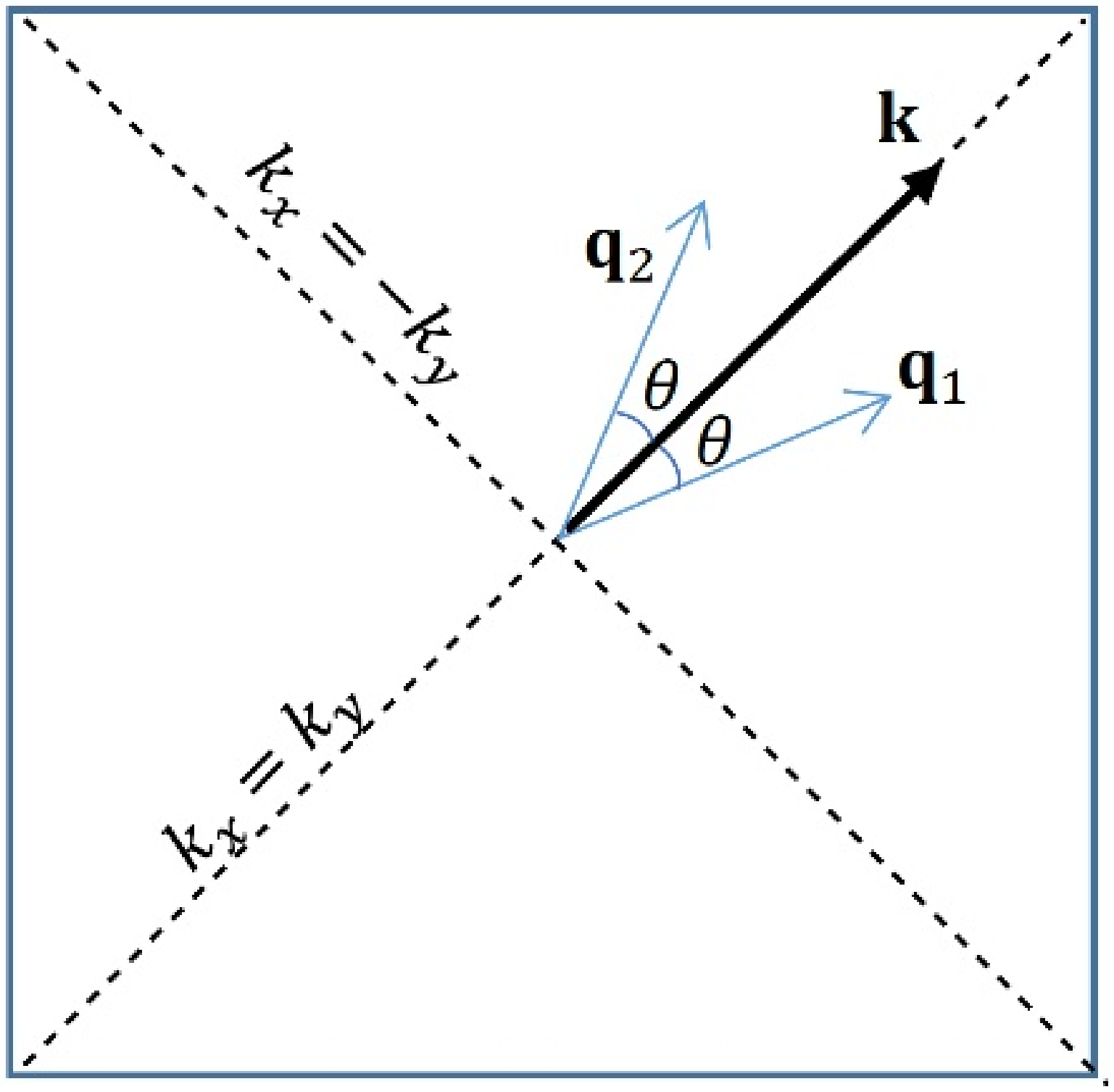}
\caption{A pair of symmetric vectors \textit{${\boldsymbol{q}}_1$} and \textit{${\boldsymbol{q}}_2$} with respect to the diagonal line \textit{$k_x=k_y$} in the 1BZ. \textit{$\left|{\boldsymbol{q}}_1\right|=\left|{\boldsymbol{q}}_2\right|$}. The thick vector $\boldsymbol {\mathrm k}$ denotes an independent, generic vector lying on the diagonal line.}
\label{fig:1}       
\end{figure}

The method of approximation adopted in present work to solve the superconducting gap equation, 
Eq.~\eqref{eq20}, is inspired by \cite{Micnas:PRB88}. To this end, we use an ansatz of the form 
\begin{eqnarray}
\label{eq32}
{\mathrm{\Delta }}_{\boldsymbol{\mathrm{q}}}={\mathrm{\Delta }}_0+{\mathrm{\Delta }}_{\eta }{\eta }_{\boldsymbol{\mathrm{q}}},\end{eqnarray} 
where ${\mathrm{\Delta }}_0,{\mathrm{\Delta }}_{\eta }\mathrm{\ }$are k-point independent. ${\mathrm{\Delta }}_{\boldsymbol{\mathrm{q}}}$ has a $d$-wave symmetry in the limit $U\to 0$. This could be shown as follows. Substitute Eq.~\eqref{eq32}
into 
Eq.~\eqref{eq20}, and employ the fact that in the limit $U\to 0,$ ${\mathrm{\Delta }}_0\to 0$, we obtain  
\begin{eqnarray}
\label{eq33}
{\mathrm{\Delta }}_{\boldsymbol{\mathrm{k}}}={\mathrm{\Delta }}_{\eta }{\eta }_{\boldsymbol{\mathrm{k}}}={\mathrm{\Delta }}_{\eta }V\sum_{\boldsymbol{\mathrm{q}}}{{\eta }^2_{\boldsymbol{\mathrm{k}}-\boldsymbol{\mathrm{q}}}{\eta }_{\boldsymbol{\mathrm{q}}}F_{\boldsymbol{\mathrm{q}}}\left(\beta ,\mu \right)}.\end{eqnarray} 
The ansatz Eq.~\eqref{eq32}
is then inserted into the RHS of Eq.~\eqref{eq33}
to compute a value of ${\mathrm{\Delta }}_{\boldsymbol{\mathrm{k}}}.$ The symmetry property of the computed ${\mathrm{\Delta }}_{\boldsymbol{\mathrm{k}}}$ can be revealed by referring to Fig.~\ref{fig:1}
where a pair of symmetric points, ${\boldsymbol{\mathrm{q}}}_1$ and ${\boldsymbol{\mathrm{q}}}_2$, symmetric about the diagonal line $k_x=k_y$ are considered. We consider a point $\boldsymbol{\mathrm{k}}$ which lies on the diagonal line such that $\boldsymbol{\mathrm{k}}=\left(k{\mathrm{sin} \frac{\pi }{4\ }\ },k{\mathrm{cos} \frac{\pi }{4\ }\ }\right)=\left(\frac{k}{\sqrt{}2\ },\frac{k}{\sqrt{}2\ }\right)$. Hence,
\begin{eqnarray}
\label{eq34}
{\eta }_{{\boldsymbol{\mathrm{k}}}^{\boldsymbol{\mathrm{'}}}}={\mathrm{cos} \left(\frac{ak}{\sqrt{}2}\right)\mathrm{\ }\ }-{\mathrm{cos} \left(\frac{ak}{\sqrt{2}}\right)\ }=0.\end{eqnarray} 
Due to the reflection symmetry of ${\boldsymbol{\mathrm{q}}}_1$ and ${\boldsymbol{\mathrm{q}}}_2$ about the diagonal line, it can be deduced that 
\begin{eqnarray}
\label{eq35}
{\eta }_{{\boldsymbol{\mathrm{q}}}_1}=-{\eta }_{{\boldsymbol{\mathrm{q}}}_2}.\end{eqnarray} 
Replacing ${\boldsymbol{\mathrm{q}}}_i$ by $\boldsymbol{\mathrm{k}}-{\boldsymbol{\mathrm{q}}}_i$, 
\begin{eqnarray}
\label{eq36}
{\eta }_{\boldsymbol{\mathrm{k}}-{\boldsymbol{\mathrm{q}}}_1}={-\eta }_{{\boldsymbol{\mathrm{k}}-\boldsymbol{\mathrm{q}}}_2}.\end{eqnarray} 
From 
Eqs.~\eqref{eq21},~\eqref{eq22},~\eqref{eq33},~\eqref{eq35} and ~\eqref{eq36},
\begin{eqnarray}
\label{eq37}
F_{{\boldsymbol{\mathrm{q}}}_1}={-F}_{{\boldsymbol{\mathrm{q}}}_2}.\end{eqnarray} 
The summation over $\boldsymbol{\mathrm{q}}$ in 
~\eqref{eq33}, namely $\sum_{\boldsymbol{\mathrm{q}}}{{\eta }^2_{\boldsymbol{\mathrm{k}}-\boldsymbol{\mathrm{q}}}{\eta }_{\boldsymbol{\mathrm{q}}}F_{\boldsymbol{\mathrm{q}}}\left(\beta ,\mu \right)}$, will vanish due to pair-wise cancellation in the form of 
\[{\eta }^2_{\boldsymbol{\mathrm{k}}-{\boldsymbol{\mathrm{q}}}_1}{\eta }_{{\boldsymbol{\mathrm{q}}}_1}F_{{\boldsymbol{\mathrm{q}}}_1}\left(\beta ,\mu \right)+{\eta }^2_{\boldsymbol{\mathrm{k}}-{\boldsymbol{\mathrm{q}}}_2}{\eta }_{{\boldsymbol{\mathrm{q}}}_2}F_{{\boldsymbol{\mathrm{q}}}_2}\left(\beta ,\mu \right) \] 
due to Eqs.~\eqref{eq35},~\eqref{eq36},~\eqref{eq37}.
The result is a gap ${\mathrm{\Delta }}_{\boldsymbol{\mathrm{k}}}$ that has $d$-wave symmetry in the limit $U\to 0$,
\begin{eqnarray}
\label{eq38}
{\mathrm{\Delta }}_{\boldsymbol{\mathrm{k}}}=0\ \mathrm{for\ }k_x={\pm k}_y.\end{eqnarray} 
Eq.~\eqref{eq38} is true because in the microscopic model, $V_{\boldsymbol{\mathrm{k}}}$ is proportional to ${\eta }^2_{\boldsymbol{\mathrm{k}}}$. If $V_{\boldsymbol{\mathrm{k}}}$ were to depend linearly on ${\eta }_{\boldsymbol{\mathrm{k}}}$, 
Eq.~\eqref{eq38} will not hold. In comparison, $V_{\boldsymbol{\mathrm{k}}}$ in the phenomenological model by \cite{Micnas:PRB88} is not known, and the above result does not follow directly, despite it is still possible to search for $s$-wave and $d$-wave gap solutions in such models. The present model agrees with that of \cite{Micnas:PRB88} for pure $d$-wave gap solutions, $U=0\Rightarrow {\mathrm{\Delta }}_0=0.$ As far as we are aware of, this is the first time an e-lattice model can show explicitly $d$-wave gap.

%

\section{Solutions for transition temperature and superconducting gap}
Superconductivity transition temperature, $T_c=\frac{1}{k_B{\beta }_c}$, can be deduced from the gap equation in the present model. The existence of the solution to ${\beta }_c$ in the limit $U\to 0$ (hence ${\mathrm{\Delta }}_0=0$) can be proven in a limiting case as discussed below. In the limit $U\to 0$, for $T$ close to $T_c$, ${\mathrm{\Delta }}_{\eta }\ll 1$, $\Rightarrow {\mathrm{\Delta }}_{\boldsymbol{\mathrm{k}}}\approx {\mathrm{\Delta }}_{\eta }{\eta }_{\mathrm {\bf k}}\ll 1.$ We expand $F_{\boldsymbol{\mathrm{q}}}$ in
Eq.~\eqref{eq33} in powers of ${\mathrm{\Delta }}_{\boldsymbol{\mathrm{q}}}$,
\begin{eqnarray}
\label{eq39}
\nonumber
F_{\boldsymbol{\mathrm{q}}}&\approx& \ f_{1\boldsymbol{\mathrm{q}}}+f_{2\boldsymbol{\mathrm{q}}}{\mathrm{\Delta }}^2_{\boldsymbol{\mathrm{q}}}+O\left({\mathrm{\Delta }}^3_{\boldsymbol{\mathrm{q}}}\right),\nonumber  \\
\mathrm{\ }f_{1\boldsymbol{\mathrm{q}}}&=&\frac{{\mathrm{tanh} \left(\frac{\beta E_{\boldsymbol{\mathrm{q}}}}{2}\right)\ }}{4NE_{\boldsymbol{\mathrm{q}}}},\ f_{2\boldsymbol{\mathrm{q}}}=\frac{\left({t^{\mathrm{'}}_{2\boldsymbol{\mathrm{q}}}- \frac{f_{1\boldsymbol{\mathrm{q}}}}{2}\ }\right)}{4NE^3_{\boldsymbol{\mathrm{q}}}},\nonumber \\
t\ensuremath{'}_{2\boldsymbol{\mathrm{q}}}&=&\left[1-{{\mathrm{tanh}}^2 \left(\frac{\beta E_{\boldsymbol{\mathrm{q}}}}{2}\right)\ }\right]\frac{\beta E_{\boldsymbol{\mathrm{q}}}}{4}.\end{eqnarray} 
From 
Eq.~\eqref{eq33} and Eq.~\eqref{eq39},
\begin{eqnarray}
\label{eq40}
\nonumber 
{\mathrm{\Delta }}_{\boldsymbol{\mathrm{k}}}&\approx& {\mathrm{\Delta }}_{\eta }{\eta }_{\boldsymbol{\mathrm{k}}}V\sum_{\boldsymbol{\mathrm{q}}}{{\eta }^2_{\boldsymbol{\mathrm{k}}-\boldsymbol{\mathrm{q}}}{\eta }_{\boldsymbol{\mathrm{q}}}F_{\boldsymbol{\mathrm{q}}}\left(\beta ,\mu \right)}\nonumber \\
&=&{\mathrm{\Delta }}_{\eta }V\sum_{\boldsymbol{\mathrm{q}}}{{\eta }^2_{\boldsymbol{\mathrm{k}}-\boldsymbol{\mathrm{q}}}{\eta }_{\boldsymbol{\mathrm{q}}}f_{1\boldsymbol{\mathrm{q}}}}+{\mathrm{\Delta }}_{\eta }V\sum_{\boldsymbol{\mathrm{q}}}{{\eta }^2_{\boldsymbol{\mathrm{k}}-\boldsymbol{\mathrm{q}}}{\eta }_{\boldsymbol{\mathrm{q}}}f_{2\boldsymbol{\mathrm{q}}}{\mathrm{\Delta }}^2_{\boldsymbol{\mathrm{q}}}}
\nonumber \\
\Rightarrow {\mathrm{\Delta }}_{\eta }{\eta }_{\boldsymbol{\mathrm{k}}}&\approx& {\mathrm{\Delta }}_{\eta }V\sum_{\boldsymbol{\mathrm{q}}}{{\eta }^2_{\boldsymbol{\mathrm{k}}-\boldsymbol{\mathrm{q}}}{\eta }_{\boldsymbol{\mathrm{q}}}f_{1\boldsymbol{\mathrm{q}}}}+{\mathrm{\Delta }}^3_{\eta }V\sum_{\boldsymbol{\mathrm{q}}}{{\eta }^2_{\boldsymbol{\mathrm{k}}-\boldsymbol{\mathrm{q}}}{\eta }^3_{\boldsymbol{\mathrm{q}}}f_{2\boldsymbol{\mathrm{q}}}}\end{eqnarray} 
Solving 
Eq.~\eqref{eq40} for ${\mathrm{\Delta }}_{\eta }$, we get ${\mathrm{\Delta }}_{\ \eta }=0$, or 
\begin{eqnarray}
\label{eq41}
{\mathrm{\Delta }}^2_{\eta }=\frac{{\eta }_k-V\sum_{\boldsymbol{\mathrm{q}}}{{\eta }^2_{\boldsymbol{\mathrm{k}}-\boldsymbol{\mathrm{q}}}{\eta }_{\boldsymbol{\mathrm{q}}}f_{1\boldsymbol{\mathrm{q}}}}}{V\sum_{\boldsymbol{\mathrm{q}}}{{\eta }^2_{\boldsymbol{\mathrm{k}}-\boldsymbol{\mathrm{q}}}{\eta }^3_{\boldsymbol{\mathrm{q}}}f_{2\boldsymbol{\mathrm{q}}}}}\mathrm{\ }.\end{eqnarray} 
As $T\to T_c,{\mathrm{\Delta }}_{\eta }\to 0$, 
Eq.~\eqref{eq41} reduces to 
\begin{eqnarray}
\label{eq42}
{\eta }_{\boldsymbol{\mathrm{k}}}\approx V\sum_{\boldsymbol{\mathrm{q}}}{{\eta }^2_{\boldsymbol{\mathrm{k}}-\boldsymbol{\mathrm{q}}}{\eta }_{\boldsymbol{\mathrm{q}}}\frac{1}{4NE_{\boldsymbol{\mathrm{q}}}}{\mathrm{tanh} \left(\frac{\beta E_{\boldsymbol{\mathrm{q}}}}{2}\right)\ }}.\end{eqnarray} 
Consider the special case of high transition temperature, i.e., ${\beta }_c=\frac{1}{k_BT_c}\ll 1$. This allows us to expand 
\begin{eqnarray}
\label{eq43}
{\mathrm{tanh} \left(\frac{{\beta }_cE_{\boldsymbol{\mathrm{q}}}}{2}\right)\ }\approx \frac{{\beta }_cE_{\boldsymbol{\mathrm{q}}}}{2}+O\left({\beta }^3_c\right).
\end{eqnarray}
Eq.~\eqref{eq42} becomes
\begin{eqnarray}
\label{eq44}
{\eta }_{\boldsymbol{\mathrm{k}}}\approx \frac{V}{8N}\sum_{\boldsymbol{\mathrm{q}}}{{\eta }^2_{\boldsymbol{\mathrm{k}}-\boldsymbol{\mathrm{q}}}{\eta }_{\boldsymbol{\mathrm{q}}}{\beta }_c+O\left({\beta }^3_c\right)}\Rightarrow {\beta }_c\approx \frac{8N{\eta }_{\boldsymbol{\mathrm{k}}}}{V\sum_{\boldsymbol{\mathrm{q}}}{{\eta }^2_{\boldsymbol{\mathrm{k}}-\boldsymbol{\mathrm{q}}}{\eta }_{\boldsymbol{\mathrm{q}}}}}.\end{eqnarray} 
We must also expand in $\beta $ for the constraint 
Eq.~\eqref{eq31} in conjunction with 
Eq.~\eqref{eq44},
\begin{eqnarray}
\label{eq45}
\nonumber
1-x &\approx& 4\sum_{\boldsymbol{\mathrm{q}}}{\left({\epsilon }_{\boldsymbol{\mathrm{q}}}-\mu \right)}\frac{\beta }{8N}\\
\Rightarrow \frac{1}{2}\beta \mu &\approx& \frac{1}{2}\frac{\beta }{N}\sum_{\boldsymbol{\mathrm{q}}}{{\epsilon }_{\boldsymbol{\mathrm{q}}}}+x-1.\end{eqnarray} 
Eq.~\eqref{eq45} solves $\mu $ directly for a given value of $x$. Note from 
Eq.~\eqref{eq44}
that ${\beta }_c$ does not depend on $\mu $ as it gets dropped out during the process of approximation. 
Eq.~\eqref{eq44} clearly shows that (inverse) transition temperature ${\beta }_c$ exists in high temperature limit. Note that the summation in the denominator of 
Eq.~\eqref{eq44}, $\sum_{\boldsymbol{\mathrm{q}}}{{\eta }^2_{\boldsymbol{\mathrm{k}}-\boldsymbol{\mathrm{q}}}{\eta }_{\boldsymbol{\mathrm{q}}}\leadstoext O\left(N\right)}$ so that the value of ${\beta }_c\sim O\left(1\right)$, which is an agreeable expectation based on order of approximation estimate. Note also that the k-dependence of the transition temperature in 
Eq.~\eqref{eq44} in general does not get cancelled off. This is a distinctive feature of the EHM compared to other phenomenological models: that each k-point contributes differently to the transition temperature. $T_c$ measured in experiments, as interpreted in the EHM, is an effective value contributed by various k-point that are being probed. 

In a more general scenario, switching on the $U$ term will break the pure $d$-wave symmetry of ${\mathrm{\Delta }}_{\boldsymbol{\mathrm{k}}}$, as is obvious from the ansatz for ${\mathrm{\Delta }}_{\boldsymbol{\mathrm{k}}}$. The solutions for ${\beta }_c$ in such a scenario are now discussed. Inclusion of the $U$ term would modify the simplified transition temperature 
Eq.~\eqref{eq44}, in addition to coupling the constraint from the dopant concentration $x$ to ${\beta }_c$. The gap equation 
Eq.~\eqref{eq20} is highly non-linear and self-iterative. Full numerical solutions could only be obtained by deploying sophisticated computational approach (e.g., genetic algorithm) but we shall not pursue along this direction here. Instead, we shall only attempt to solve ${\beta }_c$ with some simplifying assumptions. 

We begin with our assumed ansatz, 
Eq.~\eqref{eq32}. Slotting it into the LHS of the gap function 
Eq.~\eqref{eq20}, using 
Eq.~\eqref{eq25}, and distinguishing the gap equation into k-dependent and k-independent parts, the following two independent relations are obtained:
\begin{eqnarray}
\label{eq46}
{\mathrm{\Delta }}_{\eta }{\eta }_{\boldsymbol{\mathrm{k}}}&=&V\sum_{\boldsymbol{\mathrm{q}}}{{\eta }^2_{\boldsymbol{\mathrm{k}}-\boldsymbol{\mathrm{q}}}}{\mathrm{\Delta }}_{\boldsymbol{\mathrm{q}}}F_{\boldsymbol{\mathrm{q}}},\nonumber \\
{\mathrm{\Delta }}_0&=&-2U\sum_{\boldsymbol{\mathrm{q}}}{{\mathrm{\Delta }}_{\boldsymbol{\mathrm{q}}}F_{\boldsymbol{\mathrm{q}}}.}\end{eqnarray} 
Slotting ${\mathrm{\Delta }}_{\boldsymbol{\mathrm{q}}}={\mathrm{\Delta }}_0+{\mathrm{\Delta }}_{\eta }{\eta }_{\boldsymbol{\mathrm{q}}}$ back into the RHS of 
Eq.~\eqref{eq46},
\begin{eqnarray}
\label{eq47}
\nonumber 
{\mathrm{\Delta }}_0\left(1+2U\sum_{\boldsymbol{\mathrm{q}}}{F_{\boldsymbol{\mathrm{q}}}}\right)+{\mathrm{\Delta }}_{\eta }\left(2U\sum_{\boldsymbol{\mathrm{q}}}{{\eta }_{\boldsymbol{\mathrm{q}}}F_{\boldsymbol{\mathrm{q}}}}\right)=0,\end{eqnarray} 
\begin{eqnarray}\nonumber {\mathrm{\Delta }}_{\mathrm{\etaup }}\left({\eta }_{\boldsymbol{\mathrm{k}}}-V\sum_{\boldsymbol{\mathrm{q}}}{{{\eta }^2_{\boldsymbol{\mathrm{k}}-\boldsymbol{\mathrm{q}}}{\eta }_{\boldsymbol{\mathrm{q}}}F}_{\boldsymbol{\mathrm{q}}}}\right)-{\mathrm{\Delta }}_0\left(V\sum_{\boldsymbol{\mathrm{q}}}{{\eta }^2_{\boldsymbol{\mathrm{k}}-\boldsymbol{\mathrm{q}}}F_{\boldsymbol{\mathrm{q}}}}\right)=0,\end{eqnarray} 
which are then combined into a matrix form,
\begin{eqnarray} M\left( \begin{array}{c}
{\mathrm{\Delta }}_0 \\ 
{\mathrm{\Delta }}_{\eta } \end{array}
\right)=\left( \begin{array}{c}
0 \\ 
0 \end{array}
\right),\end{eqnarray} 
where
\begin{eqnarray}
\label{eq48}
M&=&\left( \begin{array}{cc}
1+2U\sum_{\boldsymbol{\mathrm{q}}}{F_{\boldsymbol{\mathrm{q}}}} & 2U\sum_{\boldsymbol{\mathrm{q}}}{{\eta }_{\boldsymbol{\mathrm{q}}}F_{\boldsymbol{\mathrm{q}}}} \\ 
-V\sum_{\boldsymbol{\mathrm{q}}}{{\eta }^2_{\boldsymbol{\mathrm{k}}-\boldsymbol{\mathrm{q}}}F_{\boldsymbol{\mathrm{q}}}} & {\eta }_{\boldsymbol{\mathrm{k}}}-V\sum_{\boldsymbol{\mathrm{q}}}{{\eta }^2_{\boldsymbol{\mathrm{k}}-\boldsymbol{\mathrm{q}}}{\eta }_{\boldsymbol{\mathrm{q}}}F_{\boldsymbol{\mathrm{q}}}} \end{array}
\right) \nonumber \\
&\equiv& \left( \begin{array}{cc}
T_1 & T_3 \\ 
-T_{4\boldsymbol{\mathrm{k}}} & T_{2\boldsymbol{\mathrm{k}}} \end{array}
\right).\end{eqnarray} 
To determine the critical temperature ${\beta }_c$, we require the determinant of the 2 by 2 matrix $M$ in 
Eq.~\eqref{eq48} to vanish. This is a sufficient condition for the existence of non-trivial solutions to 
Eq.~\eqref{eq47}: 
\begin{eqnarray}
\label{eq49}
D_{\boldsymbol{\mathrm{k}}}\left(\beta \right)&=&{\left|M\right|=\ T}_1T_{2\boldsymbol{\mathrm{k}}}+T_3T_{4\boldsymbol{\mathrm{k}}} \nonumber \\
&=&\left(1+2U\sum_{\boldsymbol{\mathrm{q}}}{F_{\boldsymbol{\mathrm{q}}}\left(\beta \right)}\right)\left({\eta }_{\boldsymbol{\mathrm{k}}}-V\sum_{\boldsymbol{\mathrm{q}}}{{\eta }^2_{\boldsymbol{\mathrm{k}}-\boldsymbol{\mathrm{q}}}{\eta }_{\boldsymbol{\mathrm{q}}}F_{\boldsymbol{\mathrm{q}}}\left(\beta \right)}\right) \nonumber  \\ &&+\left(2U\sum_{\boldsymbol{\mathrm{q}}}{{\eta }_{\boldsymbol{\mathrm{q}}}F_{\boldsymbol{\mathrm{q}}}\left(\beta \right)}\right)\left(V\sum_{\boldsymbol{\mathrm{q}}}{{\eta }^2_{\boldsymbol{\mathrm{k}}-\boldsymbol{\mathrm{q}}}F_{\boldsymbol{\mathrm{q}}}\left(\beta \right)}\right)\nonumber \\ &=&0.\end{eqnarray} 
To proceed with the abstraction of the transition temperature, the function $F_{\boldsymbol{\mathrm{q}}}\left(\beta \right)$ (as defined in 
Eq.~\eqref{eq21}) near the transition temperature ${\beta }_c$ at a k-point \textbf{q} is assumed to take the approximated form
\begin{eqnarray}
\label{eq50}
F_{\boldsymbol{\mathrm{q}}}\left(\beta \right)\approx \frac{{\mathrm{tanh} \left(\frac{\beta E_{\boldsymbol{\mathrm{q}}}}{2}\right)\ }}{4NE_{\boldsymbol{\mathrm{q}}}},\end{eqnarray} 
where, as$\ \beta $ approaches ${\beta }_c$ from above ($T$ approaches $T_c$ from below), ${\mathrm{\Delta }}_{\boldsymbol{\mathrm{q}}}$ approximately vanishes, so that  ${\lambda }_{\boldsymbol{\mathrm{q}}}\to E_{\boldsymbol{\mathrm{q}}}$. In our numerical scheme, ${\beta }_c$ at a given k-point \textbf{k} is obtained by seeking the root of 
Eq.~\eqref{eq49}, 
\begin{eqnarray}\nonumber D_{\boldsymbol{\mathrm{k}}}\left({\beta }_c\right)=0,\end{eqnarray} 
with the constraint from 
Eq.~\eqref{eq30} considered concurrently. To this end, we make the replacement ${\lambda }_{\boldsymbol{\mathrm{q}}}\to E_{\boldsymbol{\mathrm{q}}}$ in the constraint 
Eq.~\eqref{eq30}, so that it now reads
\begin{eqnarray}
\label{eq51}
x=h_{\beta }\left(\mu \right),\end{eqnarray} 
where 
\begin{eqnarray}
\label{eq52}
h_{\beta }\left(\mu \right)\equiv 1-\frac{1}{N}\sum_{\boldsymbol{\mathrm{k}}}{{\mathrm{tanh} \left[\frac{\beta \left({\epsilon }_{\boldsymbol{\mathrm{k}}}-\mu \right)}{2}\right]\ }}.\end{eqnarray} 
The first Brillouin zone (1BZ) is discretized into $N$ equally spaced sites, ${\boldsymbol{\mathrm{k}}}_j\in \left\{{\boldsymbol{\mathrm{k}}}_1,{\boldsymbol{\mathrm{k}}}_2,\dots ,{\boldsymbol{\mathrm{k}}}_N\right\}$. ${\beta }_c$, if existed, also depends on the level of dopant $x$, which is an independent variable of the model, and an experimentally controllable parameter. In our numerical scheme, a value of $N=N_k=3721$ was adopted. This corresponds to setting the k-points in the range $-\frac{\pi }{2a}\le \ k_x,k_y\le \ \frac{\pi }{2a}$ with interval $\mathrm{\Delta }k_x=\mathrm{\Delta }k_y=\frac{\pi }{N_{1\mathrm{D}}a}$, $N_{1\mathrm{D}}=60$. A ${\beta }_c\ \mathrm{vs}.\ \ x$ curve is traced out for each ${\boldsymbol{\mathrm{k}}}_j$, where $x_0\le x\le x_m$, with $x_0=0.01,x_m=0.27$. Since the transition temperature is k-point specific and $x$-dependent, we denote ${\beta }_c={\beta }_c\left({\boldsymbol{\mathrm{k}}}_j,x\right).$ The corresponding value for the superconducting gap ${\mathrm{\Delta }}_{{\boldsymbol{\mathrm{k}}}_{\boldsymbol{j}}}(x)$ at the k-point ${\boldsymbol{\mathrm{k}}}_j$ and dopant concentration $x$ exists only if the solution ${\beta }_c\left({\boldsymbol{\mathrm{k}}}_j,x\right)$ exists. If ${\beta }_c\left({\boldsymbol{\mathrm{k}}}_j,x\right)$ existed, the corresponding ${\mathrm{\Delta }}_{{\boldsymbol{\mathrm{k}}}_j}(x)$ is given by 
\begin{eqnarray}
\label{eq53}
{\mathrm{\Delta }}_{{\boldsymbol{\mathrm{k}}}_j}\left(x\right)={\mathrm{\Delta }}_0\left[1+r_{\eta }\left({\boldsymbol{\mathrm{k}}}_j,x\right){\eta }_{\boldsymbol{\mathrm{k}}_j}\right],\end{eqnarray} 
where 
\begin{eqnarray}
\label{eq54}
r_{\eta }\left({\boldsymbol{\mathrm{k}}}_j,x\right)=-\frac{1+2U\sum_{\boldsymbol{\mathrm{q}}}{F_{\boldsymbol{\mathrm{q}}}\left[{\beta }_c\left({\boldsymbol{\mathrm{k}}}_j,x\right)\right]}}{2U\sum_{\boldsymbol{\mathrm{q}}}{{\eta }_{\boldsymbol{\mathrm{q}}}F_{\boldsymbol{\mathrm{q}}}\left[{\beta }_c\left({\boldsymbol{\mathrm{k}}}_j,x\right)\right]}}.\end{eqnarray} 
Derivation of 
Eq.~\eqref{eq53} and Eq.~\eqref{eq54} is detailed in~\ref{asecA5}
of the Appendix. ${\mathrm{\Delta }}_0$ is an arbitrary constant. To be quantitative in our numerical calculation, we assigned ${\mathrm{\Delta }}_0=t$, where $t$ is the hopping parameter in 
Eq.~\eqref{eq14}. Note that ${\mathrm{\Delta }}_0,{\mathrm{\Delta }}_{\eta }$ are both k-independent per the ansatz 
Eq.~\eqref{eq32}. 

A remark is in place for the numerical solution of ${\mathrm{\Delta }}_{{\boldsymbol{\mathrm{k}}}_j}(x)$, 
Eq.~\eqref{eq53}.  The numerical values of $r_{\eta }\left({\boldsymbol{\mathrm{k}}}_j,x\right)$ as determined via 
Eq.~\eqref{eq54} will become infinite if $U$ is identically 0, despite the corresponding transition temperature (if existed) is finite. The apparent singularity in the gap value at $U=0$ is merely a numerical artifact due to the approximations made in current numerical scheme. 

\section{Numerical Results}
The numerical computation for ${\beta }_c\left({\boldsymbol{\mathrm{k}}}_j,x\right)$ and ${\mathrm{\Delta }}_{{\boldsymbol{\mathrm{k}}}_{\boldsymbol{j}}}\left(x\right)$ as predicted by the EHM is carried out in atomic units, where the following quantities assume numerical values as depicted:
\begin{eqnarray}\nonumber  a_0=e=m_e=\hslash =1,\end{eqnarray} 
\begin{eqnarray}\nonumber  1 
\AA
=1.89,\end{eqnarray} 
\begin{eqnarray}\nonumber k_B=3.17\times {10}^{-6}{\mathrm{K}}^{-1},\end{eqnarray} 
\begin{eqnarray}
\nonumber
1\ \mathrm{eV}=0.03675.\end{eqnarray} 
$a_0$ is the Bohr radius. Two parameters used for numerical calculation are the lattice constant $a$ for the CuO plane in the cuprates, and the tight-biding parameter $t$. In atomic unit, their values are respectively assigned as
\begin{eqnarray}
\nonumber
a=3.8\ 
\AA
=7.182,\end{eqnarray} 
\begin{eqnarray}
\nonumber
t=0.4\ \mathrm{eV}=0.0147.\end{eqnarray} 
Inverse temperature parameter $\beta $ is related to temperature $T$ via 
\begin{eqnarray}
\nonumber
T=\frac{1}{k_B\beta }=\frac{{10}^6\mathrm{\ K}}{3.17\beta }.\end{eqnarray} 
We shall express the value of $V$ and $U$ in unit of the hopping parameter $t$, $V=\ vt$, $U=\ ut$, where the reduced parameters are such that $u>0,v<0$. The numerical values of the gap function ${\mathrm{\Delta }}_{\boldsymbol{\mathrm{k}}}$ will also be expressed in unit of $t$. 

Transition temperature ${\beta }_c\left({\boldsymbol{\mathrm{k}}}_j,x\right)$ and superconducting gap ${\mathrm{\Delta }}_{{\boldsymbol{\mathrm{k}}}_j}\left(x\right)$, if existed, could be picked up by the numerical procedure (within the limits of the numerical resolution implemented) for any fixed values of $U,V$. However, only transition temperature of less than $T_{c\mathrm{last}}=500\ \mathrm{K}$ will be covered. Even if solutions at larger than 500 K exist, they will not be shown. Each k-point in the 1BZ can possibly generate a full $T_c$ vs. $x$ curve of their own, depending on the location of ${\boldsymbol{\mathrm{k}}}_j$, and the values of the free parameters, $U$ and $V$ assumed. These curves in general varies from k-point to k-point. In the special case of $U=0$, there shall be no solution for ${\beta }_c$ along the $k_x=\pm k_y$ line in the 1BZ due to the $d$-wave symmetry. 

\begin{figure}
\includegraphics*[width=3.3in, height=4.5in, keepaspectratio=false]{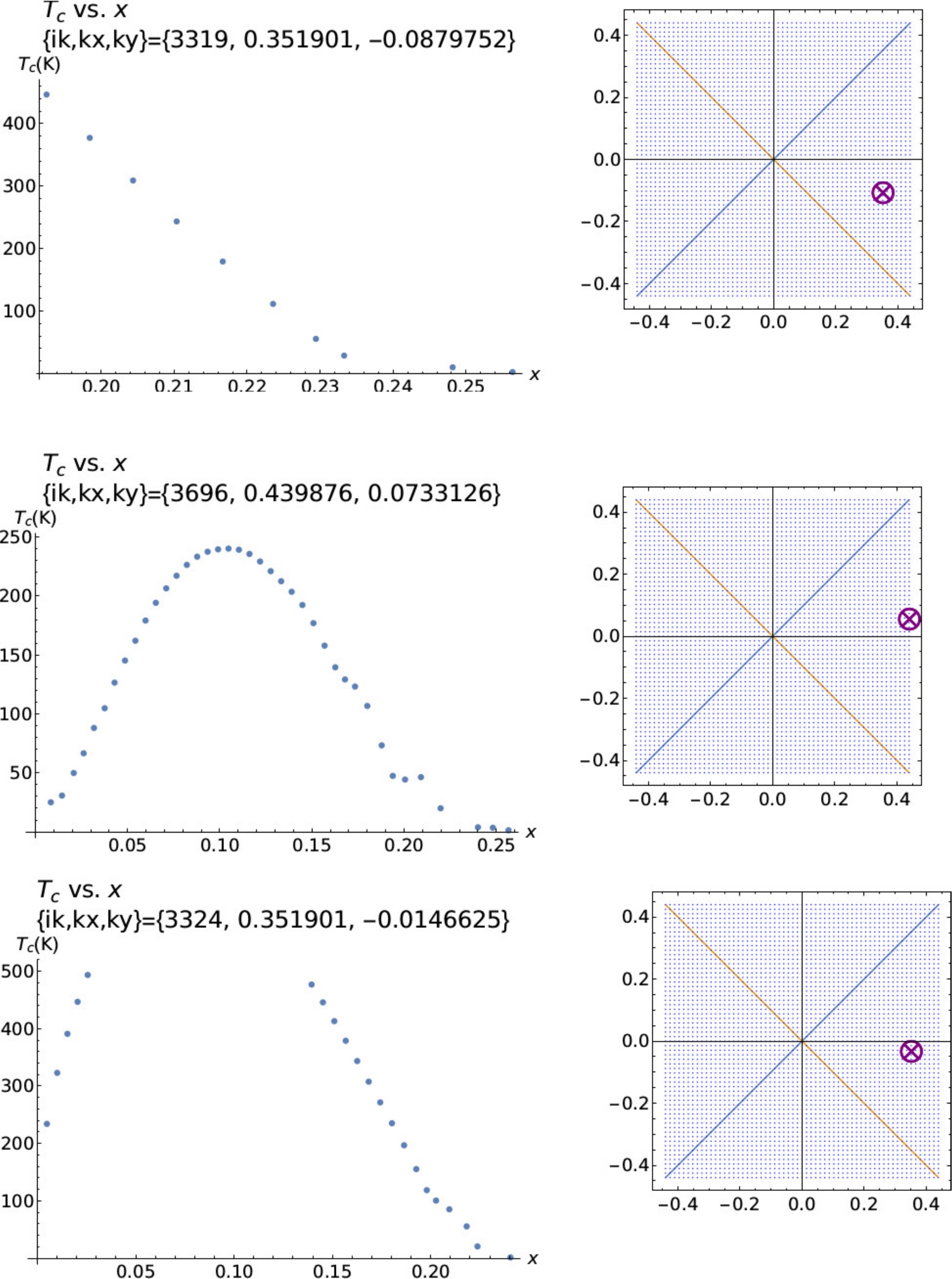}
\caption{$T_c$ vs. $x$ curves at selected k-points \{$k_x,k_y$\}=\{0.3519,-0.08798\}, \{0.3519,-0.01466\}, \{0.4399,0.07331\} for \{$u$=0.0, $v$ =-20.0\}. The cross enclosed in a circle in the 1BZ (figures in the right column) indicates the location of the corresponding k-point for the $T_c$ vs. $x$ curves in the left column.}
\label{fig:2a}       
\end{figure}

\begin{figure}
\includegraphics*[width=3.3in, height=4.5in, keepaspectratio=false]{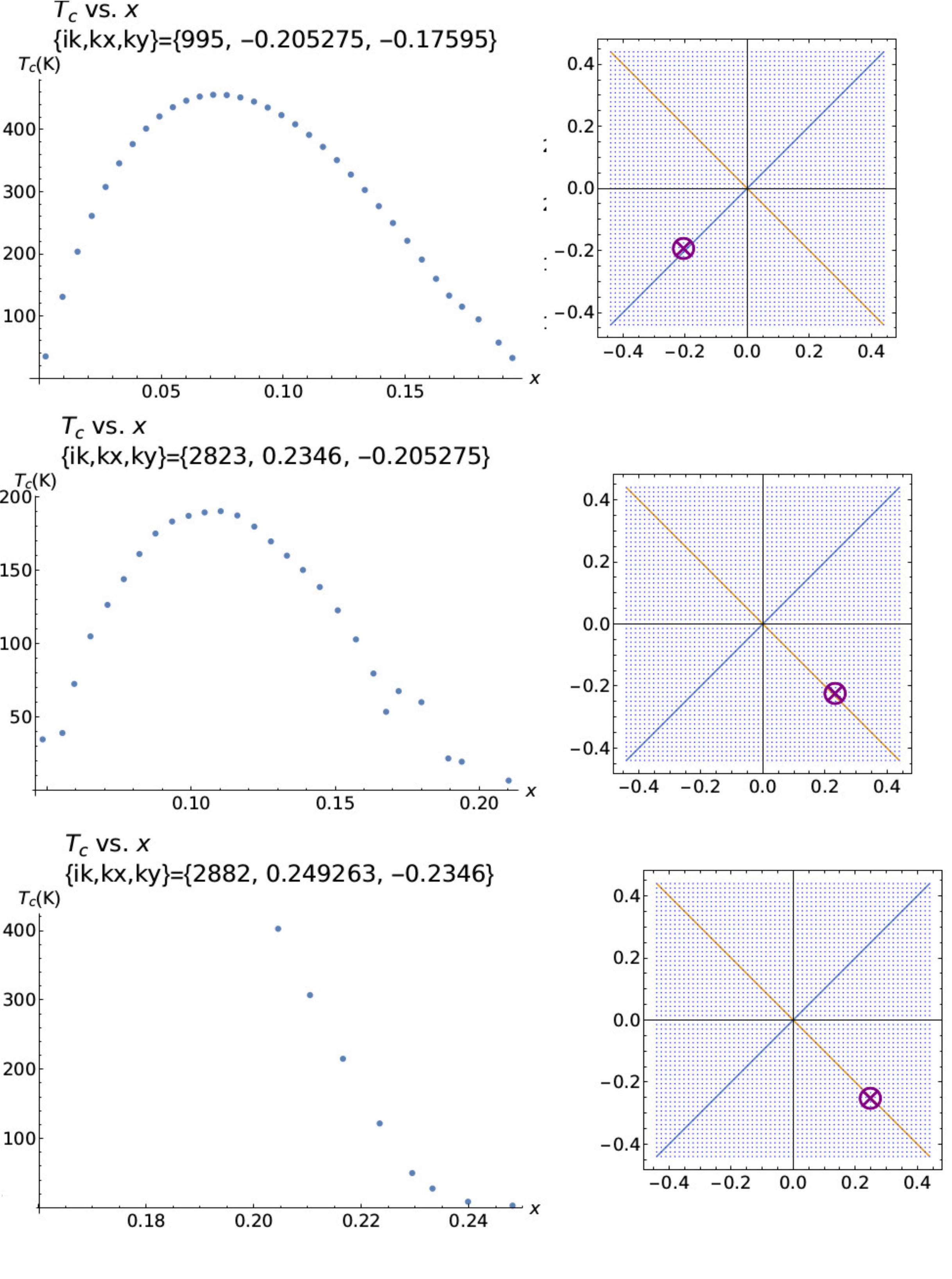}
\caption{$T_c$ vs. $x$ curves at selected k-points \{$k_x,k_y$\}=\{-0.2053,-0.176\}, \{0.2346,-0.2053\}, \{0.2493,-0.2346\} for \{$u$=0.25, $v$ =-1.0\}.}
\label{fig:2b}       
\end{figure}

As an illustration, we show in 
Fig.~\ref{fig:2a} and Fig.~\ref{fig:2b} a few $T_c\ \mathrm{vs}.\ x$ curves at  selected k-points for fixed parameters set $\left\{u=0.0,v=-20.0\right\}$ and $\left\{u=0.25,v=-1.0\right\}$ respectively. These figures provide a vivid illustration that transition temperature, as numerically solved from the superconducting gap equation, varies with dopant level $x$ that enters the solution via the constraint Eq.~\eqref{eq30}. Moreover, the shape of the $T_c \ \mathrm{vs}.\ x$ curve varies from k-point to k-point in the 1BZ. At a lower level, these $T_c \ \mathrm{vs} \ x$ solutions in turn are determined by two free parameters in the EHM, namely, $U$ and $V$.

\begin{figure*}
\includegraphics[width=1.00\textwidth]{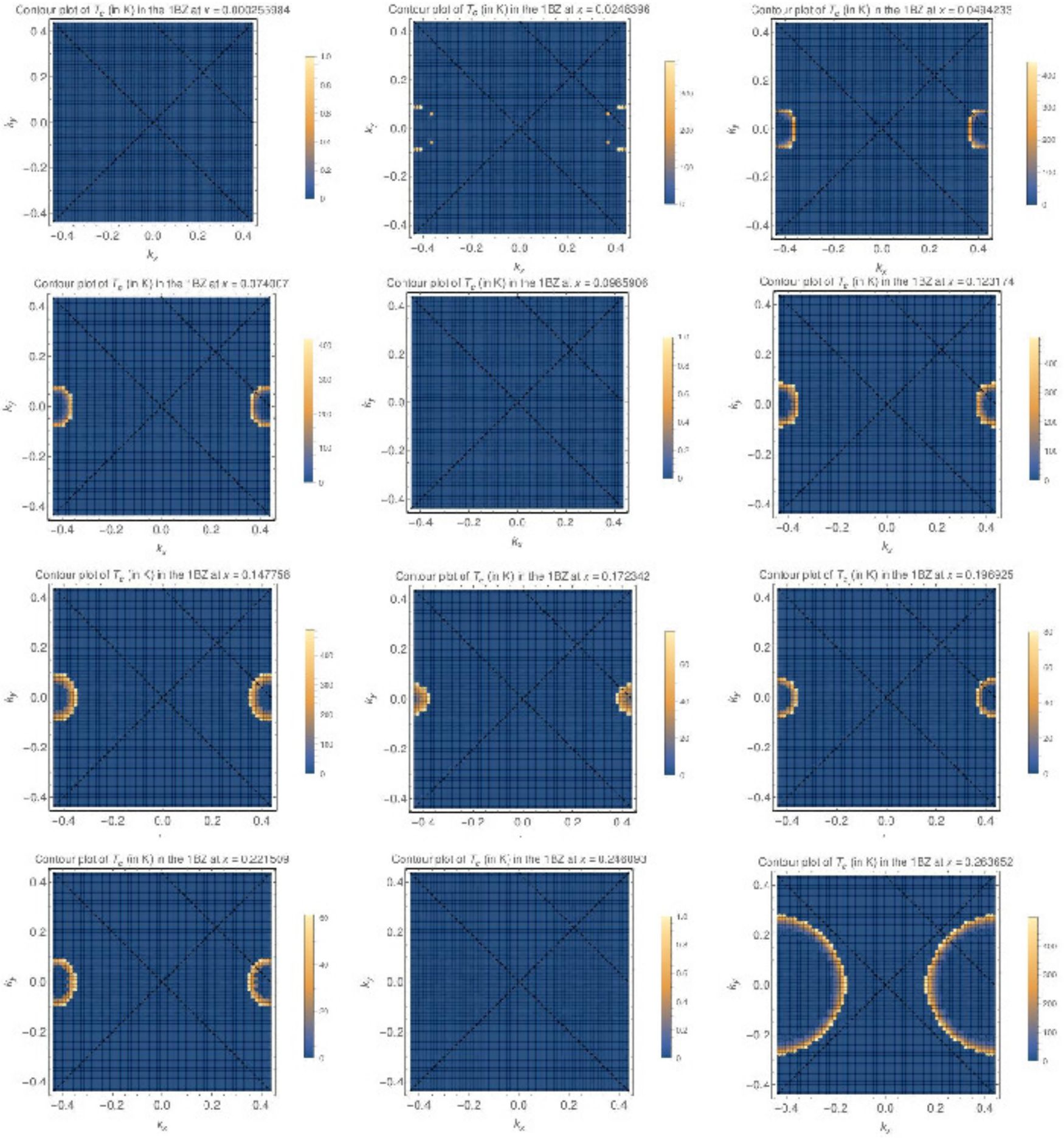}
\caption{A series of contour plots representing $T_c$ in the 1BZ for various value of $x$ at the fixed set of parameters \{$u=0.0,v=-20.0$\}}.
\label{fig:3a}       
\end{figure*}

\begin{figure*}
\includegraphics[width=1.10\textwidth]{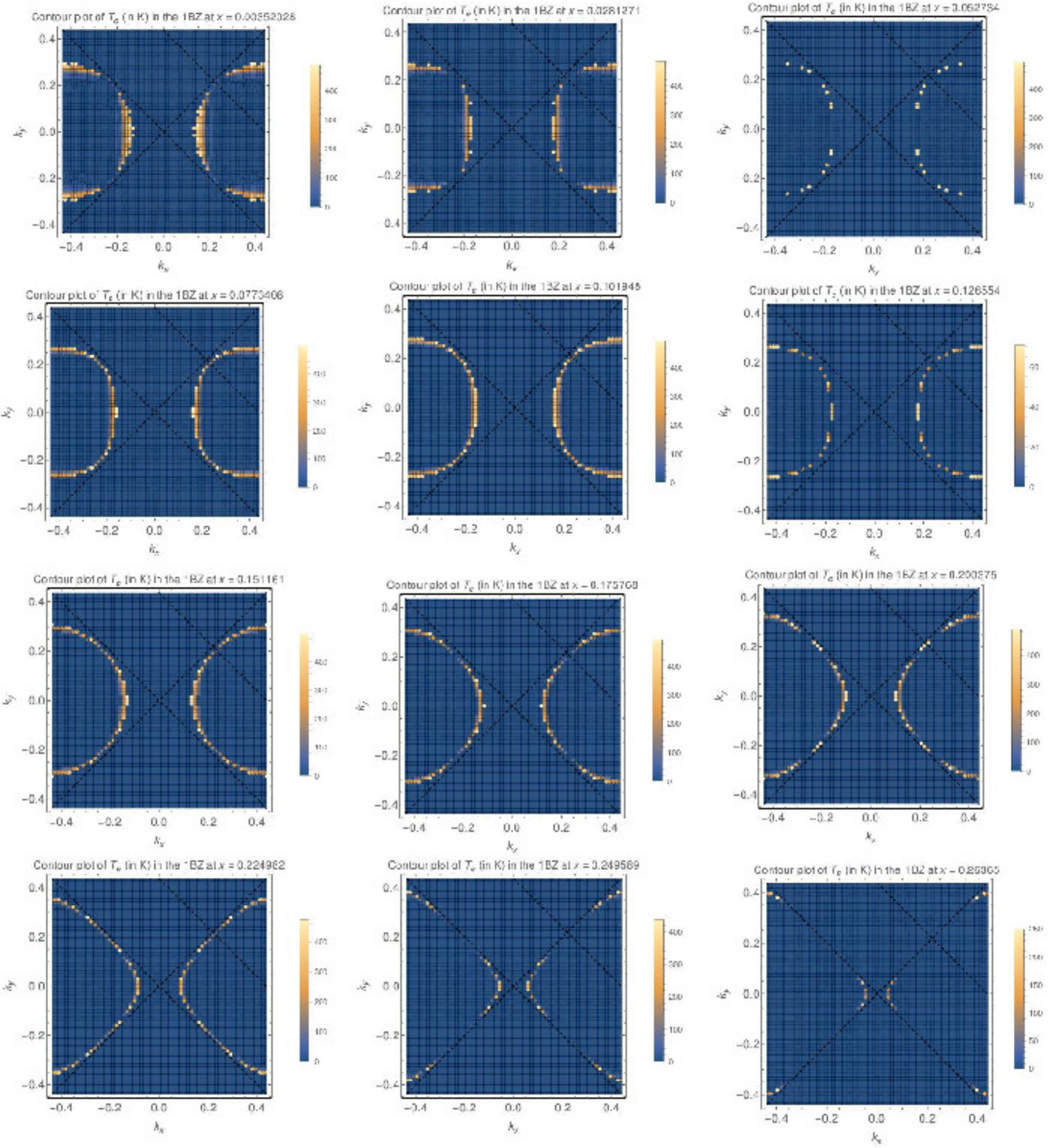}
\caption{A series of contour plots representing $T_c$ in the 1BZ for various value of $x$ at  
the fixed set of parameters \{$u=0.25,v=-1.0$\}}.
\label{fig:3b}       
\end{figure*}

For a fixed value of $x$ and parameter set \{$u,v$\}, some k-points in the 1BZ may develop ${\beta }_c$ while others don't. The collection of all solutions for ${\beta }_c$ for all k-points (including those k-points where no solutions are developed) at a fixed value of $x$ and parameter set \{$u,v$\} can be displayed in the form of a contour plot. The series of figures illustrated in 
Fig.~\ref{fig:3a}, Fig.~\ref{fig:3b} 
are the contour plots in the 1BZ for the parameters sets $\left\{u=0.0,v=\right.$
$\left.-20.0\right\}$ and $\left\{u=0.25,v=-1.0\right\}$ respectively. A collection of 12 plots, each corresponds to a specific value of $x$ ($ 0.0 \leq x \leq 0.27$) are shown in each of Fig.~\ref{fig:3a} and Fig.~\ref{fig:3b}. 
Each pixel in the contour plot is color coded to represent the value of 
$T_c({\boldsymbol{\mathrm{k}}}_j,x)$ [which is equivalent to $\beta_c({\boldsymbol{\mathrm{k}}}_j,x)$ via 
$T_c({\boldsymbol{\mathrm{k}}}_j,x)= 1/{k_B\beta_c({\boldsymbol{\mathrm{k}}}_j,x)}$] at that k-point, ${\boldsymbol{\mathrm{k}}}_{{j}}$.
The patterns of these plots, which are $x$- and $\left\{u,v\right\}$-dependent, show that solutions for ${\beta }_c$ only get developed in restricted regions in the 1BZ. In the illustrative example of the case \{$u=0.0,v= -20.0$\} in Fig.~\ref{fig:3a}, regions which develop solution for ${\beta }_c$ concentrate in the approximately semi-circular rings, with small radii and variable widths, attached to the edges of $k_x=\pm \frac{2\pi }{a}$ in the 1BZ. Meanwhile, for $\left\{u=0.25,v=\right.$
$\left.-1.0\right\}$ (Fig.~\ref{fig:3b}), the radii of these rings become larger in general, but their perimeters never cross the quarter segments confined by the diagonal and anti-diagonal lines, $k_x={\pm k}_y$. In addition, as $x$ tends towards the 0.27 limit, the semi-circles are deformed into triangular wedges with two of their common lengths pushed towards the boarders of the quarter segments (i.e., the $k_x={\pm k}_y$ lines). Each k-point (pixel) in the contour plot also has a gap value ${\mathrm{\Delta }}_{{\boldsymbol{\mathrm{k}}}_{\boldsymbol{j}}}\left({\beta }_c,x\right)$ if a solution for ${\beta }_c({\boldsymbol{\mathrm{k}}}_j,x)$ also exists there. Hence, the pattern of the contour plots for ${\mathrm{\Delta }}_{\boldsymbol{\mathrm{k}}}\left({\beta }_c,x\right)$ are exactly similar as that for ${\beta }_c(\boldsymbol{\mathrm{k}},x)$. As such we have omitted the individual figures for the contour plots of ${\mathrm{\Delta }}_{\boldsymbol{\mathrm{k}}}\left({\beta }_c,x\right)$. In general, the pattern of the solutions in the contour plot of the 1BZ not only varies with $x$, it is also sensitive to the choice of \{$u,v$\}. The contour plots selected in Fig.~\ref{fig:3a} and Fig.~\ref{fig:3b} illustrate the point that in the EHM, solutions to the  superconducting gap function are restrained to specific regions in the 1BZ. This is a unique prediction of the EHM.

\begin{figure}
\includegraphics*[width=3.25in, height=2.2in, keepaspectratio=false]{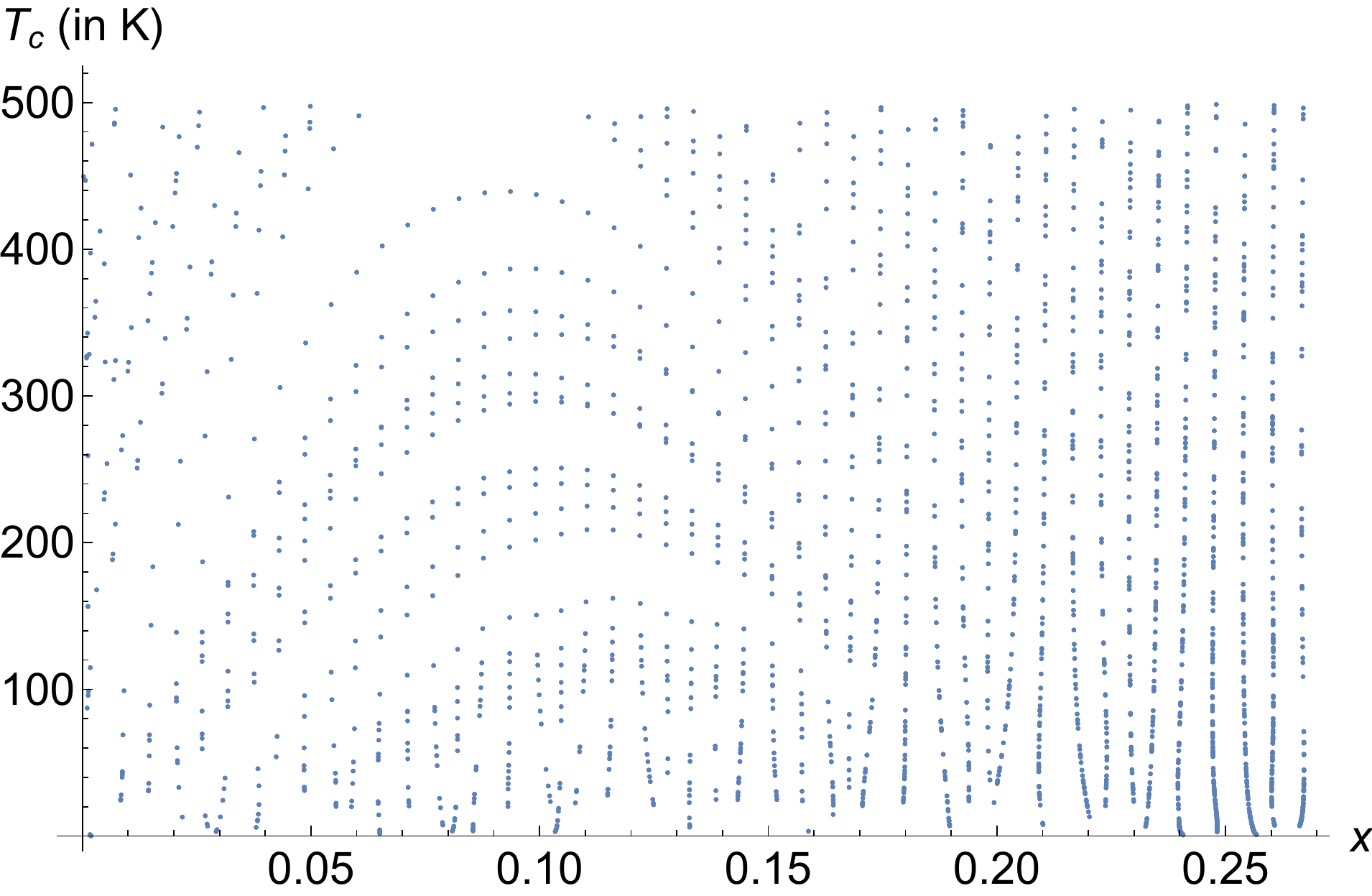}
\caption{Point plot showing the totality of all solutions $T_c$ found in the 1BZ at all values of $x$ scanned by the present numerical scheme for $u$=0.0, $v$=-20.0.}
\label{fig:5}       
\end{figure}

\begin{figure}
\includegraphics*[width=3.0in, height=3.0in, keepaspectratio=false]{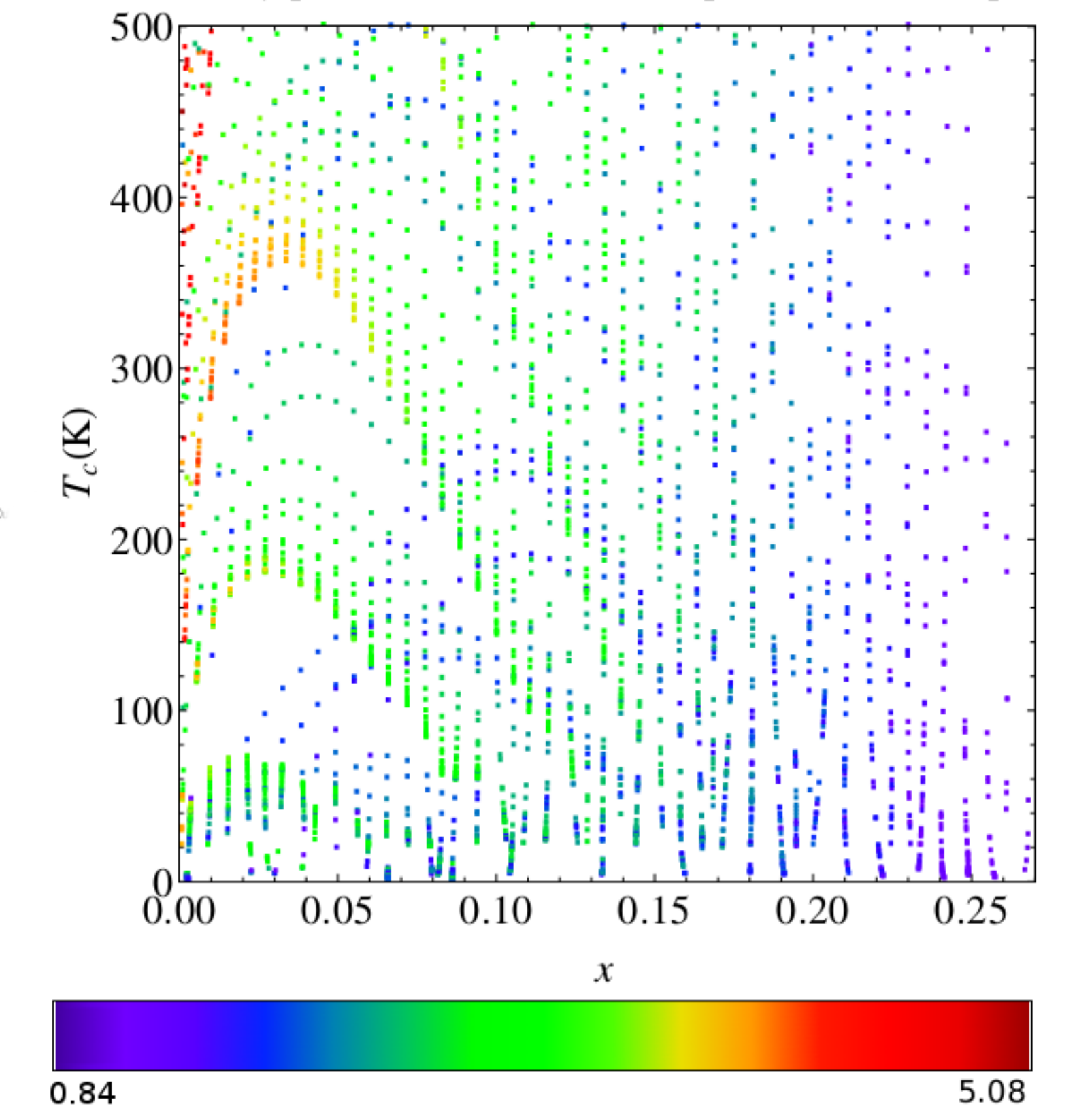}
\caption{Density plot for the collection of all $\mathrm{\Delta}_{\bf k}/ t$ values in $T_c-x$ space for 
$u$=0.25, $v$=-1.0.}
\label{fig:4}       
\end{figure}

Another illustrative way to report the solutions obtained with our numerical scheme is to present them in the form of non color-coded point plot in $T_c-x$ space. This is illustrated for the case $\left\{u=0.0,v=-20.0\right\}$ in Fig.~\ref{fig:5}.
It graphically presents the collection of all solutions ${\beta }_c$ found in the 1BZ at all values of $x$ scanned by our numerical scheme for a fixed parameter set \{$u,v$\}. The coordinates of each pixels, $\left\{x,T_c\right\}$, indicate the value of transition temperature ${\beta }_c$ obtained at the corresponding value $x$. Void pixels are those where no solution for ${\beta }_c$ (hence ${\mathrm{\Delta }}_{\boldsymbol{\mathrm{k}}}$) exists. Be noted that there is no values available for ${\mathrm{\Delta }}_{\boldsymbol{\mathrm{k}}}(\beta_c,x)$ in the case with $u=0$
because it is numerically singular when $u=0$, as explained earlier. Color-coded point plot in $T_c-x$ space is used to illustrate the numerical results of the gap functions graphically. The collection of the gaps ${\mathrm{\Delta }}_{\boldsymbol{\mathrm{k}}}\left({\beta }_c,x\right)\mathrm{/}t\mathrm{\ }$ contributed by all k-points in the 1BZ for the illustrative case \{$u$=0.25, $v$=-1.0\} is displayed in Fig.~\ref{fig:4}. 
Each pixel is color coded to reflect the value of ${\mathrm{\Delta }}_{\boldsymbol{\mathrm{k}}}\left({\beta }_c,x\right)\mathrm{/}t$ at the coordinates $\left\{x,T_c\right\}$. Both color-coded and non color-coded point plots provide a sense of how solutions for ${\beta }_c$ is distributed in the $T_c-x$ space. In general, it is found that the solution patterns illustrated in Fig.(\ref{fig:2a}) -- Fig.(\ref{fig:4})
are numerically sensitive to the choice of the parameters, $u,v$. Rigorous and exhaustive exploration of the parameter space of the model and matching the solution patterns against existing experimental findings are not within the scope of this paper but reserved for future work.

\section{Conclusion}
Our theoretical and numerical works indicate that the EHM is phenomenologically rich. Superconducting gap solutions are restrained to specific regions in the 1BZ, and transitional temperature is $k$-dependent. It is very feasible to expect that the EHM naturally inherits a huge parameter space in which experimentally measured results, such as the well-known superconducting dome and the phase diagram from electronic Raman scattering \cite{Sacuto:RPP13} can be accommodated. The EHM model hence offers a viable venue to search for or confirm any signature in k-point-sensitive experimental measurements.

%


%
\section*{Acknowledgements}
This work was begun at UKM and BSL thanks RAS for hospitality. BSL acknowledges financial support from Ministry of Science, Technology and Innovation and Malaysian Academy of Science under the Brain Gain Program Malaysia (MOSTI/BGM/B2L/ \\ 
2/5063). 
YTL wishes to acknowledge the support of Universiti Sains Malaysia RU grant (grant number 1001/PFIZIK/811240). RAS thanks the Ministry of Higher Education of Malaysia for the support (grant number ERGS/1/2011/STG/UKM/01/25). All numerical works were exclusively conducted in USM by YTL. YTL gladfully acknowledges Dr. Chan Huah Yong from the School of Computer Science, USM, for providing us computing resources to carry out part of the calculations done in this paper.

\bibliographystyle{spphys}       

\begin{thebibliography}{10}
\providecommand{\url}[1]{{#1}}
\providecommand{\urlprefix}{URL }
\expandafter\ifx\csname urlstyle\endcsname\relax
  \providecommand{\doi}[1]{DOI \discretionary{}{}{}#1}\else
  \providecommand{\doi}{DOI \discretionary{}{}{}\begingroup
  \urlstyle{rm}\Url}\fi

\bibitem{Bednorz:ZP86}
J.G. Bednorz, K.A. Muller, Zeitschrift fur Physik B Condensed Matter
  \textbf{64}(2), 189 (1986).
\newblock \doi{10.1007/bf01303701}.
\newblock \urlprefix\url{http://link.springer.com/article/10.1007/BF01303701}

\bibitem{Norman:RPP03}
M.R. Norman, C.~Pepin, Reports on Progress in Physics \textbf{66}(10), 1547
  (2003).
\newblock \urlprefix\url{http://stacks.iop.org/0034-4885/66/i=10/a=R01}

\bibitem{Lee:RPP08}
P.A. Lee, Reports on Progress in Physics \textbf{71}(1), 012501 (2008).
\newblock \urlprefix\url{http://stacks.iop.org/0034-4885/71/i=1/a=012501}

\bibitem{Lee:JSNM09}
B.S. Lee, Journal of Superconductivity and Novel Magnetism \textbf{23}(3), 333
  (2009).
\newblock \doi{doi:10.1007/s10948-009-0536-z}

\bibitem{Lee_JSNM11}
B.S. Lee, R.~Abd-Shukor, Journal of Superconductivity and Novel Magnetism
  \textbf{25}(4), 861 (2011).
\newblock \doi{10.1007/s10948-011-1369-0}

\bibitem{Lee:JSNM14}
B.S. Lee, T.L. Yoon, Journal of Superconductivity and Novel Magnetism
  \textbf{27}(12), 2673 (2014).
\newblock \doi{10.1007/s10948-014-2753-3}

\bibitem{Lawler:Nature10}
M.J. Lawler, K.~Fujita, J.~Lee, A.R. Schmidt, Y.~Kohsaka, C.K. Kim, H.~Eisaki,
  S.~Uchida, J.C. Davis, J.P. Sethna, E.A. Kim, Nature \textbf{466}(7304), 347
  (2010).
\newblock \doi{doi:10.1038/nature09169}

\bibitem{Hufner_RPP08}
S.~Hufner, M.A. Hossain, A.~Damascelli, G.A. Sawatzky, Reports on Progress in
  Physics \textbf{71}(6), 062501 (2008).
\newblock \urlprefix\url{http://stacks.iop.org/0034-4885/71/i=6/a=062501}

\bibitem{ANDERSON:Science87}
P.W. ANDERSON, Science \textbf{235}(4793), 1196 (1987).
\newblock \doi{10.1126/science.235.4793.1196}.
\newblock \urlprefix\url{http://science.sciencemag.org/content/235/4793/1196}

\bibitem{Ogata:RPP08}
M.~Ogata, H.~Fukuyama, Reports on Progress in Physics \textbf{71}(3), 036501
  (2008).
\newblock \urlprefix\url{http://stacks.iop.org/0034-4885/71/i=3/a=036501}

\bibitem{Varma:RPP16}
C.M. Varma, Reports on Progress in Physics \textbf{79}(8), 082501 (2016).
\newblock \urlprefix\url{http://stacks.iop.org/0034-4885/79/i=8/a=082501}

\bibitem{Lanzara:Nature01}
A.~Lanzara, P.V. Bogdanov, X.J. Zhou, S.A. Kellar, D.L. Feng, E.D. Lu,
  T.~Yoshida, H.~Eisaki, A.~Fujimori, K.~Kishio, J.I. Shimoyama, T.~Noda,
  S.~Uchida, Z.~Hussain, Z.X. Shen, Nature \textbf{412}(6846), 510 (2001).
\newblock \doi{10.1038/35087518}.
\newblock \urlprefix\url{http://dx.doi.org/10.1038/35087518}

\bibitem{Kamimura_JSNM12}
H.~Kamimura, H.~Ushio, Journal of Superconductivity and Novel Magnetism
  \textbf{25}(3), 677 (2012).
\newblock \doi{10.1007/s10948-012-1435-2}

\bibitem{Micnas:RMP90}
R.~Micnas, J.~Ranninger, S.~Robaszkiewicz, Rev. Mod. Phys. \textbf{62}, 113
  (1990).
\newblock \doi{10.1103/RevModPhys.62.113}.
\newblock \urlprefix\url{http://link.aps.org/doi/10.1103/RevModPhys.62.113}

\bibitem{Mott:WS96}
N.F. Mott, A.S. Alexandrov, \emph{Polarons and Bipolarons} (WORLD SCIENTIFIC
  PUB CO INC, 1996)

\bibitem{Micnas:PRB88}
R.~Micnas, J.~Ranninger, S.~Robaszkiewicz, S.~Tabor, Phys. Rev. B \textbf{37},
  9410 (1988).
\newblock \doi{10.1103/PhysRevB.37.9410}.
\newblock \urlprefix\url{http://link.aps.org/doi/10.1103/PhysRevB.37.9410}

\bibitem{Micnas:JPCM02}
R.~Micnas, B.~Tobijaszewska, Journal of Physics: Condensed Matter
  \textbf{14}(41), 9631 (2002).
\newblock \urlprefix\url{http://stacks.iop.org/0953-8984/14/i=41/a=319}

\bibitem{Kittel:QTOS87}
C.~Kittel, \emph{Quantum Theory of Solids} (JOHN WILEY \& SONS INC, 1987)

\bibitem{Haken:QFTOS83}
H.~Haken, \emph{Quantum Field Theory of Solids: An Introduction} (Elsevier
  Science Ltd, 1983).
\newblock
  \urlprefix\url{https://www.amazon.com/Quantum-Field-Theory-Solids-Introduction/dp/0444867376%3FSubscriptionId%3D0JYN1NVW651KCA56C102%26tag%3Dtechkie-20%26linkCode%3Dxm2%26camp%3D2025%26creative%3D165953%26creativeASIN%3D0444867376}

\bibitem{Sacuto:RPP13}
A.~Sacuto, Y.~Gallais, M.~Cazayous, M.A. Méasson, G.D. Gu, D.~Colson, Reports
  on Progress in Physics \textbf{76}(2), 022502 (2013).
\newblock \urlprefix\url{http://stacks.iop.org/0034-4885/76/i=2/a=022502}

\end{thebibliography}


\appendix
\section{Appendix}
\subsection{
Coefficients and their Fourier transforms of the model Hamiltonian} \label{asecA0}
The total Hamiltonian in real space $H$ is as given in 
Eq.~\eqref{eq1}. The Hamiltonian $H$ is transformed into the canonical form $H\ensuremath{'}$ via a standard displaced oscillator transformation as per 
Eq.~\eqref{eq6}. The primed terms $H\ensuremath{'}_1$ and $H\ensuremath{'}_3$ in 
Eq.~\eqref{eq6} are:
\begin{eqnarray}
\label{A1}
H\ensuremath{'}_1&=&\sum_{i,j,\sigma }{t_{ij}c^{\dagger }_{i,\sigma }c_{j,\sigma }} \nonumber \\
&\equiv& \sum_{i,j,\sigma }{t\ensuremath{'}_{ij}c^{\dagger }_{i,\sigma }c_{j,\sigma }}{\mathrm{exp} \left\{\frac{1}{\sqrt{N}}\sum_{\boldsymbol{\mathrm{k}}}{\frac{g\left(\boldsymbol{\mathrm{k}}\right)}{\hslash {\omega }_{\boldsymbol{\mathrm{k}}}}\left(a_{\boldsymbol{\mathrm{k}},\sigma }-a^{\dagger }_{\boldsymbol{\mathrm{-}}\boldsymbol{\mathrm{k}}\boldsymbol{\mathrm{,-}}\sigma }\right)\left(e^{i\boldsymbol{\mathrm{k}}\boldsymbol{\mathrm{\cdot }}{\boldsymbol{\mathrm{R}}}_i}-e^{i\boldsymbol{\mathrm{k}}\boldsymbol{\mathrm{\cdot }}{\boldsymbol{\mathrm{R}}}_j}\right)}\right\}\ },\nonumber  \\ 
H\ensuremath{'}_3&=&-\frac{1}{2}\sum_{\left\langle ij\right\rangle ,\sigma {,\sigma }\ensuremath{'}}{V_{ij}n_{i,\sigma }n_{j{,\sigma }\ensuremath{'}}} \nonumber \\
&\equiv& -\frac{1}{N}\sum_{i,j,\sigma {,\sigma }\ensuremath{'}}{\sum_{\boldsymbol{\mathrm{k}}}{\frac{g^2\left(\boldsymbol{\mathrm{k}}\right)}{\hslash {\omega }_{\boldsymbol{\mathrm{k}}}}e^{i\boldsymbol{\mathrm{k}}\cdot \left({\boldsymbol{\mathrm{R}}}_j\boldsymbol{-}{\boldsymbol{\mathrm{R}}}_i\right)}}c^{\dagger }_{i\sigma }c_{i\sigma }c^{\dagger }_{j{\sigma }\ensuremath{'}}c_{j{\sigma }\ensuremath{'}}}.\end{eqnarray} 
Referring to $H\ensuremath{'}_3$ in \eqref{A1},
we identify 
\begin{eqnarray}
\label{A2}
V_{ij}=\frac{1}{N}\sum_{\boldsymbol{\mathrm{k}}}{\frac{g^2\left(\boldsymbol{\mathrm{k}}\right)}{\hslash {\omega }_{\boldsymbol{\mathrm{k}}}}e^{i\boldsymbol{\mathrm{k}}\cdot \left({\boldsymbol{\mathrm{R}}}_j\boldsymbol{-}{\boldsymbol{\mathrm{R}}}_i\right)}}.\end{eqnarray} 
The coefficients in 
Eq.~\eqref{eq6}, namely $t_{ij},V_{ij},c_{i,\sigma },c^{\dagger }_{i,\sigma },G_i$, are related to their Fourier counterparts via
\begin{eqnarray}
\label{A3}
t_{ij}&=&\frac{1}{N}\sum_{\boldsymbol{\mathrm{k}}}{{\epsilon }_{\boldsymbol{\mathrm{k}}}e^{i\boldsymbol{\mathrm{k}}\cdot \left({\boldsymbol{\mathrm{R}}}_j-{\boldsymbol{\mathrm{R}}}_i\right)}},
\nonumber \\
V_{ij}&=&\frac{1}{N}\sum_{\boldsymbol{\mathrm{k}}}{V_{\boldsymbol{\mathrm{k}}}e^{i\boldsymbol{\mathrm{k}}\cdot \left({\boldsymbol{\mathrm{R}}}_j-{\boldsymbol{\mathrm{R}}}_i\right)}},
\nonumber \\
c_{i,\sigma \ }&=&\frac{1}{\sqrt{N}}\sum_{\boldsymbol{\mathrm{k}}}{c_{\boldsymbol{\mathrm{k}},\sigma }e^{-i\boldsymbol{\mathrm{k}}\cdot {\boldsymbol{\mathrm{R}}}_i}},
\nonumber \\
c^{\dagger }_{i,\sigma \ }&=&\frac{1}{\sqrt{N}}\sum_{\boldsymbol{\mathrm{k}}}{c^{\dagger }_{\boldsymbol{\mathrm{k}},\sigma }e^{i\boldsymbol{\mathrm{k}}\cdot {\boldsymbol{\mathrm{R}}}_i}},
\nonumber \\
G_i&=&\frac{1}{\sqrt{N}}\sum_{\boldsymbol{\mathrm{k}}}{G_{\boldsymbol{\mathrm{k}}}e^{i\boldsymbol{\mathrm{k}}\cdot {\boldsymbol{\mathrm{R}}}_i}}.\end{eqnarray} 
Comparing the expressions of $V_{ij}$ in Eq.~\eqref{A2} and Eq.~\eqref{A3},
\begin{eqnarray}
\label{A4}
V_{ij}&=&\frac{1}{N}\sum_{\boldsymbol{\mathrm{k}}}{\frac{g^2\left(\boldsymbol{\mathrm{k}}\right)}{\hslash {\omega }_{\boldsymbol{\mathrm{k}}}}e^{i\boldsymbol{\mathrm{k}}\cdot \left({\boldsymbol{\mathrm{R}}}_j\boldsymbol{-}{\boldsymbol{\mathrm{R}}}_i\right)}}=\frac{1}{N}\sum_{\boldsymbol{\mathrm{k}}}{V_{\boldsymbol{\mathrm{k}}}e^{i\boldsymbol{\mathrm{k}}\cdot \left({\boldsymbol{\mathrm{R}}}_j\boldsymbol{-}{\boldsymbol{\mathrm{R}}}_i\right)}} \nonumber \\
\Rightarrow V_{\boldsymbol{\mathrm{k}}}&=&\frac{g^2\left(\boldsymbol{\mathrm{k}}\right)}{\hslash {\omega }_{\boldsymbol{\mathrm{k}}}}.\end{eqnarray} 
This explains how 
Eq.~\eqref{eq24} is arrived at.

\subsection{
Derivation of 
Eq.~\eqref{eq9}
from 
Eq.~\eqref{eq6}
\label{asecA1}}
Each term in 
Eq.~\eqref{eq6} can be cast into Bloch representation with the aid of Fourier transformations from Eq.~\eqref{A3}.
\begin{eqnarray}
\label{A6}
H\ensuremath{'}_1&=&\sum_{i,j,\sigma }{t_{ij}c^{\dagger }_{i,\sigma }c_{j,\sigma }}=\frac{1}{N}\sum_{\sigma }{\sum_{i,j}{\sum_{\boldsymbol{\mathrm{k}}}{{\epsilon }_{\boldsymbol{\mathrm{k}}}}e^{i\boldsymbol{\mathrm{k}}\cdot \left({\boldsymbol{\mathrm{R}}}_j-{\boldsymbol{\mathrm{R}}}_i\right)}c^{\dagger }_{i,\sigma }c_{j,\sigma }}}
\nonumber \\
&=&
\sum_{\sigma }{\sum_{\boldsymbol{\mathrm{k}}}{{\epsilon }_{\boldsymbol{\mathrm{k}}}}\left(\frac{1}{\sqrt{N}}\sum_i{c^{\dagger }_{i,\sigma }e^{-i\boldsymbol{\mathrm{k}}\cdot {\boldsymbol{\mathrm{R}}}_i}}\right)\left(\frac{1}{\sqrt{N}}\sum_j{c_{j,\sigma }e^{i\boldsymbol{\mathrm{k}}\cdot {\boldsymbol{\mathrm{R}}}_j}}\right)} \nonumber \\
&=&\sum_{\boldsymbol{\mathrm{k}},\sigma }{{\epsilon }_{\boldsymbol{\mathrm{k}}}}c^{\dagger }_{\boldsymbol{\mathrm{k}},\sigma }c_{\boldsymbol{\mathrm{k}},\sigma }.\ \end{eqnarray} 
\begin{eqnarray}
\label{A7}
H\ensuremath{'}_2 &=& 
U \sum_{i,\sigma} n_{i,\sigma}n_{i,-\sigma} \nonumber \\
&=& U \sum_{\sigma } \left\{
\left(\frac{1}{\sqrt{N}}\sum_{\boldsymbol{\mathrm{k}}}{c^{\dagger }_{\boldsymbol{\mathrm{k}},\sigma }e^{i\boldsymbol{\mathrm{k}}\cdot {\boldsymbol{\mathrm{R}}}_i}}\right)\left(\frac{1}{\sqrt{N}}\sum_{{\boldsymbol{\mathrm{k}}}^{\mathrm{'}}}{c_{{\boldsymbol{\mathrm{k}}}^{\boldsymbol{\mathrm{'}}},\sigma }e^{-i{\boldsymbol{\mathrm{k}}}^{\boldsymbol{\mathrm{'}}}\cdot {\boldsymbol{\mathrm{R}}}_i}}\right) \right.
\nonumber \\
&& 
\left.
\left(\frac{1}{\sqrt{N}}\sum_{{\boldsymbol{\mathrm{k}}}^{\boldsymbol{\mathrm{''}}}}{c^{\dagger }_{\boldsymbol{\mathrm{k}},-\sigma }e^{i{\boldsymbol{\mathrm{k}}}^{\boldsymbol{\mathrm{''}}}\cdot {\boldsymbol{\mathrm{R}}}_i}}\right)\left(\frac{1}{\sqrt{N}}\sum_{{\boldsymbol{\mathrm{k}}}^{\mathrm{'''}}}{c_{{\boldsymbol{\mathrm{k}}}^{\boldsymbol{\mathrm{'''}}},-\sigma }e^{-i{\boldsymbol{\mathrm{k}}}^{\boldsymbol{\mathrm{'''}}}\cdot {\boldsymbol{\mathrm{R}}}_i}}\right) \right\}
\nonumber \\
&=&
\ \frac{U}{N^2}\sum_{\sigma }{\left(\sum_{\boldsymbol{\mathrm{k}},{\boldsymbol{\mathrm{k}}}^{\boldsymbol{\mathrm{'}}},{\boldsymbol{\mathrm{k}}}^{\boldsymbol{\mathrm{''}}},{\boldsymbol{\mathrm{k}}}^{\boldsymbol{\mathrm{'''}}}}{c^{\dagger }_{\boldsymbol{\mathrm{k}},\sigma }c_{{\boldsymbol{\mathrm{k}}}^{\boldsymbol{\mathrm{'}}},\sigma }}c^{\dagger }_{{\boldsymbol{\mathrm{k}}}^{\boldsymbol{\mathrm{''}}},-\sigma }c_{{\boldsymbol{\mathrm{k}}}^{\boldsymbol{\mathrm{'''}}},-\sigma }\right)N\delta \left(\boldsymbol{\mathrm{k}}-{\boldsymbol{\mathrm{k}}}^{\boldsymbol{\mathrm{'}}}+{\boldsymbol{\mathrm{k}}}^{\boldsymbol{\mathrm{''}}}-{\boldsymbol{\mathrm{k}}}^{\boldsymbol{\mathrm{'''}}}\right)}
\nonumber \\
&=&
\frac{U}{N^2}\sum_{\sigma ,\boldsymbol{\mathrm{k}}{,\boldsymbol{\mathrm{k}}}\ensuremath{'},\boldsymbol{\mathrm{q}}}{{c^{\dagger }_{\boldsymbol{\mathrm{k}}+\boldsymbol{\mathrm{q}},\sigma }c_{\boldsymbol{\mathrm{k}},\sigma }c}^{\dagger }_{{\boldsymbol{\mathrm{k}}}^{\boldsymbol{\mathrm{'}}}-\boldsymbol{\mathrm{q}},-\sigma }c_{{\boldsymbol{\mathrm{k}}}^{\boldsymbol{\mathrm{'}}},-\sigma }}.\end{eqnarray} 
\begin{eqnarray}
\label{A8}
H\ensuremath{'}_3&=&-\frac{1}{2}\sum_{\left\langle ij\right\rangle ,\sigma {,\sigma }\ensuremath{'}}{V_{ij}c^{\dagger }_{i,\sigma }c_{i,\sigma }c^{\dagger }_{j,{\sigma }\ensuremath{'}}c_{j,{\sigma }\ensuremath{'}}}
\nonumber \\ &=&
-\frac{1}{2}\sum_{\sigma {,\sigma }\ensuremath{'}}{\sum_{\left\langle ij\right\rangle }{V_{ij}}}
\left\{
\sum_{\boldsymbol{\mathrm{k}}}{\frac{1}{\sqrt{N}}c^{\dagger }_{\boldsymbol{\mathrm{k}},\sigma }e^{i\boldsymbol{\mathrm{k}}\cdot {\boldsymbol{\mathrm{R}}}_i}}
\sum_{{\boldsymbol{\mathrm{k}}}^{\boldsymbol{\mathrm{'}}}}{\frac{1}{\sqrt{N}}c_{{\boldsymbol{\mathrm{k}}}^{\boldsymbol{\mathrm{'}}},\sigma }e^{-i{\boldsymbol{\mathrm{k}}}^{\boldsymbol{\mathrm{'}}} {\boldsymbol{\mathrm{R}}}_i}}
\right.
 \cdot \nonumber \\ &&
\;\;\;\;\;\; 
\left.
\sum_{{\boldsymbol{\mathrm{k}}}^{\boldsymbol{\mathrm{''}}}}{\frac{1}{\sqrt{N}}c^{\dagger }_{{\boldsymbol{\mathrm{k}}}^{\boldsymbol{\mathrm{''}}},{\sigma }\ensuremath{'}}e^{i{\boldsymbol{\mathrm{k}}}^{\boldsymbol{\mathrm{''}}}\cdot {\boldsymbol{\mathrm{R}}}_j}}\sum_{{\boldsymbol{\mathrm{k}}}^{\boldsymbol{\mathrm{'''}}}}\frac{1}{\sqrt{N}}c_{{\boldsymbol{\mathrm{k}}}^{\boldsymbol{\mathrm{'''}}},{\sigma }\ensuremath{'}}e^{-i{\boldsymbol{\mathrm{k}}}^{\boldsymbol{\mathrm{'''}}}\cdot {\boldsymbol{\mathrm{R}}}_j} 
\right\}
\nonumber \\ &=&
-\frac{1}{2N^2}\sum_{\sigma {,\sigma }\ensuremath{'}}
\sum_{\boldsymbol{\mathrm{k}},{\boldsymbol{\mathrm{k}}}^{\boldsymbol{\mathrm{'}}},{\boldsymbol{\mathrm{k}}}^{\boldsymbol{\mathrm{''}}}{\boldsymbol{\mathrm{k}}}^{\boldsymbol{\mathrm{'''}}}}
\sum_{\left\langle i,j\right\rangle }
\sum_{\boldsymbol{\mathrm{q}}}
\left\{
\frac{1}{N}
V_{\boldsymbol{\mathrm{q}}} \times
\right.
\nonumber \\ &&
\left.
e^{i\boldsymbol{\mathrm{q}}\cdot \left({\boldsymbol{\mathrm{R}}}_j-{\boldsymbol{\mathrm{R}}}_i\right)}
e^{i\left(\boldsymbol{\mathrm{k}}-{\boldsymbol{\mathrm{k}}}^{\boldsymbol{\mathrm{'}}}\right)\cdot {\boldsymbol{\mathrm{R}}}_i}e^{i\left({\boldsymbol{\mathrm{k}}}^{\boldsymbol{\mathrm{''}}}-{\boldsymbol{\mathrm{k}}}^{\boldsymbol{\mathrm{'''}}}\right)\cdot {\boldsymbol{\mathrm{R}}}_j}
c^{\dagger }_{\boldsymbol{\mathrm{k}},\sigma }
c_{{\boldsymbol{\mathrm{k}}}^{\boldsymbol{\mathrm{'}}},\sigma }c^{\dagger }_{{\boldsymbol{\mathrm{k}}}^{\boldsymbol{\mathrm{''}}},{\sigma }\ensuremath{'}}
c^{\dagger }_{{\boldsymbol{\mathrm{k}}}^{\boldsymbol{\mathrm{'''}}},{\sigma }\ensuremath{'}}
\right\}
\nonumber \\ &=&
-\frac{1}{2N}\sum_{\sigma {,\sigma }\ensuremath{'}}{\sum_{\boldsymbol{\mathrm{q}}}{\sum_{\boldsymbol{\mathrm{k}},{\boldsymbol{\mathrm{k}}}^{\boldsymbol{\mathrm{'}}}}{V_{\boldsymbol{\mathrm{q}}}}}c^{\dagger }_{\boldsymbol{\mathrm{k}}+\boldsymbol{\mathrm{q}},\sigma }c_{\boldsymbol{\mathrm{k}},\sigma }c^{\dagger }_{{\boldsymbol{\mathrm{k}}}^{\boldsymbol{\mathrm{'}}}-\boldsymbol{\mathrm{q}},{\sigma }\ensuremath{'}}c^{\dagger }_{{\boldsymbol{\mathrm{k}}}^{\boldsymbol{\mathrm{'}}},{\sigma }\ensuremath{'}}}.
\end{eqnarray}
\begin{eqnarray}
\label{A9}
H\ensuremath{'}_4&=&-\sum_{i,\sigma }{G_ic^{\dagger }_{i,\sigma }c_{i,\sigma }}
\nonumber \\ &=&
-\sum_{i,\sigma }{\frac{1}{\sqrt{N}}\sum_{\boldsymbol{\mathrm{k}}}{G_{\boldsymbol{\mathrm{k}}}e^{i\boldsymbol{\mathrm{k}}\cdot {\boldsymbol{\mathrm{R}}}_i}}\left(\sum_{{\boldsymbol{\mathrm{k}}}^{\boldsymbol{\mathrm{'}}}}{\frac{1}{\sqrt{N}}c^{\dagger }_{{\boldsymbol{\mathrm{k}}}^{\boldsymbol{\mathrm{'}}},\sigma }e^{i{\boldsymbol{\mathrm{k}}}^{\boldsymbol{\mathrm{'}}}\cdot {\boldsymbol{\mathrm{R}}}_i}}\right)\left(\sum_{{\boldsymbol{\mathrm{k}}}^{\boldsymbol{\mathrm{''}}}}{\frac{1}{\sqrt{N}}c_{{\boldsymbol{\mathrm{k}}}^{\boldsymbol{\mathrm{''}}},\sigma }e^{-i{\boldsymbol{\mathrm{k}}}^{\boldsymbol{\mathrm{''}}}\cdot {\boldsymbol{\mathrm{R}}}_i}}\right)}
\nonumber \\ &=&
-\frac{1}{\sqrt{N}}\sum_{\sigma }{\sum_{\boldsymbol{\mathrm{q}},\boldsymbol{\mathrm{k}}}{G_{\boldsymbol{\mathrm{q}}}c^{\dagger }_{\boldsymbol{\mathrm{k}}-\boldsymbol{\mathrm{q}},\sigma }c_{\boldsymbol{\mathrm{k}},\sigma }}}.\end{eqnarray} 

\begin{eqnarray}
\label{a63}
H\ensuremath{'}_5&=&-\mu \sum_{i,\sigma }{c^{\dagger }_{i,\sigma }c_{i,\sigma }}\mathrm{=}-\mu \sum_{i,\sigma }{\left(\sum_{\boldsymbol{\mathrm{k}}}{\frac{1}{\sqrt{N}}c^{\dagger }_{\boldsymbol{\mathrm{k}},\sigma }e^{i\boldsymbol{\mathrm{k}}\cdot {\boldsymbol{\mathrm{R}}}_i}}\right)\left(\sum_{{\boldsymbol{\mathrm{k}}}^{\boldsymbol{\mathrm{'}}}}{\frac{1}{\sqrt{N}}c_{{\boldsymbol{\mathrm{k}}}^{\boldsymbol{\mathrm{'}}},\sigma }e^{-i{\boldsymbol{\mathrm{k}}}^{\boldsymbol{\mathrm{'}}}\cdot {\boldsymbol{\mathrm{R}}}_i}}\right)}
\nonumber \\ &=&
-\mu \sum_{\sigma ,\boldsymbol{\mathrm{k}}}{c^{\dagger }_{\boldsymbol{\mathrm{k}},\sigma }c_{\boldsymbol{\mathrm{k}},\sigma }}.
\end{eqnarray} 
Adding up Eq.~\eqref{A6} - Eq.~\eqref{a63}, 
$H\ensuremath{'}_1+H\ensuremath{'}_2+H\ensuremath{'}_3+H\ensuremath{'}_4+H\ensuremath{'}_5$ produces the Bloch representation model Hamiltonian, Eq.~\eqref{eq9}.

\subsection{
Diagonalization of trial Hamiltonian of 
Eq.~\eqref{eq10} into the form Eq.~\eqref{eq11}
\label{asecA2}
}
The trial Hamiltonian of 
Eq.~\eqref{eq10} is diagonalized by following a textbook approach, such as that discussed in pg. 164 in \cite{Kittel:QTOS87}. To this end, we also require the anti-commutative relations for the operators $c_{\boldsymbol{\mathrm{k}},\sigma },c^{\dagger }_{\boldsymbol{\mathrm{k}},\sigma }$.
We write down the Heisenberg equation of motion for $c_{\boldsymbol{\mathrm{k}},\sigma }$ in terms of the trial Hamiltonian of Eq.~\eqref{eq10},
\begin{eqnarray}
\label{A10}
i\frac{dc_{\boldsymbol{\mathrm{k}},\sigma }}{dt}\equiv i{\dot{c}}_{\boldsymbol{\mathrm{k}},\sigma }=\left[c_{\boldsymbol{\mathrm{k}},\sigma },H_t\right].\end{eqnarray} 
Expanding and simplifying the commutator in Eq.~\eqref{A10},
\begin{eqnarray}
\label{A11}
\left[c_{\boldsymbol{\mathrm{k}},\sigma },H_t\right]&=&\sum_{{\boldsymbol{\mathrm{k}}}^{\boldsymbol{'}},{\sigma }\ensuremath{'}}{\left({\epsilon }_{{\boldsymbol{\mathrm{k}}}^{\boldsymbol{'}}}-\mu -A_{{\boldsymbol{\mathrm{k}}}^{\boldsymbol{'}}}\right)}\underbrace{\left(c_{\boldsymbol{\mathrm{k}},\sigma }c^{\dagger }_{{\boldsymbol{\mathrm{k}}}^{\boldsymbol{'}},{\sigma }\ensuremath{'}}c_{{\boldsymbol{\mathrm{k}}}^{\boldsymbol{'}},{\sigma }\ensuremath{'}}-c^{\dagger }_{{\boldsymbol{\mathrm{k}}}^{\boldsymbol{\mathrm{'}}},{\sigma }\ensuremath{'}}c_{{\boldsymbol{\mathrm{k}}}^{\boldsymbol{\mathrm{'}}},{\sigma }\ensuremath{'}}c_{\boldsymbol{\mathrm{k}},\sigma }\right)}_{I}
\nonumber \\ &&
-\frac{1}{2}\sum_{{\boldsymbol{\mathrm{k}}}\ensuremath{'},{\sigma }\ensuremath{'}}
\left[ 
B_{{\boldsymbol{\mathrm{k}}}\ensuremath{'}}\underbrace{\left({c^{\dagger }_{{\boldsymbol{\mathrm{k}}}\ensuremath{'},{\sigma }\ensuremath{'}}c^{\dagger }_{-{\boldsymbol{\mathrm{k}}}\ensuremath{'},-{\sigma }\ensuremath{'}}c_{\boldsymbol{\mathrm{k}},\sigma }-c}_{\boldsymbol{\mathrm{k}},\sigma }c^{\dagger }_{{\boldsymbol{\mathrm{k}}}\ensuremath{'},{\sigma }\ensuremath{'}}c^{\dagger }_{-{\boldsymbol{\mathrm{k}}}\ensuremath{'},-{\sigma }\ensuremath{'}}\right)}_{II} 
\right.
\nonumber \\ &&
\left.
+B^*_{{\boldsymbol{\mathrm{k}}}^{\boldsymbol{\mathrm{'}}}}\underbrace{\left({c_{-{\boldsymbol{\mathrm{k}}}^{\boldsymbol{\mathrm{'}}},-{\sigma }\ensuremath{'}}c_{{\boldsymbol{\mathrm{k}}}^{\boldsymbol{\mathrm{'}}},{\sigma }\ensuremath{'}}c_{\boldsymbol{\mathrm{k}},\sigma }-c}_{\boldsymbol{\mathrm{k}},\sigma }c_{-{\boldsymbol{\mathrm{k}}}^{\boldsymbol{\mathrm{'}}},-{\sigma }\ensuremath{'}}c_{{\boldsymbol{\mathrm{k}}}^{\boldsymbol{\mathrm{'}}},{\sigma }\ensuremath{'}}\right)}_{III}
\right]
\end{eqnarray} 
We now look at the $I,II,III$ terms in Eq.~\eqref{A11}
in turn. 
The term $I$ in Eq.~\eqref{A11}
can be simplified to 
\begin{eqnarray}\nonumber
I=\delta \left(\boldsymbol{\mathrm{k}}-{\boldsymbol{\mathrm{k}}}\ensuremath{'}\right){\delta }_{\sigma ,{\sigma }\ensuremath{'}}c_{{\boldsymbol{\mathrm{k}}}^{\boldsymbol{'}},{\sigma }\ensuremath{'}}\end{eqnarray} 
by using the following anti-commutative relations
\begin{eqnarray}
\label{A12}
\left\{c_{\boldsymbol{\mathrm{k}},\sigma },c_{{\boldsymbol{\mathrm{k}}}^{\boldsymbol{\mathrm{'}}},{\sigma }\ensuremath{'}}\right\}=0;\ \left\{c_{\boldsymbol{\mathrm{k}},\sigma },c^{\dagger }_{{\boldsymbol{\mathrm{k}}}^{\boldsymbol{\mathrm{'}}},{\sigma }\ensuremath{'}}\right\}=\delta \left(\boldsymbol{\mathrm{k}}-{\boldsymbol{\mathrm{k}}}\ensuremath{'}\right){\delta }_{\sigma ,{\sigma }\ensuremath{'}}.\end{eqnarray} 
The term $II$ in Eq.~\eqref{A11}
can be simplified to 
\begin{eqnarray}
\label{A13}
\nonumber 
II=\delta \left(\boldsymbol{\mathrm{k}}+{\boldsymbol{\mathrm{k}}}\ensuremath{'}\right){\delta }_{\sigma ,-{\sigma }\ensuremath{'}}c^{\dagger }_{{\boldsymbol{\mathrm{k}}}\ensuremath{'},\sigma }-\delta \left(\boldsymbol{\mathrm{k}}-{\boldsymbol{\mathrm{k}}}\ensuremath{'}\right){\delta }_{\sigma ,{\sigma }\ensuremath{'}}c^{\dagger }_{-{\boldsymbol{\mathrm{k}}}\ensuremath{'},-{\sigma }\ensuremath{'}}\end{eqnarray} 
by using the following anti-commutative relation
\begin{eqnarray}c^{\dagger }_{{-\boldsymbol{\mathrm{k}}}^{\boldsymbol{\mathrm{'}}},-{\sigma }\ensuremath{'}}c_{\boldsymbol{\mathrm{k}},\sigma }=\delta \left(\boldsymbol{\mathrm{k}}+{\boldsymbol{\mathrm{k}}}\ensuremath{'}\right){\delta }_{\sigma ,-{\sigma }\ensuremath{'}}-c_{\boldsymbol{\mathrm{k}},\sigma }c^{\dagger }_{-{\boldsymbol{\mathrm{k}}}^{\boldsymbol{\mathrm{'}}},-{\sigma }\ensuremath{'}}.\end{eqnarray} 
The term $III$ in Eq.~\eqref{A11}
can be simplified to 
\begin{eqnarray}
\nonumber 
III={c_{\boldsymbol{\mathrm{k}},\sigma }c_{-{\boldsymbol{\mathrm{k}}}^{\boldsymbol{\mathrm{'}}},-{\sigma }\ensuremath{'}}c_{{\boldsymbol{\mathrm{k}}}^{\boldsymbol{\mathrm{'}}},\sigma }-c}_{\boldsymbol{\mathrm{k}},\sigma }c_{-{\boldsymbol{\mathrm{k}}}^{\boldsymbol{\mathrm{'}}},-{\sigma }\ensuremath{'}}c_{{\boldsymbol{\mathrm{k}}}^{\boldsymbol{\mathrm{'}}},{\sigma }\ensuremath{'}}\end{eqnarray} 
by using the following anti-commutative relation,
\begin{eqnarray}
\label{A14}
\left\{c_{\boldsymbol{\mathrm{k}},\sigma },c_{-{\boldsymbol{\mathrm{k}}}^{\boldsymbol{\mathrm{'}}},-{\sigma }\ensuremath{'}}\right\}=0.\end{eqnarray} 
Putting $I,II,III$ into Eq.~\eqref{A11},
\begin{eqnarray}
\label{A15}
 i{\dot{c}}_{\boldsymbol{\mathrm{k}},\sigma }&=&\sum_{{\boldsymbol{\mathrm{k}}}^{\boldsymbol{'}},{\sigma }\ensuremath{'}}{\left({\epsilon }_{{\boldsymbol{\mathrm{k}}}^{\boldsymbol{'}}}-\mu -A_{{\boldsymbol{\mathrm{k}}}^{\boldsymbol{'}}}\right)}\mathop{\delta \left(\boldsymbol{\mathrm{k}}-{\boldsymbol{\mathrm{k}}}\ensuremath{'}\right){\delta }_{\sigma ,{\sigma }\ensuremath{'}}c_{{\boldsymbol{\mathrm{k}}}^{\boldsymbol{'}},{\sigma }\ensuremath{'}}}_{I} 
 \nonumber \\ &&
 -\frac{1}{2}\sum_{{\boldsymbol{\mathrm{k}}}\ensuremath{'},{\sigma }\ensuremath{'}}{B_{{\boldsymbol{\mathrm{k}}}\ensuremath{'}}\left[\delta \left(\boldsymbol{\mathrm{k}}+{\boldsymbol{\mathrm{k}}}\ensuremath{'}\right){\delta }_{\sigma ,-{\sigma }\ensuremath{'}}c^{\dagger }_{{\boldsymbol{\mathrm{k}}}\ensuremath{'},\sigma }-\delta \left(\boldsymbol{\mathrm{k}}-{\boldsymbol{\mathrm{k}}}\ensuremath{'}\right){\delta }_{\sigma ,{\sigma }\ensuremath{'}}c^{\dagger }_{-{\boldsymbol{\mathrm{k}}}\ensuremath{'},-{\sigma }\ensuremath{'}}\right]}
\nonumber \\& =&
E_{\boldsymbol{\mathrm{k}}}c_{\boldsymbol{\mathrm{k}},\sigma }+{\frac{1}{2}B}_{\boldsymbol{\mathrm{k}}}c^{\dagger }_{-\boldsymbol{\mathrm{k}}\boldsymbol{\mathrm{,-}}\sigma }.\end{eqnarray}
Similarly,
\begin{eqnarray}
\label{A16}
i{\dot{c}}^{\dagger }_{\boldsymbol{\mathrm{k}},\sigma \boldsymbol{\mathrm{\ }}}={-E}_{\boldsymbol{\mathrm{k}}}c^{\dagger }_{\boldsymbol{\mathrm{k}},\sigma }-\frac{1}{2}B_{\boldsymbol{\mathrm{k}}}c_{-\boldsymbol{\mathrm{k}},-\sigma }.\end{eqnarray} 
In deriving Eq.~\eqref{A15} and Eq.~\eqref{A16}
we have defined
\begin{eqnarray}
\label{A17}
E_{\boldsymbol{\mathrm{k}}}\equiv {\epsilon }_{\boldsymbol{\mathrm{k}}}-\mu -A_{\boldsymbol{\mathrm{k}}},\end{eqnarray} 
and assumed 
\begin{eqnarray}
\label{A18}
B_{\boldsymbol{\mathrm{k}}}&=&B^*_{\boldsymbol{\mathrm{k}}}\ \left(\mathrm{i.e.,\ }B_{\boldsymbol{\mathrm{k}}}\mathrm{a\ real\ number}\right);
\nonumber \\
E_{\boldsymbol{\mathrm{k}}}&=&E_{-\boldsymbol{\mathrm{k}}},B_{\boldsymbol{\mathrm{k}}}=B_{-\boldsymbol{\mathrm{k}}}.\end{eqnarray} 
For the next step, we shall perform Bogolyubov-Valatin transformation (see, e.g., pg. 307 of Haken, \cite{Haken:QFTOS83}) on $c^{\dagger }_{-\boldsymbol{\mathrm{k}}\boldsymbol{\mathrm{,-}}\sigma }$,$c_{\boldsymbol{\mathrm{-}}\boldsymbol{\mathrm{k}}\boldsymbol{\mathrm{,-}}\sigma },c_{\boldsymbol{\mathrm{k}},\sigma }{,c}^{\dagger }_{\boldsymbol{\mathrm{k}},\sigma }$ to obtain 
\begin{eqnarray}\nonumber
{\alpha }^{\dagger }_{\boldsymbol{\mathrm{k}},\sigma \mathrm{\ }}=u_{\boldsymbol{\mathrm{k}}}c^{\dagger }_{\boldsymbol{\mathrm{k}},\sigma }-v_{\boldsymbol{\mathrm{k}}}c_{-\boldsymbol{\mathrm{k}}\boldsymbol{\mathrm{,-}}\sigma },\end{eqnarray} 
\begin{eqnarray}\nonumber
{\alpha }_{\boldsymbol{\mathrm{k}},\sigma }=u_{\boldsymbol{\mathrm{k}}}c_{\boldsymbol{\mathrm{k}},\sigma }-v_{\boldsymbol{\mathrm{k}}}c^{\dagger }_{-\boldsymbol{\mathrm{k}}\boldsymbol{\mathrm{,-}}\sigma },\end{eqnarray} 
\begin{eqnarray}
\label{A19}
\nonumber
{\alpha }_{-\boldsymbol{\mathrm{k}}\boldsymbol{\mathrm{,-}}\sigma }=u_{\boldsymbol{\mathrm{k}}}v_{\boldsymbol{\mathrm{k}}}c_{-\boldsymbol{\mathrm{k}}\boldsymbol{\mathrm{,-}}\sigma }+v_{\boldsymbol{\mathrm{k}}}c^{\dagger }_{\boldsymbol{\mathrm{k}},\sigma },\end{eqnarray} 
\begin{eqnarray}{\alpha }^{\dagger }_{-\boldsymbol{\mathrm{k}}\boldsymbol{\mathrm{,-}}\sigma }=u_{\boldsymbol{\mathrm{k}}}c^{\dagger }_{-\boldsymbol{\mathrm{k}}\boldsymbol{\mathrm{,-}}\sigma }+v_{\boldsymbol{\mathrm{k}}}c_{\boldsymbol{\mathrm{k}},\sigma },\end{eqnarray} 
with the condition $u^2_{\boldsymbol{\mathrm{k}}}+v^2_{\boldsymbol{\mathrm{k}}}=1$ and $u_{\boldsymbol{\mathrm{k}}},v_{\boldsymbol{\mathrm{k}}}$ are real numbers. In addition, we shall also assume
\begin{eqnarray}
\label{A20}
u_{-\boldsymbol{\mathrm{k}}}=u_{\boldsymbol{\mathrm{k}}},v_{-\boldsymbol{\mathrm{k}}}=v_{\boldsymbol{\mathrm{k}}}.\end{eqnarray} 
Unless otherwise specify, as a short hand notation, we shall suppress the spin subscript in the$\ c_{\boldsymbol{\mathrm{k}}},c^{\boldsymbol{\dagger }}_{\boldsymbol{\mathrm{k}}}$ and ${\alpha }^{\dagger }_{\boldsymbol{\mathrm{k}}},{\alpha }_{\boldsymbol{\mathrm{k}}}$ operators with the understanding that they are implicitly $\sigma $-bearing, i.e., 
\begin{eqnarray}
\nonumber
{\alpha }^{\dagger }_{\boldsymbol{\mathrm{k}}}\equiv {\alpha }^{\dagger }_{\boldsymbol{\mathrm{k}},\sigma \mathrm{\ }};{\alpha }_{\boldsymbol{\mathrm{k}}}\equiv {\alpha }_{\boldsymbol{\mathrm{k}},\sigma };{{\alpha }_{\boldsymbol{\mathrm{-}}\boldsymbol{\mathrm{k}}}\equiv \alpha }_{\boldsymbol{\mathrm{-}}\boldsymbol{\mathrm{k}}\boldsymbol{\mathrm{,-}}\sigma };{\alpha }^{\dagger }_{-\boldsymbol{\mathrm{k}}}\equiv {\alpha }^{\dagger }_{-\boldsymbol{\mathrm{k}}\boldsymbol{\mathrm{,-}}\sigma };\end{eqnarray} 
\begin{eqnarray}
\label{A21}
c^{\dagger }_{\boldsymbol{\mathrm{k}}}\equiv c^{\dagger }_{\boldsymbol{\mathrm{k}},\sigma };\ c_{\boldsymbol{\mathrm{-}}\boldsymbol{\mathrm{k}}}\equiv c_{-\boldsymbol{\mathrm{k}}\boldsymbol{\mathrm{,-}}\sigma };c_{-\boldsymbol{\mathrm{k}}}\equiv c_{-\boldsymbol{\mathrm{k}}\boldsymbol{\mathrm{,-}}\sigma \boldsymbol{\mathrm{\ }}};c^{\dagger }_{-\boldsymbol{\mathrm{k}}}\equiv c^{\dagger }_{-\boldsymbol{\mathrm{k}}\boldsymbol{\mathrm{,-}}\sigma .}\end{eqnarray} 
The operators in Eq.~\eqref{A21}
fulfill the following anti-commutative relations:
\begin{eqnarray}
\label{A22}
\left\{{\alpha }_{\boldsymbol{\mathrm{k}}},{\alpha }^{\dagger }_{{\boldsymbol{\mathrm{k}}}^{\boldsymbol{\mathrm{'}}}}\right\}={\delta }_{{\boldsymbol{\mathrm{k}},\boldsymbol{\mathrm{k}}}\ensuremath{'}}{\delta }_{\sigma ,{\sigma }\ensuremath{'}};\ \left\{{\alpha }_{\boldsymbol{\mathrm{k}}},{\alpha }_{{\boldsymbol{\mathrm{k}}}^{\boldsymbol{\mathrm{'}}}}\right\}=0;\ \left\{{\alpha }^{\dagger }_{\boldsymbol{\mathrm{k}}},{\alpha }^{\dagger }_{{\boldsymbol{\mathrm{k}}}^{\boldsymbol{\mathrm{'}}}}\right\}=0.\end{eqnarray} 
Inversing Eq.~\eqref{A22}
we have
\begin{eqnarray}
\label{A23}
\nonumber
c_{\boldsymbol{\mathrm{k}}}=u_{\boldsymbol{\mathrm{k}}}{\alpha }_{\boldsymbol{\mathrm{k}}}+v_{\boldsymbol{\mathrm{k}}}{\alpha }^{\dagger }_{-\boldsymbol{\mathrm{k}}},\end{eqnarray} 
\begin{eqnarray}
\nonumber
c^{\dagger }_{\boldsymbol{\mathrm{k}}}=u_{\boldsymbol{\mathrm{k}}}{\alpha }^{\dagger }_{\boldsymbol{\mathrm{k}}}+v_{\boldsymbol{\mathrm{k}}}{\alpha }_{-\boldsymbol{\mathrm{k}}},\end{eqnarray} 
\begin{eqnarray}
\nonumber
c_{-\boldsymbol{\mathrm{k}}}=u_{\boldsymbol{\mathrm{k}}}{\alpha }_{-\boldsymbol{\mathrm{k}}}-v_{\boldsymbol{\mathrm{k}}}{\alpha }^{\dagger }_{\boldsymbol{\mathrm{k}}},\end{eqnarray} 
\begin{eqnarray}c^{\dagger }_{-\boldsymbol{\mathrm{k}}}=u_{\boldsymbol{\mathrm{k}}}{\alpha }^{\dagger }_{-\boldsymbol{\mathrm{k}}}-v_{\boldsymbol{\mathrm{k}}}{\alpha }_{\boldsymbol{\mathrm{k}}}.\end{eqnarray} 
Insert Eq.~\eqref{A23}
into the trial Hamiltonian $H_t$ of Eq.~\eqref{eq10},
\begin{eqnarray}
\label{A24}
H_t&=&\underbrace{\sum_{\boldsymbol{\mathrm{k}},\sigma }{{\alpha }^{\dagger }_{\boldsymbol{\mathrm{k}}}{\alpha }_{\boldsymbol{\mathrm{k}}}\left(u^2_{\boldsymbol{\mathrm{k}}}E_{\boldsymbol{\mathrm{k}}}-\frac{1}{2}u_{\boldsymbol{\mathrm{k}}}v_{\boldsymbol{\mathrm{k}}}B_{\boldsymbol{\mathrm{k}}}-\frac{1}{2}u_{\boldsymbol{\mathrm{k}}}v_{\boldsymbol{\mathrm{k}}}B^*_{\boldsymbol{\mathrm{k}}}\right)}}_{H_{t,I}} 
\nonumber \\ &&
+\underbrace{\sum_{\boldsymbol{\mathrm{k}},\sigma }{{{\alpha }_{-\boldsymbol{\mathrm{k}}}\alpha }^{\dagger }_{-\boldsymbol{\mathrm{k}}}\left(v^2_{\boldsymbol{\mathrm{k}}}E_{\boldsymbol{\mathrm{k}}}+\frac{1}{2}u_{\boldsymbol{\mathrm{k}}}v_{\boldsymbol{\mathrm{k}}}B_{\boldsymbol{\mathrm{k}}}+\frac{1}{2}u_{\boldsymbol{\mathrm{k}}}v_{\boldsymbol{\mathrm{k}}}B^*_{\boldsymbol{\mathrm{k}}}\right)}}_{H_{t,II}} 
\nonumber \\ && 
 +\underbrace{\sum_{\boldsymbol{\mathrm{k}},\sigma }{{\alpha }^{\dagger }_{\boldsymbol{\mathrm{k}}}{\alpha }^{\dagger }_{-\boldsymbol{\mathrm{k}}}\left({u_{\boldsymbol{\mathrm{k}}}v_{\boldsymbol{\mathrm{k}}}E}_{\boldsymbol{\mathrm{k}}}+
\frac{1}{2}u^2_{\boldsymbol{\mathrm{k}}}B_{\boldsymbol{\mathrm{k}}}-\frac{1}{2}v^2_{\boldsymbol{\mathrm{k}}}B^*_{\boldsymbol{\mathrm{k}}}\right)}}_{H_{t,III}} 
\nonumber \\ &&
+\underbrace{\sum_{\boldsymbol{\mathrm{k}},\sigma }{{\alpha }_{-\boldsymbol{\mathrm{k}}}{\alpha }_{\boldsymbol{\mathrm{k}}}\left({u_{\boldsymbol{\mathrm{k}}}v_{\boldsymbol{\mathrm{k}}}E}_{\boldsymbol{\mathrm{k}}}+\frac{1}{2}u^2_{\boldsymbol{\mathrm{k}}}B^*_{\boldsymbol{\mathrm{k}}}-\frac{1}{2}v^2_{\boldsymbol{\mathrm{k}}}B_{\boldsymbol{\mathrm{k}}}\right)}}_{H_{t,IV}}
\end{eqnarray}
If $H_t$ were diagonal in the $\left\{{{\alpha }^{\dagger }_{\boldsymbol{\mathrm{k}}},\alpha }_{\boldsymbol{\mathrm{k}}}\right\}$ representation, it can be expressed in the form 
\begin{eqnarray}
\label{A25}
H_t=\sum_{\boldsymbol{\mathrm{k}},\sigma }{{\lambda }_{\boldsymbol{\mathrm{k}}}{{\alpha }^{\dagger }_{\boldsymbol{\mathrm{k}}}\alpha }_{\boldsymbol{\mathrm{k}}}}+\mathrm{constant},\end{eqnarray} 
where ${\lambda }_{\boldsymbol{\mathrm{k}}}$ the eigenvalue of $H_t$. The constant term, which does not contain any $\alpha $ operator, can be neglected when applying variation calculation on $H_t$. The equations of motion for ${\alpha }_{\boldsymbol{\mathrm{k}}}$ and ${,\ \alpha }^{\dagger }_{\boldsymbol{\mathrm{k}}}$ are respectively given by
\begin{eqnarray}
\label{A26}
i\frac{d{\alpha }^{\dagger }_{\boldsymbol{\mathrm{k}}}}{dt}&=&\left[{\alpha }^{\dagger }_{\boldsymbol{\mathrm{k}}},H_t\right]=-{\lambda }_{\boldsymbol{\mathrm{k}}}{\alpha }^{\dagger }_{\boldsymbol{\mathrm{k}}},
\nonumber \\
i\frac{d{\alpha }_{\boldsymbol{\mathrm{k}}}}{dt}&=&\left[{\alpha }_{\boldsymbol{\mathrm{k}}},H_t\right]={\lambda }_{\boldsymbol{\mathrm{k}}}{\alpha }_{\boldsymbol{\mathrm{k}}}.\end{eqnarray} 
Putting $H_t$ in 
Eq.~\eqref{A24} together with ${\alpha }^{\dagger }_{\boldsymbol{\mathrm{k}}}$ in a commutator yields 
\begin{eqnarray}
\label{A27}
\left[{\alpha }^{\dagger }_{\boldsymbol{\mathrm{k}}},H_t\right]=\left[{\alpha }^{\dagger }_{\boldsymbol{\mathrm{k}}},H_{t,I}\right]+\left[{\alpha }^{\dagger }_{\boldsymbol{\mathrm{k}}},H_{t,II}\right].\end{eqnarray} 
The terms $\left[{\alpha }^{\dagger }_{\boldsymbol{\mathrm{k}}},H_{t,III}\right]\ \mathrm{and\ }\left[{\alpha }^{\dagger }_{\boldsymbol{\mathrm{k}}},H_{t,IV}\right]$ that should otherwise appear in the RHS of Eq.~\eqref{A27}
vanish due to the following anticommutative relations, 
\begin{eqnarray}
\left\{{\alpha }^{\dagger }_{\boldsymbol{\mathrm{k}}}{,\alpha }^{\dagger }_{{\boldsymbol{\mathrm{k}}}^{\boldsymbol{\mathrm{'}}}}\right\}&=&0;
\ \left\{{\alpha }_{{\boldsymbol{\mathrm{k}}}\ensuremath{'}},{\alpha }^{\dagger }_{\boldsymbol{\mathrm{k}}}\right\}=\delta \left(\boldsymbol{\mathrm{k}}\boldsymbol{\mathrm{-}}\boldsymbol{\mathrm{k}}\boldsymbol{\mathrm{'}}\right){\delta }_{\sigma ,{\sigma }\ensuremath{'}};
\nonumber \\ 
\left\{{\alpha }_{-{\boldsymbol{\mathrm{k}}}\ensuremath{'}},{\alpha }^{\dagger }_{\boldsymbol{\mathrm{k}}}\right\}&=&\delta \left(\boldsymbol{\mathrm{k}}\boldsymbol{\mathrm{+}}\boldsymbol{\mathrm{k}}\boldsymbol{\mathrm{'}}\right){\delta }_{\sigma ,{-\sigma }\ensuremath{'}}.
\nonumber
\end{eqnarray} 
Using the following results, 
\begin{eqnarray}
\label{A28}
\left[{\alpha }^{\dagger }_{\boldsymbol{\mathrm{k}}},{\alpha }_{{\boldsymbol{\mathrm{-}}\boldsymbol{\mathrm{k}}}^{\boldsymbol{\mathrm{'}}}}{\alpha }^{\dagger }_{{-\boldsymbol{\mathrm{k}}}^{\boldsymbol{\mathrm{'}}}}\right]&=&{{{\delta }_{{\boldsymbol{\mathrm{k}}\boldsymbol{\mathrm{,-}}\boldsymbol{\mathrm{k}}}\ensuremath{'}}\delta }_{\sigma ,-{\sigma }\ensuremath{'}}\alpha }^{\dagger }_{\boldsymbol{\mathrm{-}}{\boldsymbol{\mathrm{k}}}^{\boldsymbol{\mathrm{'}}}}; \nonumber \\ 
\left[{\alpha }^{\dagger }_{\boldsymbol{\mathrm{k}}},{\alpha }^{\dagger }_{{\boldsymbol{\mathrm{k}}}^{\boldsymbol{\mathrm{'}}}}{\alpha }^{\dagger }_{{\boldsymbol{\mathrm{k}}}^{\boldsymbol{\mathrm{'}}}}\right]
&=&
-{{\delta }_{{\boldsymbol{\mathrm{k}},\boldsymbol{\mathrm{k}}}\ensuremath{'}}\delta }_{\sigma ,{\sigma }\ensuremath{'}}{\alpha }^{\dagger }_{{\boldsymbol{\mathrm{k}}}^{\boldsymbol{\mathrm{'}}}},\end{eqnarray} 
the commutators in the RHS of Eq.~\eqref{A27}
are reduced to  
\begin{eqnarray}
\label{A29}
\left[{\alpha }^{\dagger }_{\boldsymbol{\mathrm{k}}},H_{t,I}\right]=-\left(u^2_{\boldsymbol{\mathrm{k}}}E_{\boldsymbol{\mathrm{k}}}-u_{\boldsymbol{\mathrm{k}}}v_{\boldsymbol{\mathrm{k}}}B_{\boldsymbol{\mathrm{k}}}\right){\alpha }^{\dagger }_{\boldsymbol{\mathrm{k}}}\end{eqnarray} 
and  
\begin{eqnarray}
\label{A30}
\left[{\alpha }^{\dagger }_{\boldsymbol{\mathrm{k}}},H_{t,II}\right]={\alpha }^{\dagger }_{\boldsymbol{\mathrm{k}}}\left(v^2_{\boldsymbol{\mathrm{k}}}E_{\boldsymbol{\mathrm{k}}}+u_{\boldsymbol{\mathrm{k}}}v_{\boldsymbol{\mathrm{k}}}B_{\boldsymbol{\mathrm{k}}}\right).\end{eqnarray} 
By Eq.~\eqref{A27}, Eq.~\eqref{A29} and Eq.~\eqref{A30},
the commutator $\left[{\alpha }^{\dagger }_{\boldsymbol{\mathrm{k}}},H_t\right]$ in Eq.~\eqref{A26} 
becomes
\begin{eqnarray}
\label{A31}
\left[{\alpha }^{\dagger }_{\boldsymbol{\mathrm{k}}},H_t\right]=-\left[\left(u^2_{\boldsymbol{\mathrm{k}}}-v^2_{\boldsymbol{\mathrm{k}}}\right)E_{\boldsymbol{\mathrm{k}}}-2u_{\boldsymbol{\mathrm{k}}}v_{\boldsymbol{\mathrm{k}}}B_{\boldsymbol{\mathrm{k}}}\right]{\alpha }^{\dagger }_{\boldsymbol{\mathrm{k}}}.\end{eqnarray} 
Comparing Eq.~\eqref{A26} and Eq.~\eqref{A31},
the eigenvalue ${\lambda }_{\boldsymbol{\mathrm{k}}}$ is identified, namely,
\begin{eqnarray}
\label{A32}
{\lambda }_{\boldsymbol{\mathrm{k}}}\mathrm{=}\left(u^2_{\boldsymbol{\mathrm{k}}}-v^2_{\boldsymbol{\mathrm{k}}}\right)E_{\boldsymbol{\mathrm{k}}}-2u_{\boldsymbol{\mathrm{k}}}v_{\boldsymbol{\mathrm{k}}}B_{\boldsymbol{\mathrm{k}}}.\end{eqnarray} 
We next look at $u^2_{\boldsymbol{\mathrm{k}}},v^2_{\boldsymbol{\mathrm{k}}}$ in Eq.~\eqref{A32}.
Upon diagonalization, the coefficients for the off-diagonal terms of $H_t$ in Eq.~\eqref{A24}
(i.e. $H_{t,III}$, $H_{t,IV})$, which involve ${\alpha }^{\dagger }_{\boldsymbol{\mathrm{k}}}{\alpha }^{\dagger }_{\boldsymbol{\mathrm{-}}\boldsymbol{\mathrm{k}}}$,$\ {\alpha }_{-\boldsymbol{\mathrm{k}}}{\alpha }_{\boldsymbol{\mathrm{k}}}$, should vanish, i.e., 
\begin{eqnarray}
\label{A33}
{u_{\boldsymbol{\mathrm{k}}}v_{\boldsymbol{\mathrm{k}}}E}_{\boldsymbol{\mathrm{k}}}+\frac{1}{2}\left(u^2_{\boldsymbol{\mathrm{k}}}-v^2_{\boldsymbol{\mathrm{k}}}\right)B_{\boldsymbol{\mathrm{k}}}=0.\end{eqnarray} 
Since $u^2_{\boldsymbol{\mathrm{k}}}+v^2_{\boldsymbol{\mathrm{k}}}=1$, we parametrise
\begin{eqnarray}
\label{A34}
u_{\boldsymbol{\mathrm{k}}}={\mathrm{cos} \frac{{\theta }_{\boldsymbol{\mathrm{k}}}}{2}\ },\ v_{\boldsymbol{\mathrm{k}}}={\mathrm{sin} \frac{{\theta }_{\boldsymbol{\mathrm{k}}}}{2}\ }.\end{eqnarray} 
It is also assumed that the parameter ${\theta }_{\boldsymbol{\mathrm{k}}}={\theta }_{\boldsymbol{\mathrm{-}}\boldsymbol{\mathrm{k}}}$ so that Eq.~\eqref{A34}
is consistent with Eq.~\eqref{A20}.
The will cast Eq.~\eqref{A33}
into 
\begin{eqnarray}
\label{A35}
{{\mathrm{cos} \frac{{\theta }_{\boldsymbol{\mathrm{k}}}}{2}\ }{\mathrm{sin} \frac{{\theta }_{\boldsymbol{\mathrm{k}}}}{2}\ }E}_{\boldsymbol{\mathrm{k}}}+\frac{1}{2}\left({{\mathrm{cos}}^{\mathrm{2}} \frac{{\theta }_{\boldsymbol{\mathrm{k}}}}{2}\ }-{{\mathrm{sin}}^{\mathrm{2}} \frac{{\theta }_{\boldsymbol{\mathrm{k}}}}{2}\ }\right)B_{\boldsymbol{\mathrm{k}}}&=&0
\nonumber \\ 
\Rightarrow {\mathrm{tan} {\theta }_{\boldsymbol{\mathrm{k}}}\ }&=&-\frac{B_{\boldsymbol{\mathrm{k}}}}{E_{\boldsymbol{\mathrm{k}}}}.\end{eqnarray} 
Hence, from Eq.~\eqref{A33}, Eq.~\eqref{A34} and Eq.~\eqref{A35},
the eigenvalue ${\lambda }_{\boldsymbol{\mathrm{k}}}$ in Eq.~\eqref{A32}
is now simplified to
\begin{eqnarray}
\label{A36}
{\lambda }_{\boldsymbol{\mathrm{k}}}={\mathrm{cos} {\theta }_{\boldsymbol{\mathrm{k}}}\ }E_{\boldsymbol{\mathrm{k}}}-{\mathrm{sin} {\theta }_{\boldsymbol{\mathrm{k}}}\ }B_{\boldsymbol{\mathrm{k}}}=\sqrt{E^{\mathrm{2}}_{\boldsymbol{\mathrm{k}}}+B^{\mathrm{2}}_{\boldsymbol{\mathrm{k}}}}.\end{eqnarray} 
This completes the derivation of 
Eq.~\eqref{eq12}. We note that the result Eq.~\eqref{A36}
does not have 4 branches as obtained in \cite{Micnas:PRB88} because no charge ordered states are considered in the present model. 
The term $H_{t,II}$ in Eq.~\eqref{A24}
can be cast into the form
\begin{eqnarray}
\label{A37}
H_{t,II}&=&\sum_{\boldsymbol{\mathrm{k}},\sigma }{{{\alpha }_{-\boldsymbol{\mathrm{k}}}\alpha }^{\dagger }_{-\boldsymbol{\mathrm{k}}}\left(v^2_{\boldsymbol{\mathrm{k}}}E_{\boldsymbol{\mathrm{k}}}+u_{\boldsymbol{\mathrm{k}}}v_{\boldsymbol{\mathrm{k}}}B_{\boldsymbol{\mathrm{k}}}\right)}
\nonumber \\ &=&
\sum_{\boldsymbol{\mathrm{k}},\sigma }{{\alpha }^{\dagger }_{\boldsymbol{\mathrm{k}}}{\alpha }_{\boldsymbol{\mathrm{k}}}\left(v^2_{\boldsymbol{\mathrm{k}}}E_{\boldsymbol{\mathrm{k}}}+u_{\boldsymbol{\mathrm{k}}}v_{\boldsymbol{\mathrm{k}}}B_{\boldsymbol{\mathrm{k}}}\right)}+\underbrace{\sum_{\boldsymbol{\mathrm{k}},\sigma }{v^2_{\boldsymbol{\mathrm{k}}}E_{\boldsymbol{\mathrm{k}}}+u_{\boldsymbol{\mathrm{k}}}v_{\boldsymbol{\mathrm{k}}}B_{\boldsymbol{\mathrm{k}}}}}_{\mathrm{constant}}\ \end{eqnarray} 
where we have used the relation ${{\alpha }_{-\boldsymbol{\mathrm{k}}}\alpha }^{\dagger }_{-\boldsymbol{\mathrm{k}}}=1-{\alpha }^{\dagger }_{-\boldsymbol{\mathrm{k}}}{\alpha }_{-\boldsymbol{\mathrm{k}}}$ and replaced $-\boldsymbol{\mathrm{k}}\to \boldsymbol{\mathrm{k}}$. The constant term in Eq.~\eqref{A37}
is evaluated by summing over the spin degree of freedom, 
\begin{eqnarray}
\label{A38}
\mathrm{constant} &=&\sum_{\boldsymbol{\mathrm{k}},\sigma }{\left(v^2_{\boldsymbol{\mathrm{k}}}E_{\boldsymbol{\mathrm{k}}}+u_{\boldsymbol{\mathrm{k}}}v_{\boldsymbol{\mathrm{k}}}B_{\boldsymbol{\mathrm{k}}}\right)}
\nonumber \\ &=&
\sum_{\boldsymbol{\mathrm{k}},\sigma \boldsymbol{\mathrm{\ }}}{\left(\frac{E_{\boldsymbol{\mathrm{k}}}}{2}-\frac{1}{2}{\lambda }_{\boldsymbol{\mathrm{k}}}\right)=\sum_{\boldsymbol{\mathrm{k}}}{E_{\boldsymbol{\mathrm{k}}}}-\sum_{\boldsymbol{\mathrm{k}}}{{\lambda }_{\boldsymbol{\mathrm{k}}}}}\equiv C.\end{eqnarray} 
In arriving at Eq.~\eqref{A38}
the following identities, which can be derived from Eq.~\eqref{A34} and Eq.~\eqref{A35},
have been used, namely, 
\begin{eqnarray}
\label{A39}
u_{\boldsymbol{\mathrm{k}}}v_{\boldsymbol{\mathrm{k}}}&=&-\frac{1}{2}\frac{B_{\boldsymbol{\mathrm{k}}}}{\sqrt{B^2_{\boldsymbol{\mathrm{k}}}+E^2_{\boldsymbol{\mathrm{k}}}}},
\nonumber \\ 
v^2_{\boldsymbol{\mathrm{k}}}&=&\frac{1}{2}-\frac{1}{2}\frac{E_{\boldsymbol{\mathrm{k}}}}{\sqrt{B^2_{\boldsymbol{\mathrm{k}}}+E^2_{\boldsymbol{\mathrm{k}}}}},
\nonumber \\ 
{u^2_{\boldsymbol{\mathrm{k}}}-v}^2_{\boldsymbol{\mathrm{k}}}&=&\frac{E_{\boldsymbol{\mathrm{k}}}}{\sqrt{B^2_{\boldsymbol{\mathrm{k}}}+E^2_{\boldsymbol{\mathrm{k}}}}}.\end{eqnarray} 
Be noted that the expression for the constant in Eq.~\eqref{A38},
$C=\sum_{\boldsymbol{\mathrm{k}}}{E_{\boldsymbol{\mathrm{k}}}}-\sum_{\boldsymbol{\mathrm{k}}}{{\lambda }_{\boldsymbol{\mathrm{k}}}}$, is defined with summation over the variable $\boldsymbol{\mathrm{k}}$ only. Upon diagonalization, $H_{t,III}$, $H_{t,IV}$ in $H_t$ in Eq.~\eqref{A24}
vanish. Combining $H_{t,II}$ [from Eq.~\eqref{A37}]
with $H_{t,I}$ in Eq.~\eqref{A24}
gives 
Eq.~\eqref{eq11},
\begin{eqnarray}
\label{A40}
H_t=\sum_{\boldsymbol{\mathrm{k}},\sigma }{{\lambda }_{\boldsymbol{\mathrm{k}}}{\alpha }^{\dagger }_{\boldsymbol{\mathrm{k}}}{\alpha }_{\boldsymbol{\mathrm{k}}}}+C.\end{eqnarray} 
This completes the diagonalization of the trial Hamiltonian\textit{.}

\subsection{
Derivation of 
${{F}}_{{t}},\ {\left\langle {H}\right\rangle }_{{t}},{\left\langle {{H}}_{{t}}\right\rangle }_{{t}}$ 
as appear in 
Eq.~\eqref{eq15}
\label{asecA3}
}
\noindent {\bf Derivation of} ${\boldsymbol{F}}_{\boldsymbol{t}}$:\\
Partition function $Z_t$ for trial Hamiltonian $H_t$ is defined as 
\begin{eqnarray}
\label{A41}
Z_t&=&\mathrm{Tr}\ e^{-\beta H_t}\mathrm{=}e^{-\beta C}\mathrm{\cdot }\mathrm{Tr}\ {\mathrm{e}}^{-\beta \left(\sum_{\boldsymbol{\mathrm{k}},\sigma }{{\lambda }_{\boldsymbol{\mathrm{k}}}{\alpha }^{\dagger }_{\boldsymbol{\mathrm{k}}}{\alpha }_{\boldsymbol{\mathrm{k}}}}\right)}
\nonumber \\ &=&
e^{-\beta C}\mathrm{\cdot }\prod_{\boldsymbol{\mathrm{k}},\sigma }{\left[2{\mathrm{e}}^{-\frac{\beta {\lambda }_{\boldsymbol{\mathrm{k}}}}{2}}\cdot \frac{1}{2}\left({\mathrm{e}}^{\frac{\beta {\lambda }_{\boldsymbol{\mathrm{k}}}}{2}}+{\mathrm{e}}^{-\frac{\beta {\lambda }_{\boldsymbol{\mathrm{k}}}}{2}}\right)\right]}
\nonumber \\ &=&
e^{-\beta C}\mathrm{\cdot }\prod_{\boldsymbol{\mathrm{k}},\sigma }{2{\mathrm{e}}^{-\frac{\beta {\lambda }_{\boldsymbol{\mathrm{k}}}}{2}}{\mathrm{cosh} \left(\frac{\beta {\lambda }_{\boldsymbol{\mathrm{k}}}}{2}\right)\ }},\end{eqnarray} 
where the constant $C$ is as defined in Eq.~\eqref{A38}.
The trial free energy is
\begin{eqnarray}
\label{A42}
F_t&=&-k_BT{\mathrm{ln} Z_t\ }=C-k_BT{\mathrm{ln} \left\{\prod_{\boldsymbol{\mathrm{k}},\sigma }{\left[{\mathrm{e}}^{-\frac{\beta {\lambda }_{\boldsymbol{\mathrm{k}}}}{2}}\cdot 2{\mathrm{cosh} \left(\frac{\beta {\lambda }_{\boldsymbol{\mathrm{k}}}}{2}\right)\ }\right]}\right\}\ }
\nonumber \\ &=&
-k_BT\sum_{\boldsymbol{\mathrm{k}},\sigma }{\left(-\frac{\beta {\lambda }_{\boldsymbol{\mathrm{k}}}}{2}\right)}-\frac{1}{\beta }\sum_{\boldsymbol{\mathrm{k}},\sigma }{{\mathrm{ln\ } \left[\mathrm{2cosh}\left(\frac{\beta {\lambda }_{\boldsymbol{\mathrm{k}}}}{2}\right)\right]\ }}+C
\nonumber \\ &=&
\frac{1}{2}\sum_{\boldsymbol{\mathrm{k}},\sigma }{{\lambda }_{\boldsymbol{\mathrm{k}}}}-\frac{1}{\beta }\sum_{\boldsymbol{\mathrm{k}},\sigma }{{\mathrm{ln\ } \left[\mathrm{2cosh}\left(\frac{\beta {\lambda }_{\boldsymbol{\mathrm{k}}}}{2}\right)\right]\ }}+\left(\sum_{\boldsymbol{\mathrm{k}}}{E_{\boldsymbol{\mathrm{k}}}}-\sum_{\boldsymbol{\mathrm{k}}}{{\lambda }_{\boldsymbol{\mathrm{k}}}}\right)
\nonumber \\ &=&
\sum_{\boldsymbol{\mathrm{k}}}{{\lambda }_{\boldsymbol{\mathrm{k}}}}-\frac{1}{\beta }\sum_{\boldsymbol{\mathrm{k}},\sigma }{{\mathrm{ln\ } \left[\mathrm{2cosh}\left(\frac{\beta {\lambda }_{\boldsymbol{\mathrm{k}}}}{2}\right)\right]\ }}+\sum_{\boldsymbol{\mathrm{k}}}{E_{\boldsymbol{\mathrm{k}}}}-\sum_{\boldsymbol{\mathrm{k}}}{{\lambda }_{\boldsymbol{\mathrm{k}}}}
\nonumber \\ &=&
-\frac{1}{\beta }\sum_{\boldsymbol{\mathrm{k}},\sigma }{{\mathrm{ln\ } \left[\mathrm{2cosh}\left(\frac{\beta {\lambda }_{\boldsymbol{\mathrm{k}}}}{2}\right)\right]\ }}+\sum_{\boldsymbol{\mathrm{k}}}{{(\epsilon }_{\boldsymbol{\mathrm{k}}}-\mu -A_{\boldsymbol{\mathrm{k}}})}.\end{eqnarray} 

\noindent {\bf Derivation of ${\left\langle \boldsymbol{H}\right\rangle }_{\boldsymbol{t}}$}: \\ 
The derivation ${\left\langle H\right\rangle }_t$ proceeds by beginning with the definition of the model Hamiltonian $H$, Eq.~\eqref{eq9},
which is separated into three parts, 
\begin{eqnarray}\nonumber
{\left\langle H\right\rangle }_t={\left\langle h_1\right\rangle }_t+{\left\langle h_2\right\rangle }_t+{\left\langle h_3\right\rangle }_t,\end{eqnarray} 
\begin{eqnarray}
\label{A43}
h_1&=&\sum_{\boldsymbol{\mathrm{k}},\sigma }{\left({\epsilon }_{\mathrm{k}}-\mu \right)c^{\dagger }_{\boldsymbol{\mathrm{k}},\sigma }c_{\boldsymbol{\mathrm{k}},\sigma }},
\nonumber \\ 
h_2&=&-\frac{1}{2N}\sum_{\sigma ,{\sigma }\ensuremath{'}}{\sum_{\boldsymbol{\mathrm{k}},{\boldsymbol{\mathrm{k}}}\ensuremath{'},\boldsymbol{\mathrm{q}}}{\left(V_{\boldsymbol{\mathrm{q}}}-{2U\delta }_{\sigma ,{-\sigma }\ensuremath{'}}\right)c^{\dagger }_{\boldsymbol{\mathrm{k}}+\boldsymbol{\mathrm{q}},\sigma }c_{\boldsymbol{\mathrm{k}},\sigma }}c^{\dagger }_{{\boldsymbol{\mathrm{k}}}\ensuremath{'}-\boldsymbol{\mathrm{q}},{\sigma }\ensuremath{'}}c_{{\boldsymbol{\mathrm{k}}}\ensuremath{'},{\sigma }\ensuremath{'}}},
\nonumber \\ 
h_3&=&-\frac{1}{\sqrt{N}}\sum_{\sigma }{\sum_{\boldsymbol{\mathrm{k}},\boldsymbol{\mathrm{q}}}{G_{\boldsymbol{\mathrm{q}}}c^{\dagger }_{\boldsymbol{\mathrm{k}}-\boldsymbol{\mathrm{q}},\sigma }c_{\boldsymbol{\mathrm{k}},\sigma }}}.\end{eqnarray} 
In terms of the operators ${\alpha }_{\boldsymbol{\mathrm{k}},\sigma },\ {\alpha }^{\dagger }_{\boldsymbol{\mathrm{k}},\sigma }$, ${\left\langle h_1\right\rangle }_t$ reads
\begin{eqnarray}
\label{A44}
{\left\langle h_1\right\rangle }_t&=&{\left\langle \sum_{\boldsymbol{\mathrm{k}},\sigma }{\left({\epsilon }_{\boldsymbol{\mathrm{k}}}-\mu \right)\left(u_{\boldsymbol{\mathrm{k}}}{\alpha }^{\dagger }_{\boldsymbol{\mathrm{k}},\sigma }+v_{\boldsymbol{\mathrm{k}}}{\alpha }_{-\boldsymbol{\mathrm{k}},-\sigma }\right)\left(u_{\boldsymbol{\mathrm{k}}}{\alpha }_{\boldsymbol{\mathrm{k}},\sigma }+v_{\boldsymbol{\mathrm{k}}}{\alpha }^{\dagger }_{-\boldsymbol{\mathrm{k}},-\sigma }\right)}\right\rangle }_t
\nonumber \\ &=&
{\left\langle \sum_{\boldsymbol{\mathrm{k}},\sigma }{\left({\epsilon }_{\boldsymbol{\mathrm{k}}}-\mu \right)\left[u^2_{\boldsymbol{\mathrm{k}}}{\alpha }^{\dagger }_{\boldsymbol{\mathrm{k}},\sigma }{\alpha }_{\boldsymbol{\mathrm{k}},\sigma }+v^2_{\boldsymbol{\mathrm{k}}}\left(1-{\alpha }^{\dagger }_{\boldsymbol{\mathrm{-}}\boldsymbol{\mathrm{k}},-\sigma }{\alpha }_{\boldsymbol{\mathrm{-}}\boldsymbol{\mathrm{k}},-\sigma }\right)\right]}\right\rangle }_t.\end{eqnarray} 
The expectation values ${\left\langle {\alpha }^{\dagger }_{\boldsymbol{\mathrm{k}},\sigma }{\alpha }^{\dagger }_{\boldsymbol{\mathrm{-}}\boldsymbol{\mathrm{k}},-\sigma }\right\rangle }_t,\ {\left\langle {\alpha }_{\boldsymbol{\mathrm{-}}\boldsymbol{\mathrm{k}},-\sigma }{\alpha }_{\boldsymbol{\mathrm{-}}\boldsymbol{\mathrm{k}},-\sigma }\right\rangle }_t$ in Eq.~\eqref{A44}
vanish in representation in which ${\alpha }^{\dagger }_{\boldsymbol{\mathrm{k}},\sigma }{,\alpha }_{\boldsymbol{\mathrm{k}},\sigma }$ are diagonal. There are two expectation values left to be evaluated in Eq.~\eqref{A44},
i.e., \\ ${\left\langle {\alpha }^{\dagger }_{\boldsymbol{\mathrm{k}},\sigma }{\alpha }_{\boldsymbol{\mathrm{k}},\sigma }\right\rangle }_t$ and ${\left\langle {\alpha }^{\dagger }_{\boldsymbol{\mathrm{-}}\boldsymbol{\mathrm{k}},-\sigma }{\alpha }_{\boldsymbol{\mathrm{-}}\boldsymbol{\mathrm{k}},-\sigma }\right\rangle }_t$. 
${\left\langle {\alpha }_{\boldsymbol{\mathrm{k}},\sigma }{\alpha }^{\dagger }_{\boldsymbol{\mathrm{k}},\sigma }\right\rangle }_t$ is evaluated via
\begin{eqnarray}
\label{A45}
{\left\langle {\alpha }^{\dagger }_{\boldsymbol{\mathrm{k}},\sigma }{\alpha }_{\boldsymbol{\mathrm{k}},\sigma }\right\rangle }_t&=&\frac{\mathrm{Tr}\ \left({\alpha }^{\dagger }_{\boldsymbol{\mathrm{k}},\sigma }{\alpha }_{\boldsymbol{\mathrm{k}},\sigma }\right)e^{-\beta H_t}}{\mathrm{Tr}\ \left(e^{-\beta H_t}\right)}
\nonumber \\ &=&
-\frac{\partial }{\partial \left(\beta {\lambda }_{\boldsymbol{\mathrm{k}}}\right)}{\mathrm{ln} Z_t\ }=\frac{1}{2}-\frac{1}{2}{\mathrm{tanh} \left(\frac{\beta {\lambda }_{\boldsymbol{\mathrm{k}}}}{2}\right)\ }.\end{eqnarray} 
In arriving at A 45 we have ignored the contribution from constant term $C$ in $H_t$ of A 41. 

The subscrpts in ${\left\langle {\alpha }^{\dagger }_{\boldsymbol{\mathrm{-}}\boldsymbol{\mathrm{k}},-\sigma }{\alpha }_{\boldsymbol{\mathrm{-}}\boldsymbol{\mathrm{k}},-\sigma }\right\rangle }_t$ can be manipulated so that it can be expressed in terms of ${\left\langle {{\alpha }^{\dagger }_{\boldsymbol{\mathrm{k}},\sigma }\alpha }_{\boldsymbol{\mathrm{k}},\sigma }\right\rangle }_t$, 
\begin{eqnarray}
\label{A46}
-\sum_{\boldsymbol{\mathrm{k}},\sigma }{\left({\epsilon }_{\boldsymbol{\mathrm{k}}}-\mu \right)v^2_{\boldsymbol{\mathrm{k}}}{\left\langle {\alpha }^{\dagger }_{\boldsymbol{\mathrm{-}}\boldsymbol{\mathrm{k}}\boldsymbol{\mathrm{,-}}\sigma }{\alpha }_{\boldsymbol{\mathrm{-}}\boldsymbol{\mathrm{k}}\boldsymbol{\mathrm{,-}}\sigma }\right\rangle }_t}=-\sum_{\boldsymbol{\mathrm{k}},\sigma }{\left({\epsilon }_{\boldsymbol{\mathrm{k}}}-\mu \right)v^2_{\boldsymbol{\mathrm{k}},\sigma }{\left\langle {\alpha }^{\dagger }_{\boldsymbol{\mathrm{k}},\sigma }{\alpha }_{\boldsymbol{\mathrm{k}},\sigma }\right\rangle }_t},\end{eqnarray} 
where we have made use of ${\epsilon }_{-\boldsymbol{\mathrm{k}}}={\epsilon }_{\boldsymbol{\mathrm{k}}}$. Using Eq.~\eqref{A39}, Eq.~\eqref{A45}, Eq.~\eqref{A46},
the square bracket term in Eq.~\eqref{A44}
can be simplified to
\begin{eqnarray}
\label{A47}
\left[\cdots \right]=\left(u^2_{\boldsymbol{\mathrm{k}}}-v^2_{\boldsymbol{\mathrm{k}}}\right){\left\langle {\alpha }^{\dagger }_{\boldsymbol{\mathrm{k}},\sigma }{\alpha }^{\dagger }_{\boldsymbol{\mathrm{k}},\sigma }\right\rangle }_t+v^2_{\boldsymbol{\mathrm{k}}}=\frac{1}{2}-\frac{E_{\boldsymbol{\mathrm{k}}}}{2{\lambda }_{\boldsymbol{\mathrm{k}}}\ }{\mathrm{tanh} \left(\frac{\beta {\lambda }_{\boldsymbol{\mathrm{k}}}}{2}\right)\ }.\end{eqnarray} 
\begin{eqnarray}
\label{A48}
\Rightarrow {\left\langle h_1\right\rangle }_t=\frac{1}{2}\sum_{\boldsymbol{\mathrm{k}},\sigma }{\left({\epsilon }_{\boldsymbol{\mathrm{k}}}-\mu \right)\left[1-\frac{E_{\boldsymbol{\mathrm{k}}}}{{\lambda }_{\boldsymbol{\mathrm{k}}}\ }{\mathrm{tanh} \left(\frac{\beta {\lambda }_{\boldsymbol{\mathrm{k}}}}{2}\right)\ }\right]}.\end{eqnarray} 
To evaluate ${\left\langle h_2\right\rangle }_t$, we rearrange ${\left\langle c^{\dagger }_{\boldsymbol{\mathrm{k}}+\boldsymbol{\mathrm{q}},\sigma }c_{\boldsymbol{\mathrm{k}},\sigma }c^{\dagger }_{{\boldsymbol{\mathrm{k}}}\ensuremath{'}-\boldsymbol{\mathrm{q}}{,\sigma }\ensuremath{'}}c_{{\boldsymbol{\mathrm{k}}}\ensuremath{'},{\sigma }\ensuremath{'}}\right\rangle }_t$ by reverting the position of the $c^{\dagger }$operator twice and using anti-commutative relations for $c^{\dagger }_{\boldsymbol{\mathrm{k}}},c_{\boldsymbol{\mathrm{k}}}$ to arrive at 
\begin{eqnarray}
\label{A49}
{\left\langle c^{\dagger }_{\boldsymbol{\mathrm{k}}+\boldsymbol{\mathrm{q}},\sigma }c_{\boldsymbol{\mathrm{k}},\sigma }c^{\dagger }_{{\boldsymbol{\mathrm{k}}}\ensuremath{'}-\boldsymbol{\mathrm{q}}{,\sigma }\ensuremath{'}}c_{{\boldsymbol{\mathrm{k}}}\ensuremath{'},{\sigma }\ensuremath{'}}\right\rangle }_t&=&{-\left\langle c^{\dagger }_{\boldsymbol{\mathrm{k}}+\boldsymbol{\mathrm{q}},\sigma }{c^{\dagger }_{{\boldsymbol{\mathrm{k}}}\ensuremath{'}-\boldsymbol{\mathrm{q}}{,\sigma }\ensuremath{'}}c}_{\boldsymbol{\mathrm{k}},\sigma }c_{{\boldsymbol{\mathrm{k}}}\ensuremath{'},{\sigma }\ensuremath{'}}\right\rangle }_t
\nonumber \\ &=&
{\left\langle c^{\dagger }_{\boldsymbol{\mathrm{k}}+\boldsymbol{\mathrm{q}},\sigma }{c^{\dagger }_{{\boldsymbol{\mathrm{k}}}\ensuremath{'}-\boldsymbol{\mathrm{q}}{,\sigma }\ensuremath{'}}c_{{\boldsymbol{\mathrm{k}}}\ensuremath{'},{\sigma }\ensuremath{'}}c}_{\boldsymbol{\mathrm{k}},\sigma }\right\rangle }_t.\end{eqnarray} 
In view of Eq.~\eqref{A49},
$h_2$ reads (after a few manipulative steps in the subscript symbols)
\begin{eqnarray}
\label{A50}
h_2=-\frac{1}{2N}\sum_{\sigma }{\sum_{\boldsymbol{\mathrm{k}},\boldsymbol{\mathrm{q}}}{\left(V_{\boldsymbol{\mathrm{q}}\boldsymbol{-}\boldsymbol{\mathrm{k}}}-\mathrm{2}U\right){\left\langle c^{\dagger }_{\boldsymbol{\mathrm{q}},\sigma }{c^{\dagger }_{\mathrm{-}\boldsymbol{\mathrm{q}}\boldsymbol{\mathrm{,-}}\sigma }c_{\mathrm{-}\boldsymbol{\mathrm{k}}\mathrm{\ },-\sigma }c}_{\boldsymbol{\mathrm{k}},\sigma }\right\rangle }_t}}.\end{eqnarray} 
It can be shown that ${\left\langle c^{\dagger }_{\boldsymbol{\mathrm{q}},\sigma }{c^{\dagger }_{\mathrm{-}\boldsymbol{\mathrm{q}}\boldsymbol{\mathrm{,-}}\sigma }c_{\mathrm{-}\boldsymbol{\mathrm{k}}\mathrm{\ },-\sigma }c}_{\boldsymbol{\mathrm{k}},\sigma }\right\rangle }_t={\left\langle c^{\dagger }_{\boldsymbol{\mathrm{q}},\sigma }c^{\dagger }_{\mathrm{-}\boldsymbol{\mathrm{q}}\boldsymbol{\mathrm{,-}}\sigma }\right\rangle }_t{\left\langle {c_{\mathrm{-}\boldsymbol{\mathrm{k}}\mathrm{\ },-\sigma }c}_{\boldsymbol{\mathrm{k}},\sigma }\right\rangle }_t$ provided $\boldsymbol{\mathrm{q}} {\mathrm{\neq }}\boldsymbol{\mathrm{k}}$, or equivalently, $V_0$ is excluded in the summation in Eq.~\eqref{A50}.
${\left\langle c^{\dagger }_{\boldsymbol{\mathrm{q}},\sigma }c^{\dagger }_{\mathrm{-}\boldsymbol{\mathrm{q}}\boldsymbol{\mathrm{,-}}\sigma }\right\rangle }_t$ is given by the following expression [after some algebra, and making use of Eq.~\eqref{A39}, Eq.~\eqref{A45}, Eq.~\eqref{A46}, Eq.~\eqref{A51}],
\begin{eqnarray}
\label{A51}
{\left\langle c^{\dagger }_{\boldsymbol{\mathrm{q}},\sigma }c^{\dagger }_{\mathrm{-}\boldsymbol{\mathrm{q}}\boldsymbol{\mathrm{,-}}\sigma }\right\rangle }_t&=&u_{\boldsymbol{\mathrm{q}}}v_{\boldsymbol{\mathrm{q}}}\left(1-{\left\langle {{\alpha }^{\dagger }_{\boldsymbol{\mathrm{-}}\boldsymbol{\mathrm{q}},-\sigma }\alpha }_{\boldsymbol{\mathrm{-}}\boldsymbol{\mathrm{q}},-\sigma }\right\rangle }_t-{\left\langle {\alpha }^{\dagger }_{\boldsymbol{\mathrm{q}},\sigma }{\alpha }_{\boldsymbol{\mathrm{q}},\sigma }\right\rangle }_t\right)
\nonumber \\ &=&
-\frac{1}{2}\frac{B_{\boldsymbol{\mathrm{q}}}}{{\lambda }_{\boldsymbol{\mathrm{q}}}}{\mathrm{tanh} \left(\frac{\beta {\lambda }_{\boldsymbol{\mathrm{q}}}}{2}\right)\ }\end{eqnarray} 
As for the derivation of Eq.~\eqref{A51},
we can similarly obtain
\begin{eqnarray}
\label{A52}
{\left\langle {c_{\mathrm{-}\boldsymbol{\mathrm{k}}\mathrm{\ },-\sigma }c}_{\boldsymbol{\mathrm{k}},\sigma }\right\rangle }_t=-\frac{1}{2}\frac{B_{\boldsymbol{\mathrm{k}}}}{{\lambda }_{\boldsymbol{\mathrm{k}}}}{\mathrm{tanh} \left(\frac{\beta {\lambda }_{\boldsymbol{\mathrm{k}}}}{2}\right)\ }.
\end{eqnarray}
Using Eq.~\eqref{A51} and Eq.~\eqref{A52},
the expectation value of ${\left\langle h_2\right\rangle }_t$ is now obtained,
\begin{eqnarray}
\label{A53}
{\left\langle h_2\right\rangle }_t=-\frac{1}{8N}\sum_{\sigma ,\boldsymbol{\mathrm{k}}}{\sum_{\boldsymbol{\mathrm{q}}}{\left(V_{\boldsymbol{\mathrm{q}}\boldsymbol{-}\mathrm{k}}-\mathrm{2}U\right)\frac{B_{\boldsymbol{\mathrm{q}}}B_{\boldsymbol{\mathrm{k}}}}{{\lambda }_{\boldsymbol{\mathrm{q}}}{\lambda }_{\boldsymbol{\mathrm{k}}}}{\mathrm{tanh} \left(\frac{\beta {\lambda }_{\boldsymbol{\mathrm{q}}}}{2}\right)\ }{\mathrm{tanh} \left(\frac{\beta {\lambda }_{\boldsymbol{\mathrm{k}}}}{2}\right)\ }}}.\nonumber \\
\end{eqnarray} 
The $G_{\boldsymbol{\mathrm{q}}}$ term that enters ${\left\langle h_3\right\rangle }_t$ has $d$-wave symmetry and is physically significant pertaining to the pseudo-gap \cite{Lee:JSNM09}. However, as ${\left\langle h_3\right\rangle }_t$ does not contribute to the process of minimization of $F_v$ with respect to $B_{\boldsymbol{\mathrm{k}}}$ for deriving superconducting gap equation, we leave as it is without explicitly evaluating it. It is relabeled as ${\left\langle G\ \mathrm{term}\right\rangle }_t$,
\begin{eqnarray}
\label{A54}
{\left\langle h_3\right\rangle }_t\equiv {\left\langle G\ \mathrm{term}\right\rangle }_t.\end{eqnarray} 
Putting the final expression of ${\left\langle h_1\right\rangle }_t,{\left\langle h_2\right\rangle }_t$ and ${\left\langle h_3\right\rangle }_t$ together, 
\begin{eqnarray}
\label{A55}
{\left\langle H\right\rangle }_t&=&\sum_{\boldsymbol{\mathrm{k}},\sigma }{\left({\epsilon }_{\boldsymbol{\mathrm{k}}}-\mu \right){\left\langle c^{\dagger }_{\boldsymbol{\mathrm{k}},\sigma }c_{\boldsymbol{\mathrm{k}},\sigma }\right\rangle }_t}
\nonumber \\ &&
-\frac{1}{8N}\sum_{\sigma ,\boldsymbol{\mathrm{k}}}{\sum_{\boldsymbol{\mathrm{q}}}{\left(V_{\boldsymbol{\mathrm{q}}\boldsymbol{-}\mathrm{k}}-\mathrm{2}U\right)\frac{B_{\boldsymbol{\mathrm{q}}}B_{\boldsymbol{\mathrm{k}}}}{{\lambda }_{\boldsymbol{\mathrm{q}}}{\lambda }_{\boldsymbol{\mathrm{k}}}}{\mathrm{tanh} \left(\frac{\beta {\lambda }_{\boldsymbol{\mathrm{q}}}}{2}\right)\ }{\mathrm{tanh} \left(\frac{\beta {\lambda }_{\boldsymbol{\mathrm{k}}}}{2}\right)\ }}}
\nonumber \\ &&
+{\left\langle G\ \mathrm{term}\right\rangle }_t\end{eqnarray} 

\noindent {\bf Derivation of ${\left\langle {\boldsymbol{H}}_{\boldsymbol{t}}\right\rangle }_{\boldsymbol{t}}$}:\\
Referring to Eq.~\eqref{A10},
\begin{eqnarray}
\label{A56}
{\left\langle H_t\right\rangle }_t&=&\sum_{\boldsymbol{\mathrm{k}},\sigma }{\left({\epsilon }_{\boldsymbol{\mathrm{k}}}-\mu -A_{\boldsymbol{\mathrm{k}}}\right){\left\langle c^{\dagger }_{\boldsymbol{\mathrm{k}},\sigma }c_{\boldsymbol{\mathrm{k}},\sigma }\right\rangle }_t}
\nonumber \\ && +
\frac{1}{2}\sum_{\boldsymbol{\mathrm{k}},\sigma }
\left(B_{\boldsymbol{\mathrm{k}}}{\left\langle c^{\dagger }_{\boldsymbol{\mathrm{k}},\sigma }c^{\dagger }_{-\boldsymbol{\mathrm{k}},-\sigma }\right\rangle }_t
+B^*_{\boldsymbol{\mathrm{k}}}{\left\langle c_{-\boldsymbol{\mathrm{k}},-\sigma }c_{\boldsymbol{\mathrm{k}},\sigma }\right\rangle }_t\right)
.\end{eqnarray} 
Using Eq.~\eqref{A18}, Eq.~\eqref{A51}, Eq.~\eqref{A52}
we have
\begin{eqnarray}
\label{A57}
{\left\langle H_t\right\rangle }_t=\sum_{\boldsymbol{\mathrm{k}},\sigma }{\left({\epsilon }_{\boldsymbol{\mathrm{k}}}-\mu -A_{\boldsymbol{\mathrm{k}}}\right){\left\langle c^{\dagger }_{\boldsymbol{\mathrm{k}},\sigma }c_{\boldsymbol{\mathrm{k}},\sigma }\right\rangle }_t}-\frac{1}{2}\sum_{\boldsymbol{\mathrm{k}},\sigma }{\frac{B^2_{\boldsymbol{\mathrm{k}}}}{{\lambda }_{\boldsymbol{\mathrm{k}}}}{\mathrm{tanh} \left(\frac{\beta {\lambda }_{\boldsymbol{\mathrm{k}}}}{2}\right).\ }}\end{eqnarray} 
This completes the derivation of ${\left\langle H_t\right\rangle }_t.$

\noindent {\bf Derivation of ${\boldsymbol{F}}_{\boldsymbol{v}}$}:\\
Finally, putting everything together, i.e., Eq.~\eqref{A42}
(for $F_t$), Eq.~\eqref{A55}
(for ${\left\langle H\right\rangle }_t$) Eq.~\eqref{A57}
(for ${\left\langle H_t\right\rangle }_t$), the variational free energy is written as
\begin{eqnarray}
\label{A58}
F_v&=&F_t+{\left\langle H\right\rangle }_t-{\left\langle H_t\right\rangle }_t+\mu N_e
\nonumber \\ &=&
\underbrace{-\frac{1}{\beta }\sum_{\boldsymbol{\mathrm{k}},\sigma }{{\mathrm{ln\ } \left[\mathrm{2cosh}\left(\frac{\beta {\lambda }_{\boldsymbol{\mathrm{k}}}}{2}\right)\right]\ }}}_{F_{v,I}}\nonumber \\ &&
\underbrace{-\frac{1}{8N}\sum_{\sigma ,\boldsymbol{\mathrm{k}}}{\sum_{\boldsymbol{\mathrm{q}}}{\left(V_{\boldsymbol{\mathrm{q}}\boldsymbol{-}\mathrm{k}}-\mathrm{2}U\right)\frac{B_{\boldsymbol{\mathrm{q}}}B_{\boldsymbol{\mathrm{k}}}}{{\lambda }_{\boldsymbol{\mathrm{q}}}{\lambda }_{\boldsymbol{\mathrm{k}}}}{\mathrm{tanh} \left(\frac{\beta {\lambda }_{\boldsymbol{\mathrm{q}}}}{2}\right)\ }{\mathrm{tanh} \left(\frac{\beta {\lambda }_{\boldsymbol{\mathrm{k}}}}{2}\right)\ }}}}_{F_{v,II}}
\nonumber \\ &&
+\underbrace{\frac{1}{2}\sum_{\boldsymbol{\mathrm{k}},\sigma }{\frac{B^2_{\boldsymbol{\mathrm{k}}}}{{\lambda }_{\boldsymbol{\mathrm{k}}}}{\mathrm{tanh} \left(\frac{\beta {\lambda }_{\boldsymbol{\mathrm{k}}}}{2}\right)\ }}}_{F_{v,III}}
\nonumber \\ &&
\mathop{+\underbrace{{\left\langle G\ \mathrm{term}\right\rangle }_t+\sum_{\boldsymbol{\mathrm{k}}}{{\epsilon }_{\boldsymbol{\mathrm{k}}}}+\mu N\left(1-n\right)+\sum_{\boldsymbol{\mathrm{k}},\sigma }{A_{\boldsymbol{\mathrm{k}}}\left({\left\langle c^{\dagger }_{\boldsymbol{\mathrm{k}},\sigma }c_{\boldsymbol{\mathrm{k}},\sigma }\right\rangle }_t-\frac{1}{2}\right)}}}_{\mathrm{independent\ of\ }{\mathrm{B}}_{\boldsymbol{\mathrm{k}}}}.
\nonumber \\
\end{eqnarray} 
Note that the last three terms in 
\eqref{A58} are independent of $B_{\boldsymbol{\mathrm{k}}}$. Also, we have used the relation $N_e=Nn$, where $n$ denotes concentration of charge carries, defined as
\begin{eqnarray}
\label{A59}
n\equiv \frac{1}{N}\sum_{\boldsymbol{\mathrm{k}},\sigma }{n_{\boldsymbol{\mathrm{k}},\sigma }}=\frac{1}{N}\sum_{\boldsymbol{\mathrm{k}},\sigma }{{\left\langle c^{\dagger }_{\boldsymbol{\mathrm{k}},\sigma }c_{\boldsymbol{\mathrm{k}},\sigma }\right\rangle }_t}.\end{eqnarray} 
From Eq.~\eqref{A48},
we can identify $n_{\boldsymbol{\mathrm{k}},\sigma }\equiv {\left\langle c^{\dagger }_{\boldsymbol{\mathrm{k}},\sigma }c_{\boldsymbol{\mathrm{k}},\sigma }\right\rangle }_t$ in terms of the variables $E_{\boldsymbol{\mathrm{k}}}$ and ${\lambda }_{\boldsymbol{\mathrm{k}}}$, namely,
\begin{eqnarray}
\label{A60}
n_{\boldsymbol{\mathrm{k}},\sigma }\mathrm{=}\frac{1}{2}\left[1-\frac{E_{\boldsymbol{\mathrm{k}}}}{{\lambda }_{\boldsymbol{\mathrm{k}}}\ }{\mathrm{tanh} \left(\frac{\beta {\lambda }_{\boldsymbol{\mathrm{k}}}}{2}\right)\ }\right].\end{eqnarray} 
In view of Eq.~\eqref{A59} and Eq.~\eqref{A60},
we have 
\begin{eqnarray}
\label{A61}
nN=\sum_{\boldsymbol{\mathrm{k}},\sigma }{{\left\langle c^{\dagger }_{\boldsymbol{\mathrm{k}},\sigma }c_{\boldsymbol{\mathrm{k}},\sigma }\right\rangle }_t}=N-\sum_{\boldsymbol{\mathrm{k}}}{\frac{E_{\boldsymbol{\mathrm{k}}}}{{\lambda }_{\boldsymbol{\mathrm{k}}}\ }{\mathrm{tanh} \left(\frac{\beta {\lambda }_{\boldsymbol{\mathrm{k}}}}{2}\right)\ }}.\end{eqnarray} 
Note that in the RHS of Eq.~\eqref{A61},
the spin degree of freedom has already been summed over.

\subsection{
Minimization of ${\boldsymbol{F}}_{\boldsymbol{v}}$ with respect to $B_{\boldsymbol{\mathrm{k}}}$ to derive the recurrent gap equation.
\label{asecA4}
}
$F_v$ in 
Eq.~\eqref{A58} can be minimized term by term. Differentiating the first term in $F_v$ by making use of the relation 
\begin{eqnarray}
\label{A62}
\frac{\partial {\lambda }_{\boldsymbol{\mathrm{k}}}}{\partial B_{\boldsymbol{\mathrm{k}}}}=\frac{B_{\boldsymbol{\mathrm{k}}}}{{\lambda }_{\boldsymbol{\mathrm{k}}}},\end{eqnarray} 
we obtain 
\begin{eqnarray}
\label{A63}
\frac{\partial F_{v,I}}{\partial B_{\boldsymbol{\mathrm{k}}}}=-\frac{1}{2}\sum_{\boldsymbol{\mathrm{k}},\sigma }{B_{\boldsymbol{\mathrm{k}}}T_{\boldsymbol{\mathrm{k}}}},\end{eqnarray} 
where
\begin{eqnarray}
\label{A64}
T_{\boldsymbol{\mathrm{k}}}\equiv \frac{\mathrm{tanh}\left(\frac{\beta {\lambda }_{\boldsymbol{\mathrm{k}}}}{2}\right)}{{\lambda }_{\boldsymbol{\mathrm{k}}}}.\end{eqnarray} 
Differentiating the second term in $F_v$ results in   
\begin{eqnarray}
\label{A65}
\frac{\partial F_{v,II}}{\partial B_{\boldsymbol{\mathrm{k}}}}=-\frac{1}{8N}\sum_{\boldsymbol{\mathrm{k}},\sigma }{\sum_{\boldsymbol{\mathrm{q}}}{\left(V_{\boldsymbol{\mathrm{q}}\boldsymbol{-}\boldsymbol{\mathrm{k}}}-\mathrm{2}U\right)B_{\boldsymbol{\mathrm{q}}}{\mathrm{T}}_{\boldsymbol{\mathrm{q}}}\left[{\mathrm{T}}_{\boldsymbol{\mathrm{k}}}+B_{\boldsymbol{\mathrm{k}}}\frac{\partial {\mathrm{T}}_{\boldsymbol{\mathrm{k}}}}{\partial B_{\boldsymbol{\mathrm{k}}}}\right]}}.\end{eqnarray} 
Differentiating the third term in $F_v$ results in
\begin{eqnarray}
\label{A66}
\frac{\partial F_{v,III}}{\partial B_{\boldsymbol{\mathrm{k}}}}=\sum_{\boldsymbol{\mathrm{k}},\sigma }{B_{\boldsymbol{\mathrm{k}}}\left({\mathrm{T}}_{\boldsymbol{\mathrm{k}}}+\frac{1}{2}B_{\boldsymbol{\mathrm{k}}}\frac{\partial {\mathrm{T}}_{\boldsymbol{\mathrm{k}}}}{\partial B_{\boldsymbol{\mathrm{k}}}}\right)}.\end{eqnarray} 
Putting 
\eqref{A63}, \eqref{A64}, \eqref{A65}, \eqref{A66}
together into 
\eqref{A58},
\begin{eqnarray}
\label{A67}
\frac{\partial F_v}{\partial B_{\boldsymbol{\mathrm{k}}}}&=&\left[\sum_{\boldsymbol{\mathrm{k}},\sigma }{\frac{1}{2}B_{\boldsymbol{\mathrm{k}}}}-\frac{1}{8N}\sum_{\boldsymbol{\mathrm{k}},\sigma }{\sum_{\boldsymbol{\mathrm{q}}}{\left(V_{\boldsymbol{\mathrm{q}}\boldsymbol{\mathrm{-}}\boldsymbol{\mathrm{k}}}-\mathrm{2}U\right)B_{\boldsymbol{\mathrm{q}}}{\mathrm{T}}_{\boldsymbol{\mathrm{q}}}}}\right]\left({\mathrm{T}}_{\boldsymbol{\mathrm{k}}}+B_{\boldsymbol{\mathrm{k}}}\frac{\partial {\mathrm{T}}_{\boldsymbol{\mathrm{k}}}}{\partial B_{\boldsymbol{\mathrm{k}}}}\right)=0
\nonumber \\ 
\Rightarrow B_{\boldsymbol{\mathrm{k}}}&=&\frac{1}{4N}\sum_{\boldsymbol{\mathrm{q}}}{\left(V_{\boldsymbol{\mathrm{k}}\boldsymbol{\mathrm{-}}\boldsymbol{\mathrm{q}}}-\mathrm{2}U\right)B_{\boldsymbol{\mathrm{q}}}\frac{\mathrm{tanh}\left(\frac{\beta {\lambda }_{\boldsymbol{\mathrm{k}}}}{2}\right)}{{\lambda }_{\boldsymbol{\mathrm{k}}}}.}\end{eqnarray} 
 
\subsection{
Derivation of and 
Eq.~\eqref{eq53}
and 
Eq.~\eqref{eq54}
\label{asecA5}
}
In linear algebra, a homogeneous system of Eq.~\eqref{eq47}
has infinitely many non-trivial solutions if the determinant vanishes, or equivalently, the matrix $M$ in 
Eq.~\eqref{eq47} is a rank 1 matrix. The non-trivial solutions $\left( \begin{array}{c}
{\mathrm{\Delta }}_0 \\ 
{\mathrm{\Delta }}_{\eta } \end{array}
\right)$ has only one linearly independent arbitrary variable, which we will choose as ${\mathrm{\Delta }}_0$. ${\mathrm{\Delta }}_{\eta }$is parametrized in terms of ${\mathrm{\Delta }}_0$ via $r_{\eta }=\frac{{\mathrm{\Delta }}_{\eta }}{{\mathrm{\Delta }}_0}$. If the root for $D_{\boldsymbol{\mathrm{k}}}\left(\beta \right)=\left|M\right|=T_1T_{2\boldsymbol{\mathrm{k}}}+T_3T_{4\boldsymbol{\mathrm{k}}}\ =0$ [i.e., Eq.~\eqref{eq49}]
exists, the matrix $M\ $is reduced into row-reduced echelon form (which is always possible because it is a rank 1 matrix with vanishing determinant),
\begin{eqnarray}M\leadstoext \left( \begin{array}{cc}
1 & \frac{T_3}{T_1} \\ 
0 & 0 \end{array}
\right).\end{eqnarray} 
Thus, 
\begin{eqnarray}\left( \begin{array}{cc}
1 & \frac{T_3}{T_1} \\ 
0 & 0 \end{array}
\right)\left( \begin{array}{c}
{\mathrm{\Delta }}_0 \\ 
{\mathrm{\Delta }}_{\eta } \end{array}
\right)=\left( \begin{array}{c}
0 \\ 
0 \end{array}
\right)\Rightarrow r_{\eta }=\frac{{\mathrm{\Delta }}_{\eta }}{{\mathrm{\Delta }}_0}=-\frac{T_1}{T_3}=-\frac{1+2U\sum_{\boldsymbol{\mathrm{q}}}{F_{\boldsymbol{\mathrm{q}}}}}{2U\sum_{\boldsymbol{\mathrm{q}}}{{\eta }_{\boldsymbol{\mathrm{q}}}F_{\boldsymbol{\mathrm{q}}}}}.\end{eqnarray} 
The numerical value of the gap ${\mathrm{\Delta }}_{\boldsymbol{\mathrm{k}}}$\textbf{ }at k-point $\boldsymbol{\mathrm{k}}$ is given as per
\begin{eqnarray}{\mathrm{\Delta }}_{\boldsymbol{\mathrm{k}}}={\mathrm{\Delta }}_0+{\mathrm{\Delta }}_{\eta }{\eta }_{\boldsymbol{\mathrm{k}}}={\mathrm{\Delta }}_0\left(1+r_{\eta }{\eta }_{\boldsymbol{\mathrm{k}}}\right).\end{eqnarray} 
This completes the derivation of Eq.~\eqref{eq53} and Eq.~\eqref{eq54}.

\end{document}